\documentclass[11pt,fleqn]{article}

\usepackage{a4}
\usepackage{amstext}
\usepackage{amsfonts}
\usepackage{amssymb}
\usepackage{amsmath}
\usepackage{color}
\usepackage{subfig}
\usepackage{epsfig}
\usepackage{booktabs}
\usepackage{caption}
\usepackage{multirow}
\usepackage{footnote}
\usepackage{dsfont}
\usepackage{longtable}
\usepackage{xcolor,colortbl}

\newcommand{\gtapprox}{\raisebox{-0.5ex}{$\,\stackrel{>}{\scriptstyle\sim}\,$}}
\newcommand{\ltapprox}{\raisebox{-0.5ex}{$\,\stackrel{<}{\scriptstyle\sim}\,$}}

\begin{document}


\begin{center}

{\Large {\bf{} Masses of }$D${\bf{} mesons, }$D_s${\bf{} mesons and charmonium}}

{\Large {\bf{} states from twisted mass lattice QCD}}

\vspace{0.5cm}

\textbf{Martin Kalinowski, Marc Wagner}

Goethe-Universit\"at Frankfurt am Main, Institut f\"ur Theoretische Physik, \\ Max-von-Laue-Stra{\ss}e 1, D-60438 Frankfurt am Main, Germany

\vspace{0.7cm}

\begin{picture}(0,0)%
\includegraphics{Logo.pstex}%
\end{picture}%
\setlength{\unitlength}{4144sp}%
\begingroup\makeatletter\ifx\SetFigFont\undefined%
\gdef\SetFigFont#1#2#3#4#5{%
  \reset@font\fontsize{#1}{#2pt}%
  \fontfamily{#3}\fontseries{#4}\fontshape{#5}%
  \selectfont}%
\fi\endgroup%
\begin{picture}(1800,1835)(1,-996)
\end{picture}%

\vspace{0.4cm}

September 8, 2015

\end{center}

\vspace{0.1cm}

\begin{tabular*}{16cm}{l@{\extracolsep{\fill}}r} \hline \end{tabular*}

\vspace{-0.40cm}
\begin{center} \textbf{Abstract} \end{center}
\vspace{-0.40cm}

We compute masses of $D$ mesons, $D_s$ mesons and charmonium states using Wilson twisted mass lattice QCD. We present results for spin $J = 0,1,2,3$, parity $\mathcal{P} = -,+$ and in case of charmonium also charge conjugation $\mathcal{C} = -,+$. Computations are based on quark-antiquark creation operators and performed at three different unphysically heavy $u/d$ quark masses allowing an extrapolation to the physical $u/d$ quark mass. Within combined statistical and systematic errors, which are around $2 \% \ldots 3 \%$, our results agree with available experimental results. Particular focus is put on the $J^\mathcal{P} = 1^+$ mesons $D_1(2430)$ and $D_1(2420)$. We separate and classify these close-by states according to the total angular momentum of their light degrees of freedom, $j \approx 1/2$ and $j \approx 3/2$. This is a first important step to study decays $B^{(\ast)} \rightarrow D_1 + l + \nu$, for which a long-standing conflict between theory and experiment exists.

\begin{tabular*}{16cm}{l@{\extracolsep{\fill}}r} \hline \end{tabular*}

\thispagestyle{empty}


\newpage

\setcounter{page}{1}

\section{Introduction}

More than fifty $D$ meson, $D_s$ meson and charmonium states have been observed in experiments \cite{PDG}.
Several of them seem to be theoretically well-understood, e.g.\ the $J^\mathcal{P} = 0^-$ pseudoscalar and $J^\mathcal{P} = 1^-$ vector ground states. There are, however, open questions regarding some of the more recently found excitations. A prominent example is the charmonium-like state $X(3872)$ first observed by Belle \cite{Choi:2003ue}, which is most likely not an ordinary quark-antiquark state and whose structure is under debate. The situation is similar for several other charmonium-like $X$ states. In particular the electrically charged states, e.g.\ $X(3900)^\pm$ or $X(4020)^\pm$, cannot be simple $c \bar{c}$ pairs. One rather assumes a four quark structure, e.g.\ a mesonic molecule or a diquark-antidiquark pair. The situation for open charm mesons is related, even though fewer $D$ and $D_s$ states have been observed. For example the positive parity mesons $D_{s0}^\ast(2317)$ and $D_{s1}(2460)$ first reported by BaBar \cite{Aubert:2003fg} and CLEO \cite{Besson:2003cp}, respectively, are unexpectedly light. Again this could be an indication that these states are not just quark-antiquark pairs, but are composed of two quarks and two antiquarks, a scenario at the moment neither established nor ruled out.

There are many interesting and to some extent successful approaches to study $D$ mesons, $D_s$ mesons and charmonium states theoretically, e.g.\ quark models \cite{Ebert:2009ua}, effective theories respecting QCD symmetries \cite{Eshraim:2014eka} or Dyson-Schwinger and Bethe-Salpeter equations \cite{Fischer:2014cfa}, to just name a few. Of course, it would be highly desirable to understand these mesons and their properties starting from first principles, i.e.\ the QCD Lagrangian, without any assumptions, model simplifications or truncations. The corresponding and commonly used method to achieve that goal is lattice QCD, a numerical method to compute QCD observables. In principle, it allows to investigate and to quantify all possible sources of systematic error. Lattice meson spectroscopy is, however, a challenging task, where many problems have currently only partly been solved. On the one hand there are issues concerning lattice QCD in general. For example simulations with physically light $u/d$ quarks are extremely demanding with respect to high performance computing resources. Similarly, to remove discretization errors one has to study the continuum limit, which requires simulations at several different lattice spacings, again a very challenging task with respect to computational resources. On the other hand there are problems specific to lattice hadron spectroscopy. An example is the investigation of states, which can decay into lighter multi-particle states. Such states should theoretically be treated as resonances and not as stable quark-antiquark states, which is technically extremely difficult, even for simple cases, where only a single decay channel exists. Examples are $D_0^\ast(2400)$ and $D_1(2430)$ with quantum numbers $J^P = 0^+$ and $J^P = 1^+$. Similarly, it is very challenging to study mesons, which might have a structure more complicated than a simple quark-antiquark pair, e.g.\ candidates for tetraquarks or hybrid mesons. While there has been a lot of impressive progress regarding lattice hadron spectroscopy within the last couple of years, there is certainly still a lot of room for improvement. Simple states, in particular pseudoscalar and vector ground state mesons, have, meanwhile, been studied very accurately, including simulations at or extrapolations to physically light $u/d$ quark masses and the continuum limit. On the other hand, the majority of studies concerned with parity, radial and orbital excitations are still at a more exploratory stage, i.e.\ have quite often been performed at unphysically heavy quark masses or at a single finite lattice spacing. Recent reviews discussing the status of lattice QCD computations of $D$ and $D_s$ mesons and of charmonium are \cite{Mohler:2015zsa} and \cite{DeTar:2011nn,Prelovsek:2015fra}, respectively.

The most common approach to compute meson masses using lattice QCD is to employ meson creation operators, which are composed of a quark and an antiquark operator, and to extract meson masses from the exponential decay of corresponding correlation functions\footnote{For a basic introduction to lattice hadron spectroscopy cf.\ \cite{Weber:2013eba}.}. This strategy yields accurate and solid results for mesons, which resemble quark-antiquark pairs and which are quite stable, i.e.\ many of the low-lying states in the $D$ meson, $D_s$ meson and charmonium sector. Recent lattice QCD papers following this strategy to compute masses and spectra of $D$ and $D_s$ mesons and of charmonium are \cite{Dudek:2007wv,Dong:2009wk,Burch:2009az,Dudek:2010wm,Mohler:2011ke,Namekawa:2011wt,
Bali:2011dc,Bali:2011rd,Liu:2012ze,Yang:2012mya,Dowdall:2012ab,Bali:2012ua,
Moir:2013ub,Galloway:2014tta,Bali:2015lka}. Rigorous treatments of more complicated mesonic systems like the previously mentioned unstable $D_0^\ast(2400)$ and $D_1(2430)$ mesons or any of the tetraquark candidates, e.g.\ $X(3900)^\pm$ or $X(4020)^\pm$, require more advanced techniques, in particular the implementation of meson creation operators composed of two quark and two antiquark operators and possibly studies of the volume dependence of the masses of corresponding scattering states\footnote{For a basic introduction on how to study resonances using lattice QCD cf.\ \cite{Prelovsek:2011nk}.}. Examples of recent lattice papers exploring and using such techniques to study specific $D$, $D_s$ or charmonium states are \cite{Gong:2011nr,Mohler:2012na,Liu:2012zya,Prelovsek:2013cra,
Prelovsek:2013xba,Mohler:2013rwa,Ikeda:2013vwa,Lang:2014yfa,Prelovsek:2014swa,Guerrieri:2014nxa}.

The goal of this paper is to compute the masses of several low-lying $D$ meson, $D_s$ meson and charmonium states using Wilson twisted mass lattice QCD with 2+1+1 dynamical quark flavors. One of the main advantages of this particular discretization of QCD is automatic $\mathcal{O}(a)$ improvement, i.e.\ discretization errors appear only quadratically in the small lattice spacing $a$ and are, hence, quite small. From a technical point of view we employ a large variety of quark-antiquark meson creation operators and are, hence, able to study total angular momentum $J = 0,1,2,3$, parity $\mathcal{P} = -,+$ and in case of charmonium charge conjugation $\mathcal{C} = -,+$.
Computations are performed at several unphysically heavy $u/d$ quark masses (corresponding pion masses $m_\pi  \approx 276 \, \textrm{MeV} \, , \, 315 \, \textrm{MeV} \, , \, 443 \, \textrm{MeV}$), which allow extrapolations to the physical point. At the moment computations are, however, restricted to a single lattice spacing $a \approx 0.0885 \, \textrm{fm}$, i.e.\ we are currently not able to perform a continuum extrapolation. Nevertheless, by using two different Wilson twisted mass discretizations of the meson creation operators we are able to crudely estimate the magnitude of discretization errors associated with our resulting meson masses. Computations at smaller lattice spacings and corresponding continuum extrapolations are planned for the near future and will be part of an upcoming publication.

As mentioned above some $D$ meson, $D_s$ meson and charmonium states are quite unstable or might have a structure much different from a quark-antiquark pair. Even though we present results for these states in the following, a rigorous treatment might require more advanced techniques, in particular the inclusion of four-quark creation operators as discussed above. We are in the process of developing such techniques using a similar lattice QCD setup \cite{Alexandrou:2012rm,Abdel-Rehim:2014zwa,Berlin:2015faa}. The techniques and results presented in this paper are an important prerequisite for such more advanced computations. Our long-term goal is the computation of the low-lying $D$ and $D_s$ meson and charmonium spectra as fully as possible with all sources of systematic error removed or quantified (in particular computations at physically light $u/d$ quark masses including continuum extrapolations), using quark-antiquark creation operators supplemented, whenever necessary, by four-quark creation operators.

In this work we also study the structure of the two lightest $D$ mesons with $J^\mathcal{P} = 1^+$, $D_1(2430)$ and $D_1(2420)$. These states are similar in mass, but their structure is quite different. One of them has $j \approx 1/2$, while the other has $j \approx 3/2$, where $j$ denotes the total angular momentum of the light quark and of gluons. We demonstrate, how to resolve and classify both states from computations based on a single $J^\mathcal{P} = 1^+$ correlation matrix. The eigenvector components obtained by solving a generalized eigenvalue problem provide linear combinations of meson creation operators suited to specifically excite $D_1(2430)$ and $D_1(2420)$. This is an important first step to study decays $B^{(\ast)} \rightarrow D^{\ast \ast} + l + \nu$, for which there is a persistent conflict between theory and experiment regarding branching ratios (cf.\ \cite{Bigi:2007qp} for a detailed discussion). Our techniques and results can be used to extend recent lattice QCD computations of these decays, where $D^{\ast \ast}$ has been restricted to $J^\mathcal{P} = 0^+, 2^+$, but did not include $D_1(2430)$ and $D_1(2420)$ \cite{Atoui:2013sca,Atoui:2013ksa}.

Parts of this work have been presented at recent conferences \cite{Kalinowski:2012re,Kalinowski:2013wsa,Wagner:2013laa}.

This paper is structured as follows. In section~\ref{sec.setup} we introduce the 2+1+1 flavor Wilson twisted mass lattice setup. In section~\ref{sec.op} and section~\ref{sec.correl.matrices} we discuss the technical aspects of our computations, in particular the employed meson creation operators, their corresponding quantum numbers and how we compute correlation matrices and meson masses. In section~\ref{sec.results} we present our results, first for $D$ and $D_s$ mesons in section~\ref{opencharm.A1}, then for charmonium in section~\ref{charmonium}. These results are summarized in plots and tables in section~\ref{SEC669}, where we also give a brief outlook.



\section{\label{sec.setup}Lattice QCD setup}


\subsection{\label{SEC432}Gauge link ensembles, sea quarks}

We use gauge link configurations generated with 2+1+1 dynamical quark flavors by the European Twisted Mass Collaboration (ETMC) \cite{Baron:2008xa,Baron:2009zq,Baron:2010bv,Baron:2011sf}. The gluonic action is the Iwasaki gauge action \cite{Iwasaki:1985we}. For the light degenerate $(u,d)$ quark doublet the standard Wilson twisted mass action
\begin{eqnarray}
\label{EQN001} S_{\textrm{light}}[\chi^{(l)},\bar{\chi}^{(l)},U] \ \ = \ \ \sum_x \bar{\chi}^{(l)}(x) \Big(D_\textrm{W}(m_0) + i \mu \gamma_5 \tau_3\Big) \chi^{(l)}(x)
\end{eqnarray}
has been used \cite{Frezzotti:2000nk}, for the heavy $(c,s)$ sea quark doublet the Wilson twisted mass formulation for non-degenerate quarks
\begin{eqnarray}
\label{EQN002} S_{\textrm{heavy}}[\chi^{(h)},\bar{\chi}^{(h)},U] \ \ = \ \ \sum_x \bar{\chi}^{(h)}(x) \Big(D_\textrm{W}(m_0) + i \mu_\sigma \gamma_5 \tau_1 + \mu_\delta \tau_3\Big) \chi^{(h)}(x)
\end{eqnarray}
\cite{Frezzotti:2003xj}. $D_\mathrm{W}$ denotes the Wilson Dirac operator,
\begin{eqnarray}
\label{EQN302} D_\mathrm{W}(m_0) \ \ = \ \ \frac{1}{2} \Big(\gamma_\mu \Big(\nabla_\mu + \nabla^\ast_\mu\Big) - \nabla^\ast_\mu \nabla_\mu\Big) + m_0 ,
\end{eqnarray}
$\chi^{(l)} = (\chi^{(u)},\chi^{(d)})$ and $\chi^{(h)} = (\chi^{(c)},\chi^{(s)})$ are the quark fields in the so-called twisted basis and $\tau_1$ and $\tau_3$ denote the first and third Pauli matrix acting in flavor space. At maximal twist physical quantities, e.g.\ meson masses, are automatically $\mathcal{O}(a)$ improved \cite{Frezzotti:2003ni}. The tuning has been done by adjusting $m_0$ such that the PCAC quark mass in the light quark sector vanishes (cf.\ \cite{Baron:2010bv} for details). For a review on Wilson twisted mass lattice QCD we refer to \cite{Shindler:2007vp}.

In this work we use three ensembles with gauge coupling $\beta = 1.90$, which amounts to a lattice spacing $a \approx 0.0885 \, \textrm{fm}$ (scale setting via the pion mass and the pion decay constant \cite{Carrasco:2014cwa}). The ensembles differ in the unphysically heavy $u/d$ quark mass $\mu = 0.0030 \, , \, 0.0040 \, , \, 0.0080$ corresponding to $m_\pi  \approx 276 \, \textrm{MeV} \, , \, 315 \, \textrm{MeV} \, , \, 443 \, \textrm{MeV}$. The $s$ and the $c$ quark masses are represented by $\mu_\sigma = 0.150$ and $\mu_\delta = 0.190$. These values have been chosen such that the lattice QCD results for $2 m_K^2 - m_\pi^2$ and for $m_D$, quantities, which depend only weakly on the light $u/d$ quark mass, are close to the corresponding physical values \cite{Baron:2010bv,Baron:2010th,Baron:2010vp}. Details of these gauge link ensembles are collected in Table~\ref{tab.ensembles}.

\begin{table}[htb]
\centering
\begin{tabular}{|c|c|c|c|c|c|c|c|c|}
      \hline
      \multirow{2}*{ensemble} & \multirow{2}*{$\beta$} & \multirow{2}*{$(L/a)^3 \times T/a$} & \multirow{2}*{$\mu$} & \multirow{2}*{$\mu_\sigma$} & \multirow{2}*{$\mu_\delta$} & $a$ & $m_\pi$ & \# of \\
      & & & & & & $(\textrm{fm})$ & $(\textrm{MeV})$ & configurations \\
      \hline
      A30.32 & $1.90$ & $32^3 \times 64$ & $0.0030$ & $0.150$ & $0.190$ & $0.0885$  & $276$ & $1200$ \\
      A40.32 &	    & $32^3 \times 64$ & $0.0040$ &         &         & 	    & $315$ & $\phantom{0}800$\\
      A80.24 &	    & $24^3 \times 48$ & $0.0080$ &         &         &	    & $443$ & $1700$ \\
      \hline
\end{tabular}
\caption{\label{tab.ensembles}Gauge link ensembles ($(L/a)^3 \times T/a$: number of lattice sites; \# of configurations: number of gauge link configurations used).}
\end{table}


\subsection{\label{SEC586}Valence quarks}

For the light degenerate $(u,d)$ valence quark doublet we use the same action, which was used to simulate the corresponding sea quarks, i.e.\ the action (\ref{EQN001}).

For the heavy $s$ and $c$ valence quarks we use twisted mass doublets of degenerate quarks, i.e.\ a different discretization as for the corresponding sea quarks. We use the action (\ref{EQN001}) with the replacements $\chi^{(l)} \rightarrow \chi^{(s)} = (\chi^{(s^+)} , \chi^{(s^-)})$, $\mu \rightarrow \mu_s$ and $\chi^{(l)} \rightarrow \chi^{(c)} = (\chi^{(c^+)} , \chi^{(c^-)})$, $\mu \rightarrow \mu_c$, respectively. We do this, to avoid mixing of strange and charm quarks, which inevitably takes place in a unitary non-degenerate Wilson twisted mass setup, and which is particularly problematic for observables containing charm quarks, e.g.\ masses of $D$ and $D_s$ mesons and of charmonium (cf.\ \cite{Baron:2010th,Baron:2010vp} for a detailed discussion of these problems).

The degenerate valence doublets allow two realizations for strange as well as for charm quarks, either with a twisted mass term $+i \mu_{s,c} \gamma_5$ (i.e.\ $\chi^{(s^+)}$ or $\chi^{(c^+)}$) or $-i \mu_{s,c} \gamma_5$ (i.e.\ $\chi^{(s^-)}$ or $\chi^{(c^-)}$). For a quark-antiquark meson creation operator, e.g.\ $\bar{\chi}^{(1)} \gamma_5 \chi^{(2)}$, the sign combinations $(+,-)$ and $(-,+)$ for the antiquark $\bar{\chi}^{(1)}$ and the quark $\chi^{(2)}$ are related by symmetry, i.e.\ the corresponding correlation functions are identical. These correlation functions differ, however, from their counterparts with sign combinations $(+,+)$ and $(-,-)$ due to different discretization errors. In section~\ref{sec.results} we will show for each computed meson mass both the $(+,-) \equiv (-,+)$ and the $(+,+) \equiv (-,-)$ result. The mass differences are $\mathcal{O}(a^2)$, because of automatic $\mathcal{O}(a)$ improvement at maximal twist. These differences provide a first estimate of the magnitude of discretization errors at our currently used lattice spacing $a \approx 0.0885 \, \textrm{fm}$.

The tuning of the valence quark masses $\mu_s$ and $\mu_c$ is discussed in section~\ref{subsec.massdetermination}.



\section{\label{sec.op}Meson creation operators and trial states}


\subsection{Meson creation operators in the continuum}

In the continuum a quark-antiquark operator creating a mesonic trial state with definite quantum numbers $J^{\mathcal{P}\mathcal{C}}$ (total angular momentum $J$, parity $\mathcal{P}$, charge conjugation $\mathcal{C}$), when applied to the vacuum $| \Omega \rangle$, is
\begin{eqnarray}
\label{EQN101} O_{\Gamma,\bar{\psi}^{(1)} \psi^{(2)}}^\textrm{physical} \ \ \equiv \ \ \frac{1}{\sqrt{V}} \int d^3r \, \bar{\psi}^{(1)}(\mathbf{r}) \int_{|\Delta \mathbf{r}| = R} d^3\Delta r \, U(\mathbf{r};\mathbf{r} + \Delta \mathbf{r}) \Gamma(\Delta \mathbf{r}) \psi^{(2)}(\mathbf{r} + \Delta \mathbf{r}) .
\end{eqnarray}
$(1/\sqrt{V}) \int d^3r$ projects to vanishing total momentum ($V$ is the spatial volume), i.e.\ realizes a meson at rest. To allow for orbital angular momentum between the antiquark $\bar{\psi}^{(1)}$ and the quark $\psi^{(2)}$, they are spatially separated. $\int_{|\Delta \mathbf{r}| = R} d^3\Delta r$ denotes an integration over a sphere of radius $R$, which is the distance between the antiquark and the quark. $\Gamma(\Delta \mathbf{r})$ is a suitable combination of spherical harmonics and $\gamma$ matrices (cf.\ Table~\ref{tab.operators}, column ``$\Gamma(\mathbf{n})$, pb''), which combines orbital angular momentum and the two quark spins to total angular momentum $J$ and determines parity $\mathcal{P}$ and in case of identical quark flavors charge conjugation $\mathcal{C}$. $U(\mathbf{r};\mathbf{r} + \Delta \mathbf{r})$ is a straight gluonic parallel transporter connecting the antiquark and the quark in a gauge invariant way. For $D$ mesons e.g.\ $\bar{\psi}^{(1)} \psi^{(2)} = \bar{u} c$, for $D_s$ mesons e.g.\ $\bar{\psi}^{(1)} \psi^{(2)} = \bar{s} c$ and for charmonium $\bar{\psi}^{(1)} \psi^{(2)} = \bar{c} c$.

\begin{table}[p]
\centering
\begin{tabular}{|c|c|c|c|c|c|c|c|c|c|c|}
\hline
\multicolumn{1}{|c|}{}&  \multicolumn{3}{c|}{continuum}        &  \multicolumn{4}{c|}{twisted mass lattice QCD} \\
\cline{2-8} 
 \multicolumn{1}{|c|}{index}&  \multicolumn{1}{ c|}{\multirow{2}{*}{$\Gamma(\mathbf{n})$, pb}}        &   \multicolumn{1}{c|}{\multirow{2}{*}{$J$}}       &   \multicolumn{1}{c|}{\multirow{2}{*}{$\mathcal{P}\mathcal{C}$}} &  \multicolumn{1}{c|}{\multirow{2}{*}{tb, $(\pm,\mp)$}} &  \multicolumn{1}{c|}{\multirow{2}{*}{tb, $(\pm,\pm)$}} &\multicolumn{2}{c|}{\multirow{2}{*}{$\mathrm{O}^S \otimes  \mathrm{O}^L \rightarrow \mathrm{O}^J$}} \\
 \multicolumn{1}{|c|}{}&                                                                        &                                                    &                                                                  &                                                        &                                                        &\multicolumn{2}{c|}{}\\
\hline
\hline
1&$\gamma_5$					&\multirow{8}{*}{0}&$- +$&pb&$\pm i\gamma_5\times$&\multirow{4}{*}{$A_1\otimes A_1$ }&\multirow{8}{*}{$A_1$}  \\  
2&$\gamma_0\gamma_5$				&&$- +$&$\pm i\gamma_5\times$&pb&&  \\  
3&$\mathds{1}$					&&$+ +$&pb&$\pm i\gamma_5\times$&&  \\  
4&$\gamma_0$					&&$+ -$&$\pm i\gamma_5\times$&pb&&  \\ 
\cline{2-2}\cline{4-7} 
5&$        \gamma_5\gamma_j\mathbf{n}_j$&&$- -$&$\pm i\gamma_5\times$&pb&\multirow{4}{*}{$T_1\otimes T_1$}&  \\  
6&$\gamma_0\gamma_5\gamma_j\mathbf{n}_j$&&$- +$&pb&$\pm i\gamma_5\times$&&  \\  
7&$                \gamma_j\mathbf{n}_j$&&$+ +$&$\pm i\gamma_5\times$&pb&&  \\  
8&$\gamma_0        \gamma_j\mathbf{n}_j$&&$+ +$&pb&$\pm i\gamma_5\times$&&  \\ 
\hline\hline
1&$                \gamma_1$&\multirow{16}{*}{1}&$- -$&$\pm i\gamma_5\times$&pb&\multirow{4}{*}{$T_1\otimes A_1$}&\multirow{16}{*}{$T_1$}  \\
2&$\gamma_0        \gamma_1$&&$- -$&pb&$\pm i\gamma_5\times$&&  \\ 
3&$        \gamma_5\gamma_1$&&$+ +$&$\pm i\gamma_5\times$&pb&&  \\ 
4&$\gamma_0\gamma_5\gamma_1$&&$+ -$&pb&$\pm i\gamma_5\times$&&  \\
\cline{2-2}\cline{4-7} 
5&$\mathbf{n}_1$			&&$- -$&pb&$\pm i\gamma_5\times$&\multirow{4}{*}{$A_1\otimes T_1$}&  \\
6&$\gamma_0\mathbf{n}_1$		&&$- +$&$\pm i\gamma_5\times$&pb&&  \\ 
7&$\gamma_5\mathbf{n}_1$		&&$+ -$&pb&$\pm i\gamma_5\times$&&  \\ 
8&$\gamma_0\gamma_5\mathbf{n}_1$	&&$+ -$&$\pm i\gamma_5\times$&pb&&  \\
\cline{2-2}\cline{4-7} 
 9&$                (\mathbf{n}\times\vec{\gamma})_1$&&$+ +$&$\pm i\gamma_5\times$&pb&\multirow{4}{*}{$T_1\otimes T_1$}&  \\
10&$\gamma_0        (\mathbf{n}\times\vec{\gamma})_1$&&$+ +$&pb&$\pm i\gamma_5\times$&&  \\ 
11&$\gamma_5        (\mathbf{n}\times\vec{\gamma})_1$&&$- -$&$\pm i\gamma_5\times$&pb&&  \\ 
12&$\gamma_0\gamma_5(\mathbf{n}\times\vec{\gamma})_1$&&$- +$&pb&$\pm i\gamma_5\times$&&  \\
\cline{2-2}\cline{4-7} 
13&$                \gamma_1(2\mathbf{n}^2_1-\mathbf{n}^2_2-\mathbf{n}^2_3)$&&$- -$&$\pm i\gamma_5\times$&pb&\multirow{4}{*}{$T_1\otimes E$}&  \\
14&$\gamma_0        \gamma_1(2\mathbf{n}^2_1-\mathbf{n}^2_2-\mathbf{n}^2_3)$&&$- -$&pb&$\pm i\gamma_5\times$&&  \\ 
15&$        \gamma_5\gamma_1(2\mathbf{n}^2_1-\mathbf{n}^2_2-\mathbf{n}^2_3)$&&$+ +$&$\pm i\gamma_5\times$&pb&&  \\ 
16&$\gamma_0\gamma_5\gamma_1(2\mathbf{n}^2_1-\mathbf{n}^2_2-\mathbf{n}^2_3)$&&$+ -$&pb&$\pm i\gamma_5\times$&&  \\
\hline\hline
1&$                (\mathbf{n}_1^2+\mathbf{n}_2^2-2\mathbf{n}_3^2)$&\multirow{8}{*}{2}&$+ +$&pb&$\pm i\gamma_5\times$&\multirow{4}{*}{$A_1\otimes E$}&\multirow{8}{*}{$E$} \\
2&$\gamma_0        (\mathbf{n}_1^2+\mathbf{n}_2^2-2\mathbf{n}_3^2)$&&$+ -$&$\pm i\gamma_5\times$&pb&&  \\ 
3&$        \gamma_5(\mathbf{n}_1^2+\mathbf{n}_2^2-2\mathbf{n}_3^2)$&&$- +$&pb&$\pm i\gamma_5\times$&&  \\ 
4&$\gamma_0\gamma_5(\mathbf{n}_1^2+\mathbf{n}_2^2-2\mathbf{n}_3^2)$&&$- +$&$\pm i\gamma_5\times$&pb&&  \\
\cline{2-2}\cline{4-7}
5&$                (\gamma_1\mathbf{n}_1+\gamma_2\mathbf{n}_2-2\gamma_3\mathbf{n}_3)$&&$+ +$&$\pm i\gamma_5\times$&pb&\multirow{4}{*}{$T_1\otimes T_1$}&  \\
6&$\gamma_0        (\gamma_1\mathbf{n}_1+\gamma_2\mathbf{n}_2-2\gamma_3\mathbf{n}_3)$&&$+ +$&pb&$\pm i\gamma_5\times$&&  \\ 
7&$        \gamma_5(\gamma_1\mathbf{n}_1+\gamma_2\mathbf{n}_2-2\gamma_3\mathbf{n}_3)$&&$- -$&$\pm i\gamma_5\times$&pb&&  \\ 
8&$\gamma_0\gamma_5(\gamma_1\mathbf{n}_1+\gamma_2\mathbf{n}_2-2\gamma_3\mathbf{n}_3)$&&$- +$&pb&$\pm i\gamma_5\times$&&  \\
\hline\hline
1&$                (\gamma_3\mathbf{n}_2+\gamma_2\mathbf{n}_3)$&\multirow{8}{*}{2}&$+ +$&$\pm i\gamma_5\times$&pb&\multirow{4}{*}{$T_1\otimes T_1$}&\multirow{8}{*}{$T_2$}  \\
2&$\gamma_0        (\gamma_3\mathbf{n}_2+\gamma_2\mathbf{n}_3)$&&$+ +$&pb&$\pm i\gamma_5\times$&&  \\ 
3&$        \gamma_5(\gamma_3\mathbf{n}_2+\gamma_2\mathbf{n}_3)$&&$- -$&$\pm i\gamma_5\times$&pb&&  \\ 
4&$\gamma_0\gamma_5(\gamma_3\mathbf{n}_2+\gamma_2\mathbf{n}_3)$&&$- +$&pb&$\pm i\gamma_5\times$&&  \\
\cline{2-2}\cline{4-7}
5&$                \gamma_1(\mathbf{n}_2^2-\mathbf{n}_3^2)$&&$- -$&$\pm i\gamma_5\times$&pb&\multirow{4}{*}{$T_1\otimes E$}&  \\
6&$\gamma_0        \gamma_1(\mathbf{n}_2^2-\mathbf{n}_3^2)$&&$- -$&pb&$\pm i\gamma_5\times$&&  \\ 
7&$        \gamma_5\gamma_1(\mathbf{n}_2^2-\mathbf{n}_3^2)$&&$+ +$&$\pm i\gamma_5\times$&pb&&  \\ 
8&$\gamma_0\gamma_5\gamma_1(\mathbf{n}_2^2-\mathbf{n}_3^2)$&&$+ -$&pb&$\pm i\gamma_5\times$&&  \\ 
\hline
\end{tabular}
\caption{\label{tab.operators}Meson creation operators.}
\end{table}


\subsection{\label{subsec.operators}Meson creation operators in Wilson twisted mass lattice QCD}

Our lattice meson creation operators are of similar form,
\begin{eqnarray}
\label{EQN507} O_{\Gamma,\bar{\chi}^{(1)} \chi^{(2)}}^\textrm{twisted} \ \ \equiv \ \ \frac{1}{\sqrt{V/a^3}} \sum_\mathbf{n} \bar{\chi}^{(1)}(\mathbf{n}) \sum_{\Delta \mathbf{n} = \pm \mathbf{e}_x , \pm \mathbf{e}_y , \pm \mathbf{e}_z} U(\mathbf{n};\mathbf{n} + \Delta \mathbf{n}) \Gamma(\Delta \mathbf{n}) \chi^{(2)}(\mathbf{n} + \Delta \mathbf{n}) ,
\end{eqnarray}
where the integration over a sphere with center at $\mathbf{r}$ has been replaced by the sum over the six neighboring lattice sites of $\mathbf{n}$ and $U(\mathbf{n};\mathbf{n} + \Delta \mathbf{n})$ denotes the link between $\mathbf{n}$ and $\mathbf{n} + \Delta \mathbf{n}$. Moreover, physical basis quark operators $\bar{\psi}^{(1)}$, $\psi^{(2)}$ have been replaced by their twisted basis counterparts $\bar{\chi}^{(1)}$, $\chi^{(2)}$.


\subsubsection{\label{subsubsec.basis}Physical basis and twisted basis}

In the continuum the relation between the physical and the twisted basis is given by the twist rotation
\begin{eqnarray}
\label{EQN589} \psi^{(f)} \ \ = \ \ \exp\Big(i \gamma_5 \tau_3 \omega / 2\Big) \chi^{(f)} \quad , \quad \bar{\psi}^{(f)} \ \ = \ \ \bar{\chi}^{(f)} \exp\Big(i \gamma_5 \tau_3 \omega / 2\Big)
\end{eqnarray}
with the twist angle $\omega$, where $\omega = \pi / 2$ at maximal twist. $\chi^{(f)}$ denotes either the light doublet $\chi^{(l)} = (\chi^{(u)} , \chi^{(d)})$, the strange doublet $\chi^{(s)} = (\chi^{(s^+)} , \chi^{(s^-)})$ or the charm doublet $\chi^{(c)} = (\chi^{(c^+)} , \chi^{(c^-)})$ (cf.\ also section~\ref{SEC586}).

When transforming a twisted basis quark bilinear $\bar{\chi}^{(1)} \Gamma \chi^{(2)}$ as e.g.\ appearing in (\ref{EQN507}) to the physical basis or vice versa, the result depends not only on $\Gamma$, but also on the flavor combination, i.e.\ whether $\bar{\chi}^{(1)}$ and $\chi^{(2)}$ are upper components (twisted mass term $+i \mu \gamma_5$) or lower components (twisted mass term $-i \mu \gamma_5$) of twisted basis doublets. For example
\begin{eqnarray}
\label{EQN767}\bar{\psi}^{(u)} \gamma_5 \psi^{(c^-)} \ \ = \ \ \bar{\chi}^{(u)} \exp\Big(+i \gamma_5 \omega / 2\Big) \gamma_5 \exp\Big(-i \gamma_5 \omega / 2\Big) \chi^{(c^-)} \ \ = \ \ \bar{\chi}^{(u)} \gamma_5 \chi^{(c^-)} ,
\end{eqnarray}
while
\begin{eqnarray}
\nonumber & & \hspace{-0.7cm} \bar{\psi}^{(u)} \gamma_5 \psi^{(c^+)} \ \ = \ \ \bar{\chi}^{(u)} \exp\Big(+i \gamma_5 \omega / 2\Big) \gamma_5 \exp\Big(+i \gamma_5 \omega / 2\Big) \chi^{(c^+)} \ \ = \\
\label{EQN768} & & = \ \ \bar{\chi}^{(u)} \exp\Big(+i \gamma_5 \omega\Big) \gamma_5 \chi^{(c^+)} \ \ \stackrel{\omega = \pi/2}{=} \ \ \bar{\chi}^{(u)} (+i \gamma_5) \gamma_5 \chi^{(c^+)} \ \ = \ \ +i \bar{\chi}^{(u)} \chi^{(c^+)} .
\end{eqnarray}
In the columns ``tb, $(\pm,\mp)$'' and ``tb, $(\pm,\pm)$'' of Table~\ref{tab.operators} we list for all flavor combinations ($+$ and $-$ denote the signs in front of the twisted mass terms for $\bar{\chi}^{(1)}$ and $\chi^{(2)}$) and all $\Gamma$ combinations of our meson creation operators, how physical and twisted basis are related. ``pb'' indicates that the twisted basis $\Gamma$ is the same as the physical basis $\Gamma$ (cf.\ e.g.\ (\ref{EQN767})), while ``$\pm i \gamma_5 \times$'' denotes that the physical basis $\Gamma$ has to be multiplied from the left with $\pm i \gamma_5$ to obtain the corresponding twisted $\Gamma$ (cf.\ e.g.\ (\ref{EQN768})).

At finite lattice spacing the twist rotation only holds for renormalized operators, i.e.\ for bare lattice quark operators the twist rotation (\ref{EQN589}) is only an approximate relation. Nevertheless, it is possible to unambiguously interpret states obtained from correlation functions of twisted basis meson creation operators in terms of QCD quantum numbers as we will explain and demonstrate in section \ref{sec.quantumnumbers}.


\subsubsection{\label{subsubsec.isospin}Isospin, parity and charge conjugation}

Isospin $I$ and parity $\mathcal{P}$ are symmetries of QCD. While in Wilson twisted mass lattice QCD the $z$ component of isospin $I_z$ is still a quantum number, $I$ and $\mathcal{P}$ are broken by $\mathcal{O}(a)$ due to the Wilson term $-\bar{\chi}^{(l)} (a/2) \nabla^\ast_\mu \nabla_\mu \chi^{(l)}$ appearing in the twisted mass actions (\ref{EQN001}) and (\ref{EQN302}). Only a specific combination of both symmetries, light flavor exchange $u \leftrightarrow d$ combined with parity, is still a symmetry. We denote this symmetry by $\mathcal{P}^{(\textrm{tm})}$ acting on the light twisted basis quark doublet $\chi^{(l)} = (\chi^{(u)} , \chi^{(d)})$ according to $\mathcal{P}^{(\textrm{tm})} \chi^{(l)} = \gamma_0 \tau_1 \chi^{(l)}$. Similarly, $\mathcal{P}^{(\textrm{tm})} \chi^{(s)} = \gamma_0 \tau_1 \chi^{(s)}$ and $\mathcal{P}^{(\textrm{tm})} \chi^{(c)} = \gamma_0 \tau_1 \chi^{(c)}$. Note that $[I_z , \mathcal{P}^{(\textrm{tm})}] \neq 0$. In general, it is, therefore, not possible to classify states according to $I_z$ and $\mathcal{P}^{(\textrm{tm})}$ at the same time.

For $D$ mesons we use trial states $O_{\Gamma,\bar{\chi}^{(1)} \chi^{(2)}}^\textrm{twisted} | \Omega \rangle$ with defined $I_z$, e.g.\ $\bar{\chi}^{(1)} \chi^{(2)} = \bar{\chi}^{(d)} \chi^{(c^+)}$ is suited for $D$ mesons with $I_z = +1/2$. There are eight appropriate flavor combinations for $D$ mesons, where the four with opposite signs in front of the twisted mass terms (denoted by $(\pm,\mp)$ throughout the paper),
\begin{eqnarray}
\label{EQN348} \bar{\chi}^{(d)} \chi^{(c^+)} \ \ , \ \ \bar{\chi}^{(u)} \chi^{(c^-)} \ \ , \ \ \bar{\chi}^{(c^-)} \chi^{(u)} \ \ , \ \ \bar{\chi}^{(c^+)} \chi^{(d)} ,
\end{eqnarray}
are related by symmetry and yield identical correlation functions. Similarly the four flavor combinations with identical signs in front of the twisted mass terms (denoted by $(\pm,\pm)$),
\begin{eqnarray}
\label{EQN349} \bar{\chi}^{(u)} \chi^{(c^+)} \ \ , \ \ \bar{\chi}^{(d)} \chi^{(c^-)} \ \ , \ \ \bar{\chi}^{(c^+)} \chi^{(u)} \ \ , \ \ \bar{\chi}^{(c^-)} \chi^{(d)} ,
\end{eqnarray}
also yield identical correlation functions. However, at finite lattice spacing $(\pm,\mp)$ and $(\pm,\pm)$ correlation functions slightly differ. As a consequence $D$ mesons computed on the one hand with a $(\pm,\pm)$ and on the other hand with a $(\pm,\mp)$ flavor combination, but which are otherwise identical, will differ in mass. Due to automatic $\mathcal{O}(a)$ improvement of Wilson twisted mass lattice QCD at maximal twist, this mass splitting will be proportional to $a^2$, i.e.\ is expected to be rather small and will vanish quadratically, when approaching the continuum limit. Even though we consider only a single lattice spacing $a \approx 0.0885 \, \textrm{fm}$ throughout this work, the splitting between $(\pm,\mp)$ and $(\pm,\pm)$ flavor combinations will provide a crude estimate of the magnitude of lattice discretization errors associated with the resulting meson masses.

Since parity is not a symmetry, there is no rigorous separation between $\mathcal{P} = +$ and $\mathcal{P} = -$ states in Wilson twisted mass lattice QCD. Corresponding correlation functions, e.g.\ between the meson creation operators $O_{\gamma_5,\bar{\chi}^{(d)} \chi^{(c^+)}}^\textrm{twisted}$ (continuum quantum numbers $J^{\mathcal{P}} = 0^-$) and $O_{\mathds{1},\bar{\chi}^{(d)} \chi^{(c^+)}}^\textrm{twisted}$ (continuum quantum numbers $J^{\mathcal{P}} = 0^+$), are $\mathcal{O}(a)$, but do not vanish.

Identical considerations apply for $D_s$ mesons, when replacing $(u,d) \rightarrow (s^+,s^-)$.

For charmonium states there is an additional symmetry, charge conjugation $\mathcal{C}$, where $\mathcal{C} \chi^{(f)} = \gamma_0 \gamma_2 (\bar{\chi}^{(f)})^T$. For charmonium creation operators there are two appropriate flavor combinations,
\begin{eqnarray}
\bar{\chi}^{(c^+)} \chi^{(c^+)} \ \ , \ \ \bar{\chi}^{(c^-)} \chi^{(c^-)} ,
\end{eqnarray}
which are again related by symmetry. One might also consider flavor combinations with opposite signs in front of the twisted mass terms,
\begin{eqnarray}
\label{EQN350} \bar{\chi}^{(c^+)} \chi^{(c^-)} \ \ , \ \ \bar{\chi}^{(c^-)} \chi^{(c^+)} .
\end{eqnarray}
In this case $\mathcal{C}$ is not a symmetry, but $\mathcal{C}$ combined with $\mathcal{P}^{(\textrm{tm})}$ denoted by $\mathcal{C} \circ \mathcal{P}^{(\textrm{tm})}$. Note, however, that a rigorous treatment of charmonium states with flavor combinations (\ref{EQN350}) is not possible, because disconnected diagrams are excluded by construction. Since disconnected diagrams are expected to be negligible compared to statistical errors and, therefore, ignored in this work (cf.\ also the discussion in section~\ref{charmonium}), computations using (\ref{EQN350}) are still useful to get a first estimate of lattice discretization errors (as discussed above, the differences between $(\pm,\pm)$ and $(\pm,\mp)$ charmonium masses are proportional to $a^2$).


\subsubsection{Rotational symmetry and total angular momentum}

On a cubic lattice rotational symmetry is reduced to symmetry with respect to cubic rotations. There are only five different irreducible representations of the cubic group $\mathrm{O}$ (labeled by $A_1$, $T_1$, $E$, $T_2$, $A_2$), each corresponding to an infinite number of $\mathrm{SO}(3)$ irreducible representations in the continuum (labeled by a non-negative integer referring to e.g.\ spin, orbital angular momentum or total angular momentum $S,L,J = 0,1,2,\ldots$): \\
\begin{tabular}{ccl}
$A_1$ & $\rightarrow$ & angular momenta $0, 4, \ldots$ \\
$T_1$ & $\rightarrow$ & angular momenta $1, 3, 4, \ldots$ \\
$E$   & $\rightarrow$ & angular momenta $2, 4, \ldots$ \\
$T_2$ & $\rightarrow$ & angular momenta $2, 3, 4, \ldots$ \\
$A_2$ & $\rightarrow$ & angular momenta $3, \ldots$
\end{tabular} \\

The two quark spins of the meson creation operators (\ref{EQN507}) can be combined via $\gamma$ matrices to $S = 0, 1$, which corresponds to the $A_1$ (singlet) and the $T_1$ (triplet) representation. The spin is coupled to orbital angular momentum $L$, where we can access with our choice of summing over six neighboring lattice sites in (\ref{EQN507}) the cubic representations $A_1$, $T_1$ and $E$. Therefore, the total angular momentum $\mathrm{O}$ representations of our lattice meson creation operators are
\begin{eqnarray}
\nonumber & & \hspace{-0.7cm} \underbrace{\Big(A_1 \oplus T_1\Big)}_{\textrm{spin } S} \otimes \underbrace{\Big(A_1 \oplus T_1 \oplus E\Big)}_{\textrm{orbital angular momentum } L} \ \ = \ \ \underbrace{A_1 \oplus A_1 \oplus T_1 \oplus T_1 \oplus T_1 \oplus T_1 \oplus E \oplus E \oplus T_2 \oplus T_2}_{\textrm{total angular momentum } J} . \\
 & &
\end{eqnarray}
The meson creation operators listed in Table~\ref{tab.operators} are sorted and organized according to these $10$ multiplets (cf.\ the column ``$\mathrm{O}^S \otimes \mathrm{O}^L \rightarrow \mathrm{O}^J$'').


\subsection{\label{subsec.smearing}Smearing of gauge links and quark fields}

To enhance the overlap of trial states $O_{\Gamma,\bar{\chi}^{(1)} \chi^{(2)}}^\textrm{twisted} | \Omega \rangle$ to low lying meson states, we use standard smearing techniques. This allows to read off meson masses from the exponential decay of correlation functions at rather small temporal separations, where the signal-to-noise ratio is favorable.


Smearing is done in two steps. First we replace spatial gauge links by their APE smeared counterparts. Then we use Gaussian smearing on the quark fields $\chi^{(l)}$, $\chi^{(s)}$ and $\chi^{(c)}$, which resorts to the APE smeared spatial links. The parameters we have chosen are $N_\textrm{APE} = 10$, $\alpha_\textrm{APE} = 0.5$, $N_\textrm{Gauss} = 30$ and $\kappa_\textrm{Gauss} = 0.5$. This corresponds to a Gaussian width of the smeared quark fields of approximately $2.7 \times a \approx 0.24 \, \textrm{fm}$ (cf.\ \cite{Jansen:2008si} for detailed equations).

Smearing is a symmetric operation with respect to cubic rotations and spatial reflections. Therefore, it does not change the quantum numbers $J$ and $\mathcal{P}$ generated by the corresponding meson creation operators as listed in Table~\ref{tab.operators}.



\section{\label{sec.correl.matrices}Computation and analysis of correlation matrices}


\subsection{\label{sub.computation}Computation of correlation matrices}

For each twisted mass sector characterized by flavor $\bar{\chi}^{(1)} \chi^{(2)}$, the cubic representation $\mathrm{O}^J$ and, in case of charmonium, either $\mathcal{C}$ (for twisted mass signs $(\pm,\pm)$) or $\mathcal{C} \circ \mathcal{P}^{(\textrm{tm})} \quad$\footnote{The $\mathcal{C} \circ \mathcal{P}^{(\textrm{tm})}$ quantum number associated with a twisted basis meson creation operator from Table~\ref{tab.operators}, column ``tb, $(\pm,\mp)$'' is the product of the $\mathcal{P}$ and $\mathcal{C}$ quantum numbers also listed in Table~\ref{tab.operators}, column ``$\mathcal{P C}$''.} (for twisted mass signs $(\pm,\mp)$), we compute temporal correlation matrices of meson creation operators
\begin{eqnarray}
\label{EQN698} C_{\Gamma_j;\Gamma_k;\bar{\chi}^{(1)} \chi^{(2)}}(t) \ \ \equiv \ \ \langle \Omega | \Big(S(O_{\Gamma_j,\bar{\chi}^{(1)} \chi^{(2)}}^\textrm{twisted})\Big)^\dagger(t) \Big(S(O_{\Gamma_k,\bar{\chi}^{(1)} \chi^{(2)}}^\textrm{twisted})\Big)(0) | \Omega \rangle .
\end{eqnarray}
$j$ and $k$ label the rows and columns of a correlation matrix or, equivalently, are indices of the meson creation operators entering a correlation matrix (cf.\ Table~\ref{tab.operators}, column ``index''). $S(\ldots)$ indicates that APE smeared gauge links and Gaussian smeared quark fields are used for the meson creation operators (cf.\ section~\ref{subsec.smearing}). For the computations we use a generalization of the one-end trick, which is explained in detail in appendix~\ref{sec.computation}.

Since parity is only an approximate symmetry in twisted mass lattice QCD, we consider correlation matrices of meson creation operators with both $\mathcal{P} = +$ and $\mathcal{P} = -$.
\begin{itemize}
\item For $D$ and $D_s$ mesons:
\begin{itemize}
\item $A_1$, $E$, $T_2$: $8 \times 8$ correlation matrices.

\item $T_1$: $16 \times 16$ correlation matrix.
\end{itemize}

\item For charmonium and twisted mass signs $(\pm,\pm)$, i.e.\ $\bar{\chi}^{(c^+)} \chi^{(c^+)}$ and $\bar{\chi}^{(c^-)} \chi^{(c^-)}$:
\begin{itemize}
\item 
$A_1$, $E$ and $\mathcal{C} = +$: $6 \times 6$ correlation matrices; \\
$A_1$, $E$ and $\mathcal{C} = -$: $2 \times 2$ correlation matrices.

\item 
$T_1$ and $\mathcal{C} = +$: $6 \times 6$ correlation matrix; \\
$T_1$ and $\mathcal{C} = -$: $10 \times 10$ correlation matrix.

\item 
$T_2$ and $\mathcal{C} = +$: $4 \times 4$ correlation matrix; \\
$T_2$ and $\mathcal{C} = -$: $4 \times 4$ correlation matrix.
\end{itemize}

\item For charmonium and twisted mass signs $(\pm,\mp)$, i.e.\ $\bar{\chi}^{(c^+)} \chi^{(c^-)}$ and $\bar{\chi}^{(c^-)} \chi^{(c^+)}$:
\begin{itemize}
\item 
$A_1$, $E$ and $\mathcal{C} \circ \mathcal{P}^{(\textrm{tm})} = +$: $4 \times 4$ correlation matrices; \\
$A_1$, $E$ and $\mathcal{C} \circ \mathcal{P}^{(\textrm{tm})} = -$: $4 \times 4$ correlation matrices.

\item 
$T_1$ and $\mathcal{C} \circ \mathcal{P}^{(\textrm{tm})} = +$: $10 \times 10$ correlation matrix; \\
$T_1$ and $\mathcal{C} \circ \mathcal{P}^{(\textrm{tm})} = -$: $6 \times 6$ correlation matrix.

\item 
$T_2$ and $\mathcal{C} \circ \mathcal{P}^{(\textrm{tm})} = +$: $6 \times 6$ correlation matrix; \\
$T_2$ and $\mathcal{C} \circ \mathcal{P}^{(\textrm{tm})} = -$: $2 \times 2$ correlation matrix.
\end{itemize}
\end{itemize}

In Table~\ref{tab.ensembles} we list for each ensemble the number of gauge link configurations used for the computation of the correlation matrices $C_{\Gamma_j;\Gamma_k;\bar{\chi}^{(1)} \chi^{(2)}}$. The four stochastic sources needed for the one-end trick (cf.\ eq.\ (\ref{EQN677})) are located on a timeslice, which is randomly chosen for every gauge link configuration. We only use a single set of four stochastic timeslice sources, i.e.\ a single sample, for each gauge link configuration.

We have checked the computation of the correlation matrices $C_{\Gamma_j;\Gamma_k;\bar{\chi}^{(1)} \chi^{(2)}}(t)$ by testing the symmetries twisted mass $\gamma_5$ hermiticity, twisted mass parity $\mathcal{P}^{(\textrm{tm})}$, twisted mass time reversal, charge conjugation $\mathcal{C}$ and cubic rotations on the level of correlation matrix elements. In a second step we have averaged the elements related by these symmetries, to improve the signal-to-noise ratio.


\subsection{\label{sec.quantumnumbers}Determination of meson masses and assignment of parity}

We determine meson masses from the correlation matrices $C_{\Gamma_j;\Gamma_k;\bar{\chi}^{(1)} \chi^{(2)}}$ specified in the previous subsection (in the following denoted by $C_{j k} \equiv C_{\Gamma_j;\Gamma_k;\bar{\chi}^{(1)} \chi^{(2)}}$ for simplicity).

In a first step we replace
\begin{eqnarray}
C_{j k}(t) \ \ \rightarrow \ \ \hat{C}_{j k}(t) \ \ \equiv \ \ \frac{C(t)}{\sqrt{C_{j j}(t=a) C_{k k}(t=a)}} .
\end{eqnarray}
This amounts to a correlation matrix $\hat{C}(t)$ with meson creation operators $\hat{O}_j \equiv O_j / \sqrt{C_{j j}(t=a)}$, i.e.\ operators, which are normalized differently than operators (\ref{EQN507}), but are otherwise identical. Such a normalization clearly does not change the exponential decay of correlation matrix elements, i.e.\ the meson masses we are interested in. However, it corresponds to trial states $\hat{O}_j | \Omega \rangle$, which have a similar norm. This is important both for a correct assignment of parity and for a meaningful interpretation of the structure of the state associated with an extracted meson mass.

We solve generalized eigenvalue problems
\begin{eqnarray}
\label{eqn.GEP} \hat{C}(t) \vec{v}^{(n)}(t) \ \ = \ \ \lambda^{(n)}(t) \hat{C}(t_0) \vec{v}^{(n)}(t)
\end{eqnarray}
with $t_0 = a \quad$\footnote{Theoretical arguments given in \cite{Blossier:2009kd} suggest to choose $t_0 \geq t/2$, since then unwanted contributions of excited states are strongly suppressed, in particular, when using large correlation matrices. In practice we find effective masses, which are essentially independent of $t_0$, but with statistical errors increasing for $t_0 > a$.} (for a detailed discussion of this generalized eigenvalue problem cf.\ \cite{Blossier:2009kd} and references therein). For an $N \times N$ correlation matrix $\hat{C}(t)$ one obtains $N$ eigenvalues $\lambda^{(n)}(t)$ and $N$ corresponding eigenvectors $\vec{v}^{(n)}(t)$, $n=0,\ldots,N-1$.

From the eigenvalues we compute $N$ effective masses $m_n^{\textrm{eff}}(t)$ by solving
\begin{eqnarray}
\label{eqn.effm} \frac{\cosh(m_n^{\textrm{eff}}(t) (T/2 - t))}{\cosh(m_n^{\textrm{eff}}(t) (T/2 - (t+a)))} \ \ \equiv \ \ \frac{\lambda^{(n)}(t)}{\lambda^{(n)}(t+a)} .
\end{eqnarray}
While for $m_n (T/2 - t) \gg 1$ this equation is equivalent to the commonly known definition
\begin{eqnarray}
\label{EQN599} m_n^{\textrm{eff}}(t) \ \ \equiv \ \ \frac{1}{a} \ln\bigg(\frac{\lambda^{(n)}(t)}{\lambda^{(n)}(t+a)}\bigg) ,
\end{eqnarray}
(\ref{eqn.effm}) yields in contrast to (\ref{EQN599}) plateau-like effective masses $m_n^{\textrm{eff}}(t) \approx \textrm{const}$ also for large temporal separations $t$ in the region $t \approx T/2$. The low-lying meson masses $m_n$ in the sector defined by the operators of the correlation matrix $\hat{C}$, i.e.\ by $\Gamma_j$ and by $\bar{\chi}^{(1)} \chi^{(2)}$, are then determined by performing uncorrelated $\chi^2$-minimizing fits of constants $m_n$ to the plateau-like regions of $m_n^{\textrm{eff}}(t)$ at sufficiently large $t$.

\begin{figure}[htb]
\centering
\includegraphics[width=5.55cm,angle=-90]{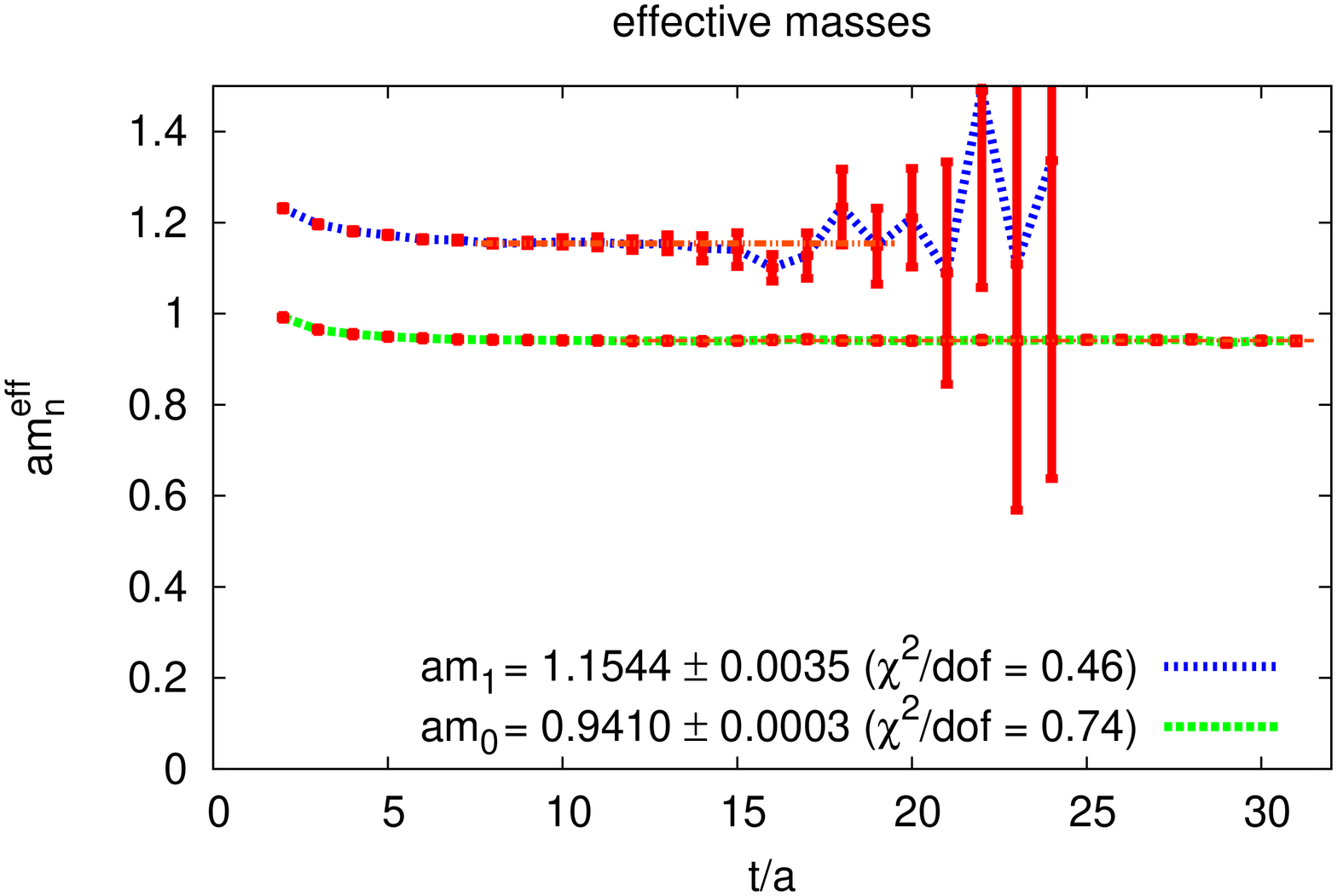} \\
\includegraphics[width=5.55cm,angle=-90]{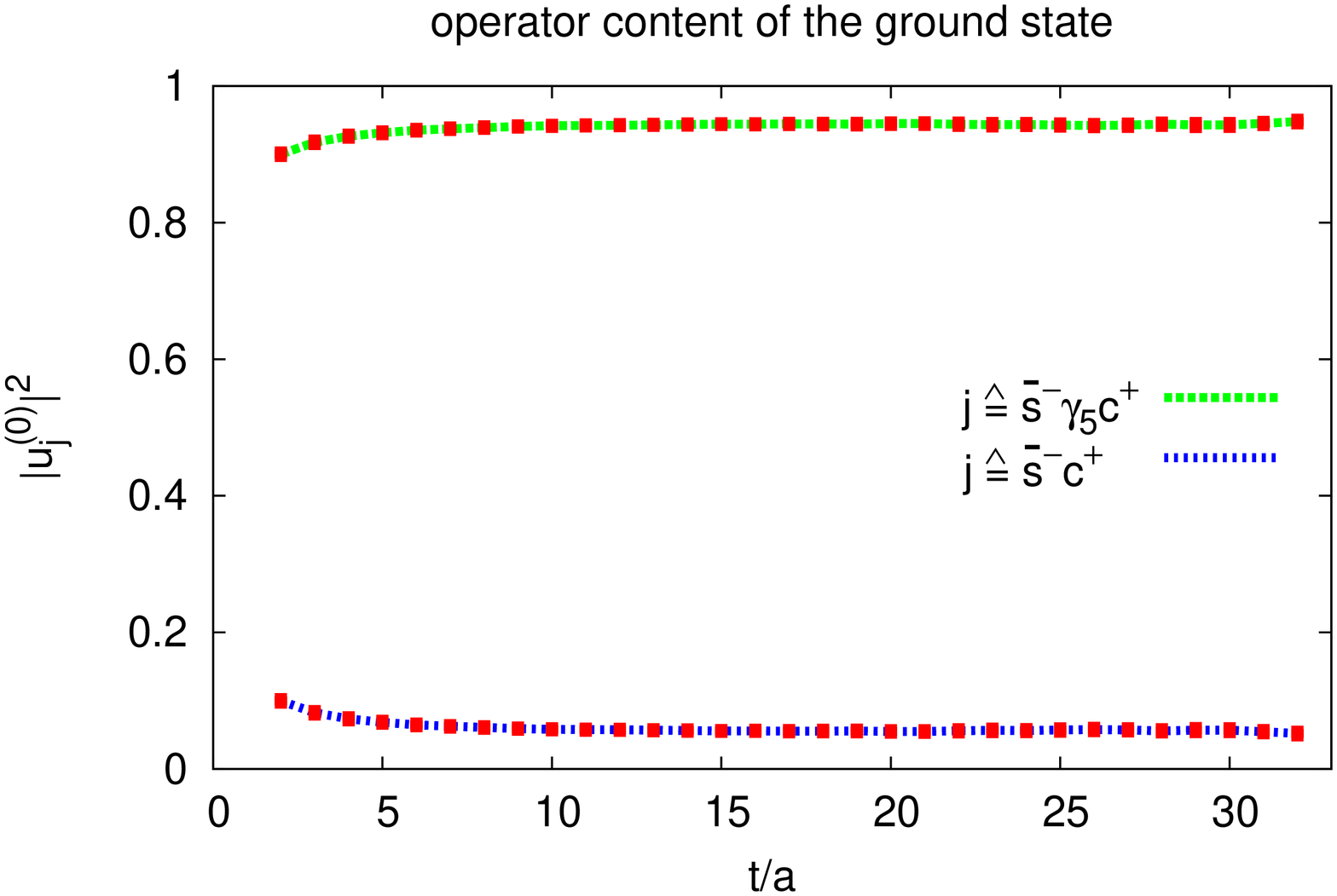}
\includegraphics[width=5.55cm,angle=-90]{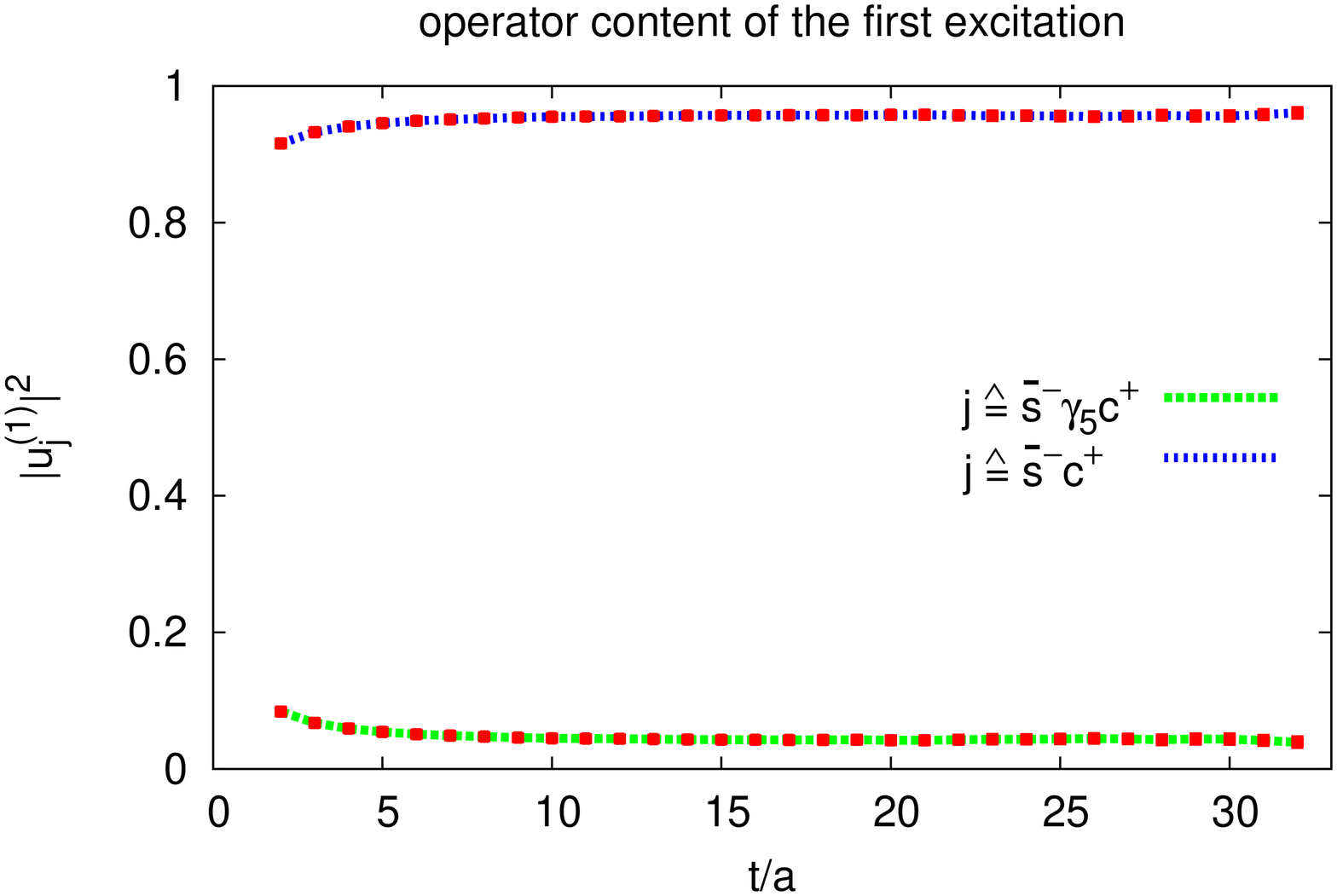}
\caption{\label{fig.example.gep}Determination of meson masses and assignment of parity for $D_s$ mesons with $J=0$: $2 \times 2$ correlation matrix with quark flavors $\bar{\chi}^{(1)} \chi^{(2)} = \bar{\chi}^{(s^-)} \chi^{(c^+)}$, $\mathrm{O}^J$ representation $A_1$ and $\Gamma_j \in \{ \gamma_5 , \mathds{1} \}$ (A30.32 ensemble).}
\end{figure}

As an example we consider a $2 \times 2$ correlation matrix with quark flavors $\bar{\chi}^{(1)} \chi^{(2)} = \bar{\chi}^{(s^-)} \chi^{(c^+)}$ (i.e.\ twisted mass signs $(-,+)$), $\mathrm{O}^J$ representation $A_1$ and $\Gamma_0 = \gamma_5$ (operator index $1$ in Table~\ref{tab.operators}) and  $\Gamma_1 = \mathds{1}$ (operator index $3$ in Table~\ref{tab.operators}). The two resulting effective masses $m_n^{\textrm{eff}}(t)$ are plotted in Figure~\ref{fig.example.gep} (top). Clearly, there are plateaus at large temporal separations $t$. The fitting ranges for the constants $m_n$ are indicated by straight orange lines. The resulting values for $m_n$ and the corresponding $\chi^2/\textrm{dof} \ltapprox 1$ are also included in the plot.

The eigenvectors $\vec{v}^{(n)}(t)$ allow to assign parity to the extracted meson masses and to make qualitative statements about the structure of the corresponding states. The absolute values of the $j$-th entry of the normalized vector
\begin{eqnarray}
\label{eqn.Cv} \vec{u}^{(n)}(t) \ \ \equiv \ \ \frac{\hat{C}(t_0) \vec{v}^{(n)}(t)}{|\hat{C}(t_0) \vec{v}^{(n)}(t)|}
\end{eqnarray}
indicates, to which extent the meson creation operator $\hat{O}_j$ associated with the $j$-th line and $j$-th column of the correlation matrix $\hat{C}(t)$ excites the state corresponding to the $n$th extracted meson mass $m_n$.

In Figure~\ref{fig.example.gep} (bottom, left) the ``operator content'' $|u_j^{(0)}|^2$ of the ground state (corresponding to the extracted mass $m_0$) is plotted as a function of the temporal separation $t$. At large $t$ the meson creation operator with $\Gamma = \gamma_5$ is clearly dominating ($\approx 95 \%$), while the meson creation operator with $\Gamma = \mathds{1}$ contributes on a negligible level ($\approx 5 \%$). Since $\Gamma = \gamma_5$ corresponds to negative parity, we assign the quantum number $\mathcal{P} = -$ to the ground state (i.e.\ identify that state as the $D_s$ meson). From a similar plot (Figure~\ref{fig.example.gep}, [bottom, right]) one can infer that the first excitation, which is dominated by $\Gamma = \mathds{1}$, has $\mathcal{P} = +$, i.e.\ should correspond to $D_{s0}^\ast$.


\subsection{\label{subsec.massdetermination}Extrapolating meson masses to physical strange and charm valence quark masses}

\begin{figure}[htb]
\centering
\includegraphics[width=5.55cm,angle=-90]{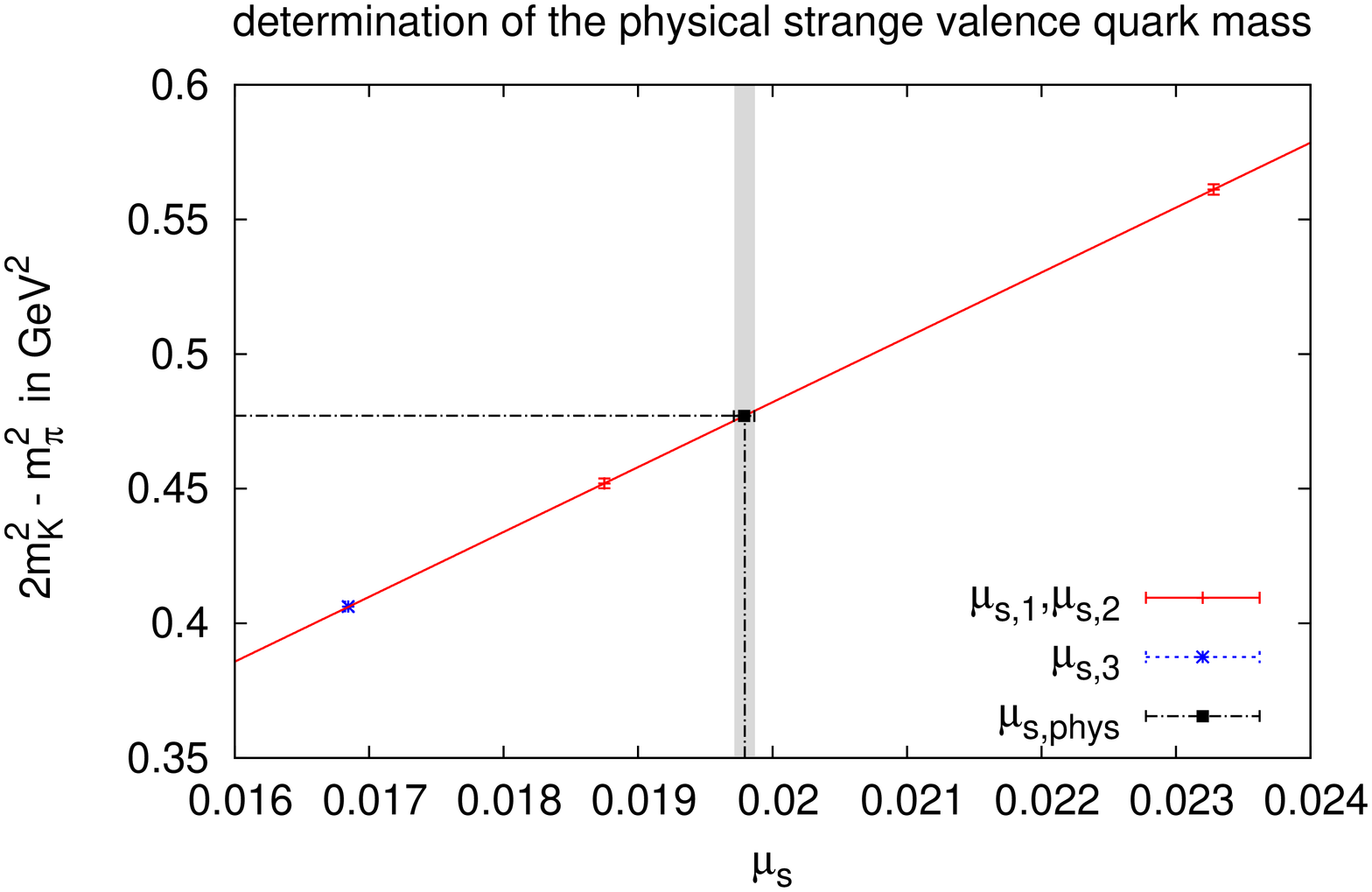}
\includegraphics[width=5.55cm,angle=-90]{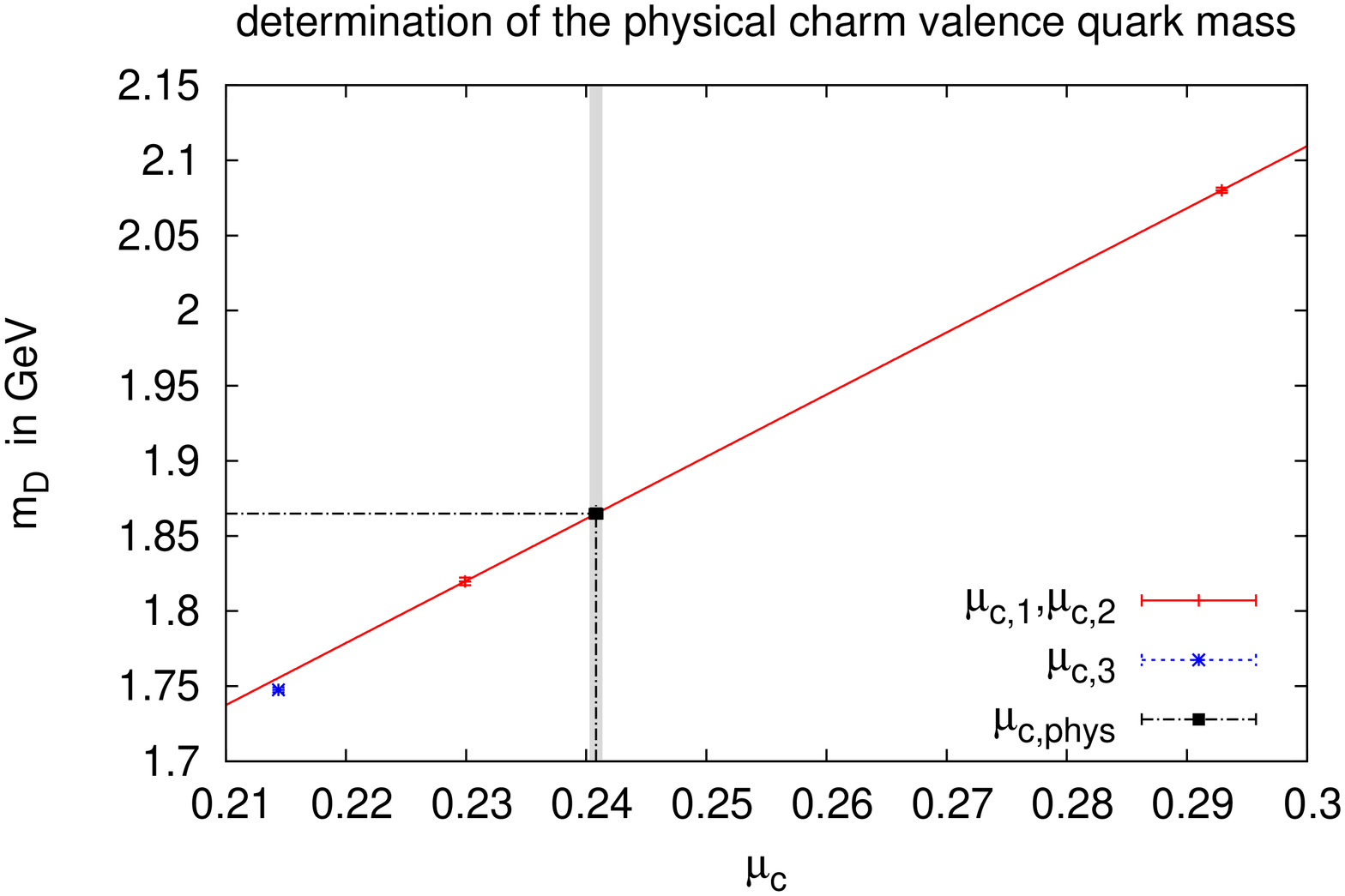}\\
\includegraphics[width=5.55cm,angle=-90]{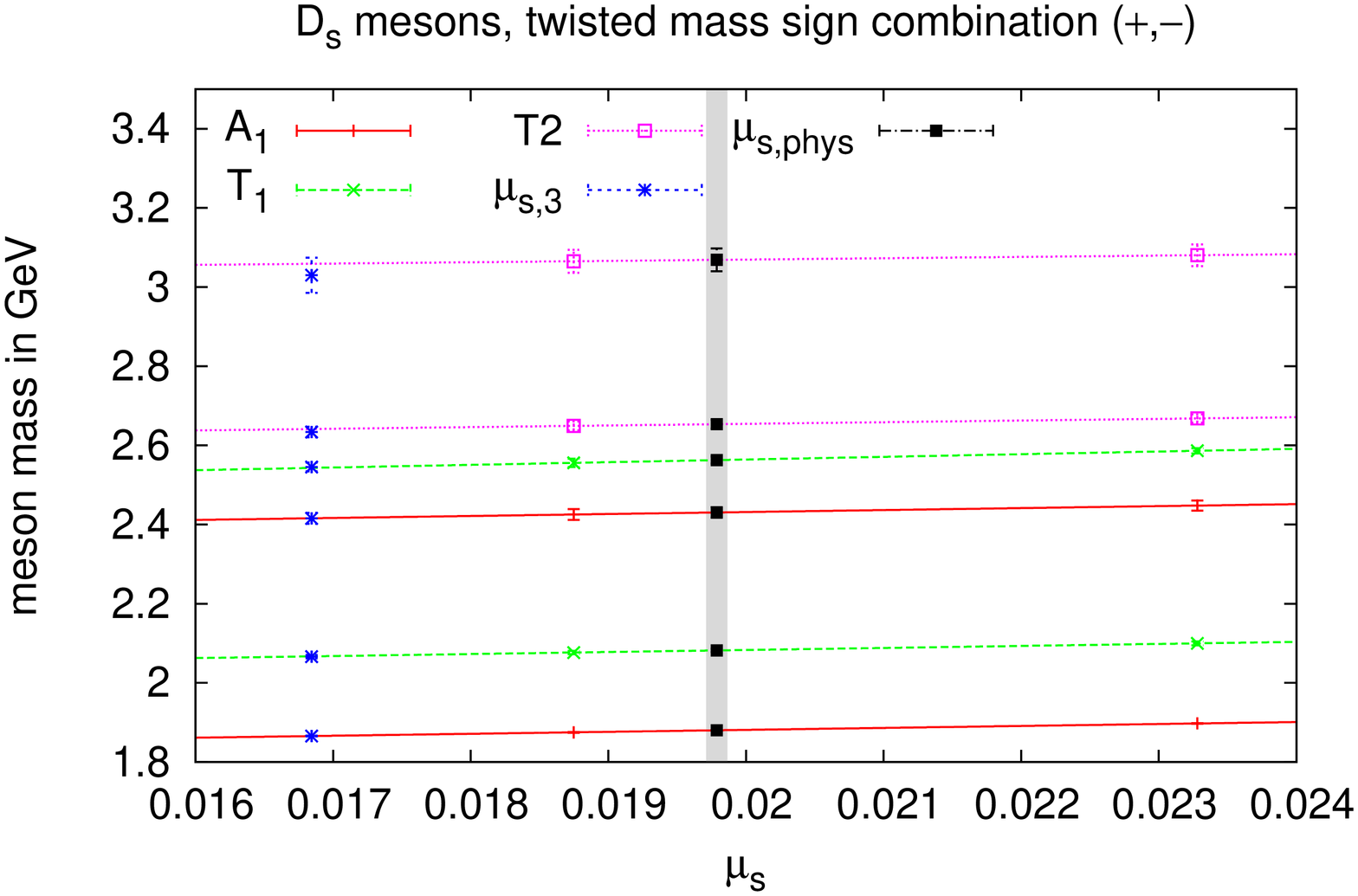}
\includegraphics[width=5.55cm,angle=-90]{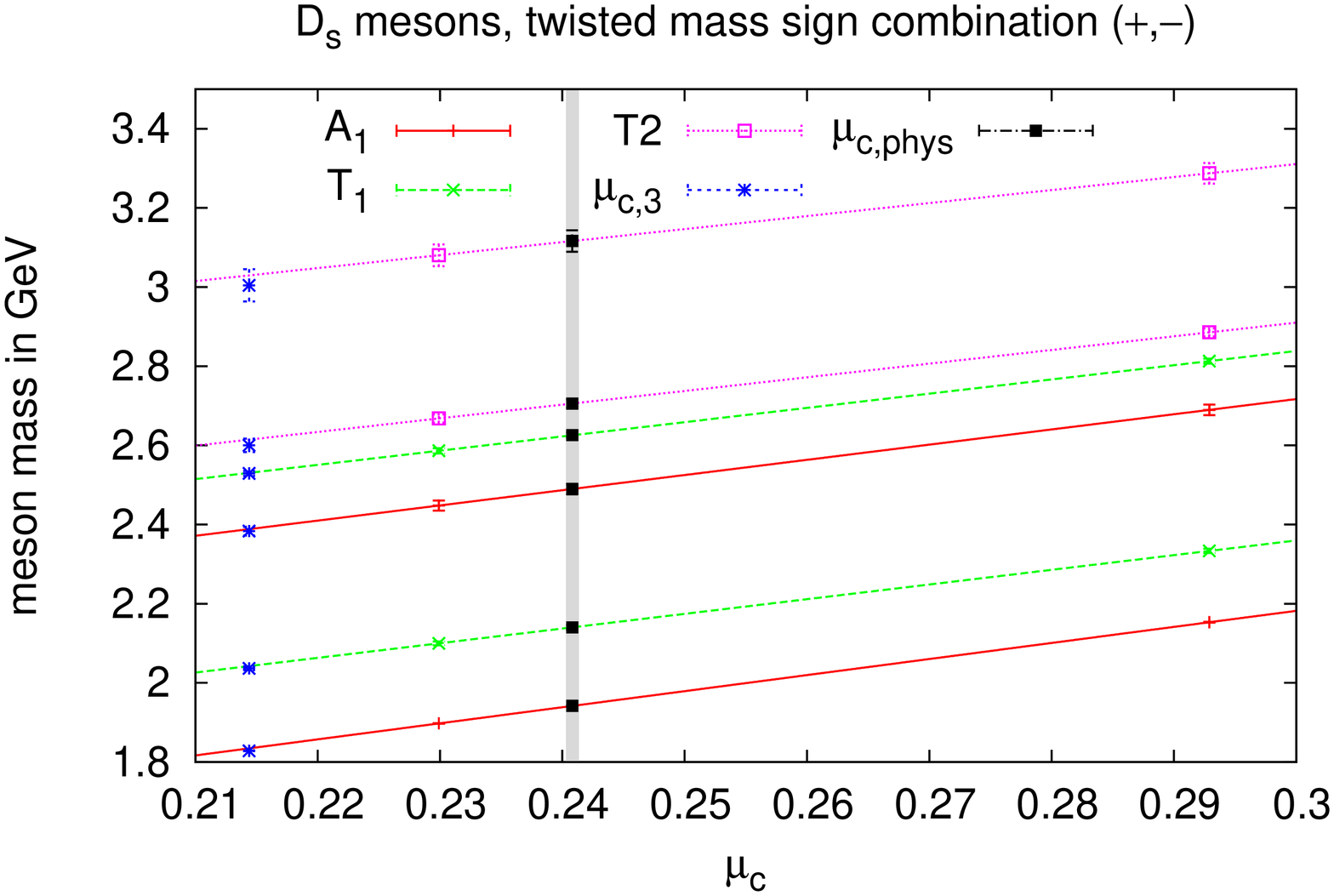}
\caption{\label{FIG001}\textbf{(top):} Determining physical strange and charm valence quark masses by means of linear extrapolations (A80.24 ensemble). \textbf{(bottom):} Linear extrapolations of various $D_s$ meson masses both in $\mu_s$ and $\mu_c$ to physical strange and charm valence quark masses.}
\end{figure}

For each ensemble we compute $m_K$ at two different values of the strange valence quark mass, $\mu_{s,1}$ and $\mu_{s,2}$, both in the region of the physical value. By means of a linear extrapolation we then determine $\mu_{s,\textrm{phys}}$ such that $2 m_K(\mu_s)^2 - m_\pi^2|_{\mu_s = \mu_{s,\textrm{phys}}}$  agrees with the experimental result $2 m_{K^0}^2 - m_{\pi^0}^2 = 0.477 \, \textrm{GeV}^2$ \cite{PDG}:
\begin{eqnarray}
\mu_{s,\textrm{phys}} \ \ = \ \ \mu_{s,2} + (\mu_{s,1} - \mu_{s,2}) \frac{X_{\textrm{phys}} - X(\mu_{s,2})}{X(\mu_{s,1}) - X(\mu_{s,2})} \quad , \quad X(\mu_s) \ \ \equiv \ \ 2 m_K(\mu_s)^2 - m_\pi^2
\end{eqnarray}
(for $m_K(\mu_s)$ and $m_\pi$ we use $(\pm,\mp)$ twisted mass sign combinations, which are known to yield smaller discretization errors \cite{Urbach:2007rt,Frezzotti:2007qv}). Since in leading order chiral perturbation theory $2 m_K^2 - m_\pi^2$ is independent of the light $u/d$ quark mass, $\mu_{s,\textrm{phys}}$ should be very close to the physical strange quark mass. Results are collected in Table~\ref{TAB002}. For ensemble A80.24 the procedure is illustrated in Figure~\ref{FIG001} (top left; red points correspond to $\mu_{s,1}$ and $\mu_{s,2}$, the black dashed lines to $\mu_{s,\textrm{phys}}$). To verify the validity of these linear extrapolations, we performed for ensemble A80.24 additional computations with a third value of the strange valence quark mass, $\mu_{s,3}$. The corresponding result $2 m_K(\mu_s)^2 - m_\pi^2|_{\mu_s = \mu_{s,3}}$ (the blue point in Figure~\ref{FIG001} [top left]) is consistent with the linear extrapolation.

\begin{table}[htb]
\centering
\begin{tabular}{|c|c||c|c||c|c|}
\hline
 & & & & & \vspace{-0.4cm} \\
ensemble & $x$ & $\mu_{s,x}$ & $2 m_K^2 - m_\pi^2$ in $\textrm{GeV}^2$ & $\mu_{c,x}$ & $m_D$ in $\textrm{GeV}$ \\
 & & & & & \vspace{-0.4cm} \\
\hline
 & & & & & \vspace{-0.4cm} \\
\hline
 & & & & & \vspace{-0.4cm} \\
A30.32        & $1$             & $0.018750\phantom{()}$ & $0.454(2)\phantom{^\ast}$ & $0.22700\phantom{()}$ & $1.782(2)\phantom{^\ast}$ \\
              & $2$             & $0.022800\phantom{()}$ & $0.551(2)\phantom{^\ast}$ & $0.27720\phantom{()}$ & $2.002(2)\phantom{^\ast}$ \\
 & & & & & \vspace{-0.30cm} \\
              & $\textrm{phys}$ & $0.01969(8)$           & $0.477^\ast\phantom{(0)}$ & $0.2459(4)$           & $1.865^\ast\phantom{(0)}$ \\
 & & & & & \vspace{-0.4cm} \\
\hline
 & & & & & \vspace{-0.4cm} \\
A40.32        & $1$             & $0.018750\phantom{()}$ & $0.449(1)\phantom{^\ast}$ & $0.23887\phantom{()}$ & $1.841(2)\phantom{^\ast}$ \\
              & $2$             & $0.023220\phantom{()}$ & $0.555(1)\phantom{^\ast}$ & $0.27678\phantom{()}$ & $2.007(3)\phantom{^\ast}$ \\
 & & & & & \vspace{-0.30cm} \\
              & $\textrm{phys}$ & $0.01994(5)$ & $0.477^\ast\phantom{(0)}$ & $0.2443(5)$           & $1.865^\ast\phantom{(0)}$ \\
 & & & & & \vspace{-0.4cm} \\
\hline
 & & & & & \vspace{-0.4cm} \\
A80.24        & $1$             & $0.018749\phantom{()}$ & $0.452(2)\phantom{^\ast}$ & $0.22999\phantom{()}$ & $1.820(3)\phantom{^\ast}$ \\
              & $2$             & $0.023280\phantom{()}$ & $0.561(2)\phantom{^\ast}$ & $0.29299\phantom{()}$ & $2.080(2)\phantom{^\ast}$ \\
 & & & & & \vspace{-0.30cm} \\
              & $3$             & $0.016844\phantom{()}$ & $0.406(2)\phantom{^\ast}$ & $0.21438\phantom{()}$ & $1.748(2)\phantom{^\ast}$ \\
 & & & & & \vspace{-0.30cm} \\
              & $\textrm{phys}$ & $0.01979(8)$           & $0.477^\ast\phantom{(0)}$ &$0.2408(5)$              & $1.865^\ast\phantom{(0)}$\vspace{-0.4cm}\\

 & & & & & \\
\hline
\end{tabular}
\caption{\label{TAB002}Determining physical strange and charm valence quark masses ($^\ast$: experimental results for $2 m_{K^0}^2 - m_{\pi^0}^2$ and $m_{D^0}$ from \cite{PDG}).}
\end{table}

Proceeding in an analogous way we determine a charm valence quark mass $\mu_{c,\textrm{phys}}$ for each ensemble, which is close to the physical charm quark mass, this time using $m_D$:
\begin{eqnarray}
\mu_{c,\textrm{phys}} \ \ = \ \ \mu_{c,2} + (\mu_{c,1} - \mu_{c,2}) \frac{m_{D,\textrm{phys}} - m_D(\mu_{c,2})}{m_D(\mu_{c,1}) - m_D(\mu_{c,2})}
\end{eqnarray}
with $m_{D,\textrm{phys}} \equiv m_{D^0} = 1.865 \, \textrm{GeV}$ \cite{PDG} (cf.\ Table~\ref{TAB002} and Figure~\ref{FIG001} [top right]). Again we tested the quality of the linear extrapolation for ensemble A80.24 with a third value of the charm valence quark mass, $\mu_{c,3}$. This time, there is a visible discrepancy between $m_D(\mu_c)|_{\mu_c = \mu_{c,3}}$ and the corresponding linear extrapolation. This difference, however, is less than $10 \, \textrm{MeV}$, i.e.\ significantly smaller than the estimated magnitude of lattice discretization errors (cf.\ section~\ref{sec.results}). Moreover, the obtained valence charm quark mass $\mu_{c,\textrm{phys}}$, which is rather close to $\mu_{c,1}$, is essentially identical to a valence charm quark mass one would obtain, when using $\mu_{c,1},\mu_{c,3}$ instead of $\mu_{c,1},\mu_{c,2}$.

Note that there is a mismatch of around $10 \% \ldots 20 \%$ of the strange and charm sea quark masses of our ensembles (represented by $\mu_\sigma= 0.150$ and $\mu_\delta = 0.190$; cf.\ section~\ref{SEC432}) and the corresponding physical values: $2 m_{K,\textrm{sea}}^2 - m_{\pi,\textrm{sea}}^2 \approx 0.59 \, \textrm{GeV}^2$ and $m_{D,\textrm{sea}} \approx 2.1 \, \textrm{GeV}$. Since strange and charm sea quarks are expected to have a rather small effect on hadron masses this slight mismatch should have negligible influence on the resulting meson spectra.

Now that physical strange and charm valence quark masses are known, the linear extrapolation procedure is reversed: each $D$ meson, $D_s$ meson and charmonium mass $m$ is computed at two or three pairs of valence quark masses, $(\mu_{s,1} , \mu_{c,2})$, $(\mu_{s,2} , \mu_{c,1})$ and $(\mu_{s,2} , \mu_{c,2})$, and the linear extrapolation is then performed to $(\mu_{s,\textrm{phys}} , \mu_{c,\textrm{phys}})$ according to
\begin{eqnarray}
\label{EQN704} m_{\textrm{phys}} \ \ = \ \ m(\mu_{c,2}) + \Big(m(\mu_{c,1}) - m(\mu_{c,2})\Big) \frac{\mu_{c,\textrm{phys}} - \mu_{c,2}}{\mu_{c,1} - \mu_{c,2}}
\end{eqnarray}
for $D$ mesons and charmonium and according to
\begin{eqnarray}
\nonumber & & \hspace{-0.7cm} m_{\textrm{phys}} \ \ = \ \ m(\mu_{s,2},\mu_{c,2}) + \Big(m(\mu_{s,1},\mu_{c,2}) - m(\mu_{s,2},\mu_{c,2})\Big) \frac{\mu_{s,\textrm{phys}} - \mu_{s,2}}{\mu_{s,1} - \mu_{s,2}} \\
\label{EQN705} & & \hspace{0.675cm} + \Big(m(\mu_{s,2},\mu_{c,1}) - m(\mu_{s,2},\mu_{c,2})\Big) \frac{\mu_{c,\textrm{phys}} - \mu_{c,2}}{\mu_{c,1} - \mu_{c,2}}
\end{eqnarray}
for $D_s$ mesons. Of course, it would have been possible to perform computations of these meson spectra using directly strange and charm valence quark masses $(\mu_{s,\textrm{phys}} , \mu_{c,\textrm{phys}})$. We consider this linear extrapolation procedure, eqs.\ (\ref{EQN704}) and (\ref{EQN705}), however, superior, because it allows a lot more flexibility during the final analysis. For example, one can easily change the value of the lattice spacing to investigate a possible source of systematic error\footnote{There appear to be unresolved inconsistencies between different collaborations regarding scale setting and standard non-perturbative scales like $r_0$ \cite{Sommer:2014mea}. Even within the ETM Collaboration there exist two values of the lattice spacing for the ensembles we are using: $a = 0.0885(36) \, \textrm{fm}$ obtained from the pion decay constant \cite{Carrasco:2014cwa} and $a = 0.0920(21) \, \textrm{fm}$ obtained from the nucleon mass \cite{Alexandrou:2013joa}.} without the need for redoing propagator computations and contractions. Alternatively, one can even determine the lattice spacing in physical units by matching the obtained meson spectra with corresponding experimental data. Another application would be to crudely estimate systematic errors due to isospin breaking and electromagnetic effects using e.g.\ $2 m_{K^\pm}^2 - m_{\pi^\pm}^2$ and $m_{D^\pm}$ instead of $2 m_{K^0}^2 - m_{\pi^0}^2$ and $m_{D^0}$, when determining the strange and charm valence quark masses $\mu_{s,\textrm{phys}}$ and $\mu_{c,\textrm{phys}}$. We plan to investigate such issues and use this flexibility for an evolved error analysis in an upcoming publication, when we have meson masses available for several values of the lattice spacing.

In Figure~\ref{FIG001} (bottom) we show examples of linear extrapolations of various $D_s$ meson masses ($(\pm,\mp)$ twisted mass sign combinations, $O^J$ representations $A_1, T_1, T_2$ and $\mathcal{P} = \pm$) both in $\mu_s$ and in $\mu_c$ to physical strange and charm valence quark masses (ensemble A80.24). Again we compare with computations performed with a third strange and charm valence quark mass (blue points). We find excellent agreement demonstrating once more the validity of the linear extrapolations. Similar consistent results have been obtained for $D$ meson and for charmonium masses and for both $(\pm,\mp)$ and $(\pm,\pm)$ twisted mass sign combinations.


\subsection{Determination of statistical errors}

Statistical errors for $D$ and $D_s$ meson and charmonium masses (extrapolated to physical strange and charm valence quark masses as explained in the previous subsection) are determined on each ensemble via an evolved jackknife analysis starting at the level of the correlation matrices (\ref{EQN698}). To exclude statistical correlations between gauge link configurations, which are close in Monte Carlo simulation time, we performed a suitable binning of these configurations.



\section{\label{sec.results}Results}

We compute $D$ meson, $D_s$ meson and charmonium masses on each of the three ensembles listed in Table~\ref{tab.ensembles}. These ensembles correspond to a single lattice spacing $a \approx 0.0885 \, \textrm{fm}$, but differ in the unphysically heavy $u/d$ quark mass ($m_\pi \approx 276 \, \textrm{MeV} \, , \, 315 \, \textrm{MeV} \, , \, 443 \, \textrm{MeV}$). Each meson mass is extrapolated linearly in $m_\pi^2$ to the physical value of the $u/d$ quark mass ($m_\pi = m_{\pi^0} = 135 \, \textrm{MeV}$). Strange and charm valence quark masses correspond to their physical values (cf.\ section~\ref{subsec.massdetermination}).

Since at the moment computations are only available for a single lattice spacing, we are not in a position to perform a continuum extrapolation. Lattice discretization errors are, however, expected to be small, because the lattice spacing we use, $a \approx 0.0885 \, \textrm{fm}$, is rather fine and meson masses are automatically $\mathcal{O}(a)$ improved due to our specific quark discretization, Wilson twisted mass lattice QCD at maximal twist \cite{Frezzotti:2003ni}. In this formulation each meson mass can be computed in two slightly different ways, either using $(\pm,\mp)$ or $(\pm,\pm)$ twisted mass sign combinations (cf.\ section~\ref{subsubsec.basis}). The corresponding two results differ by lattice discretization errors. We use these differences to crudely estimate the magnitude of the lattice discretization errors associated with our resulting meson masses.

The continuum limit will be studied in a future publication. There we also plan to include a more detailed discussion and analysis of systematic errors, including e.g.\ isospin breaking and and electromagnetic effects.


\subsection{\label{opencharm.A1}$D$ and $D_s$ mesons}

Since the $D$ and the $D_s$ meson spectrum are qualitatively very similar, these two sectors are discussed in parallel in the following. The charmonium sector is presented separately in section~\ref{charmonium}.


\subsubsection{\label{SEC983}$A_1$ representation (spin $J=0$)}

We compute the masses of the two lowest states in the $A_1$ sector (both for $D$ and for $D_s$ mesons), one with $\mathcal{P} = -$, the other with $\mathcal{P} = +$, for each of the three ensembles (cf.\ Table~\ref{tab.ensembles}) and both twisted mass sign combinations $(\pm,\mp)$ and $(\pm,\pm)$. In this process we solve the generalized eigenvalue problem (\ref{eqn.GEP}) using $2 \times 2$ correlation matrices with creation operator indices $1,3$ (cf.\ Table~\ref{tab.operators}), i.e.\ $\Gamma = \gamma_5 , \mathds{1}$. Note again that parity is not an exact symmetry in Wilson twisted mass lattice QCD at finite lattice spacing. The positive and negative parity states form a single sector, where the ground state has $\mathcal{P} = -$ (the $D/D_s$ meson) and the first excited state has $\mathcal{P} = +$ (the $D_0^\ast/D_{s0}^\ast$ meson). One has to determine the masses of both states at the same time from a single correlation matrix (cf.\ the detailed discussion in section~\ref{sec.quantumnumbers}).

Even though we have implemented eight different creation operators for the $A_1$ representation (cf.\ Table~\ref{tab.operators}), the above mentioned rather small $2 \times 2$ correlation matrices turn out to be an optimal choice: statistical errors are quite small and long and stable plateaus are obtained both for effective masses and for operator contents (cf.\ e.g.\ Figure~\ref{fig.example.gep}).


\subsubsection*{$\mathcal{P} = -$: $D$ and $D_s$}

\begin{figure}[htb]
\centering
\includegraphics[width=5.55cm,angle=-90]{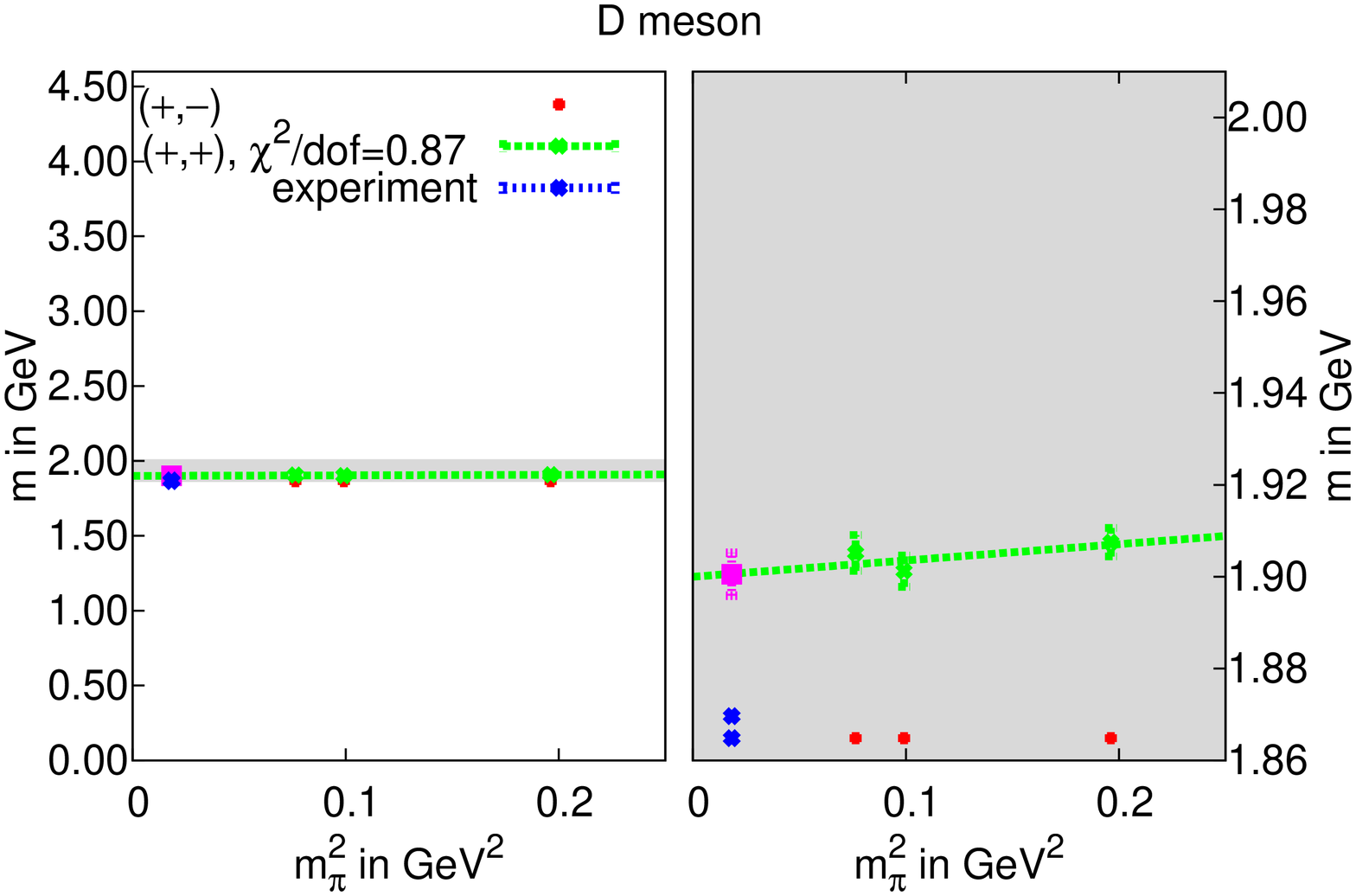}
\includegraphics[width=5.55cm,angle=-90]{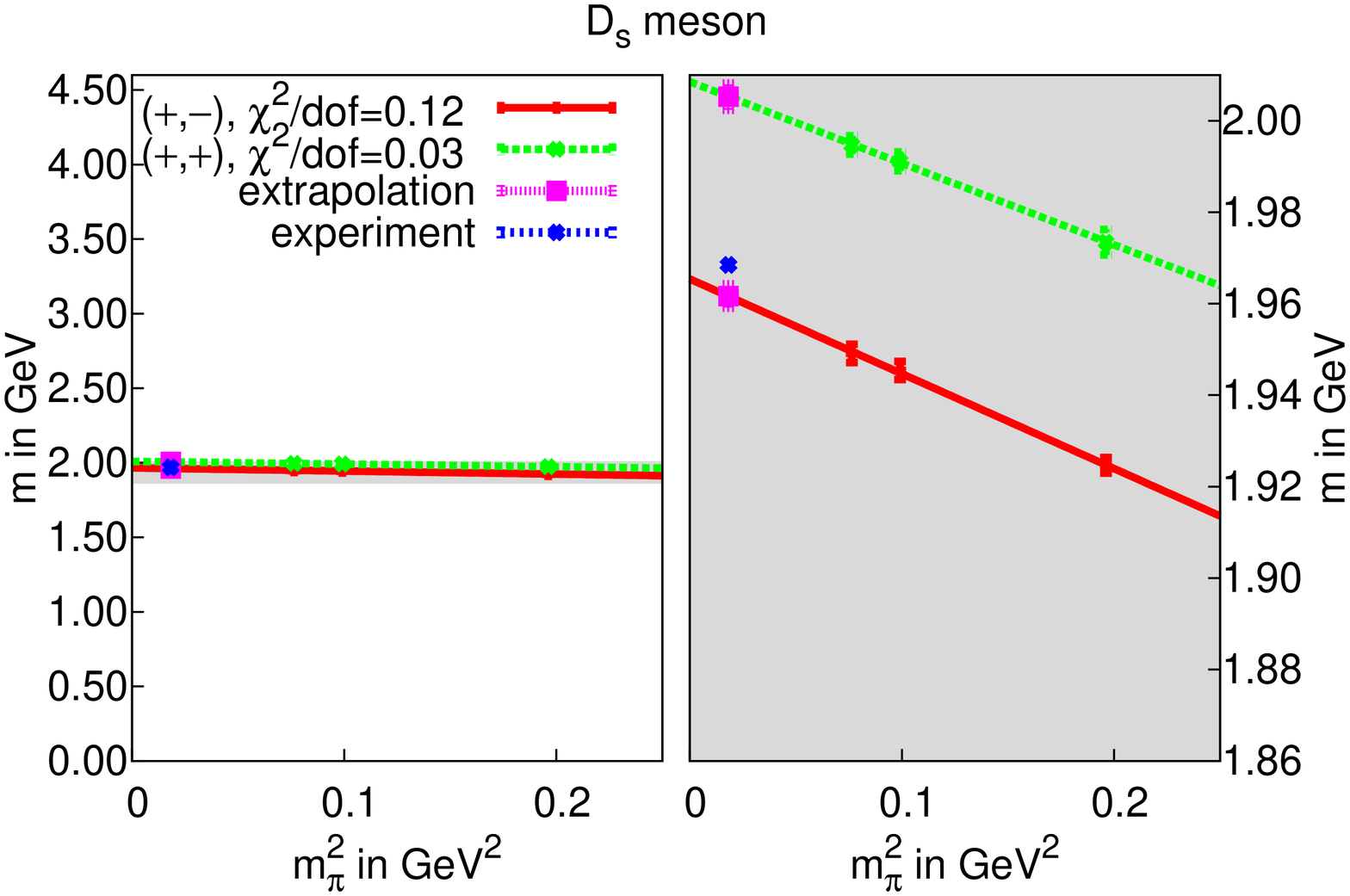}
\caption{\label{fig.A1.0}$A_1$ representation (spin $J=0$), $\mathcal{P} = -$. \textbf{(left):} $D$ meson. \textbf{(right):} $D_s$ meson.}
\end{figure}

As already mentioned above and as expected the ground states in the $A_1$ sectors have negative parity, i.e.\ correspond to the $D$ meson and the $D_s$ meson. In Figure~\ref{fig.A1.0} the computed $D$ meson masses (left) and $D_s$ meson masses (right) are shown for all three ensembles (i.e.\ \\ $m_\pi \approx 276 \, \textrm{MeV} \, , \, 315 \, \textrm{MeV} \, , \, 443 \, \textrm{MeV}$) and for both twisted mass sign combinations, $(\pm,\mp)$ (red points) and $(\pm,\pm)$ (green points). The straight red and green lines are linear extrapolations in $m_\pi^2$ to physical $u/d$ quark masses corresponding to $m_\pi = m_{\pi^0} = 135 \, \textrm{MeV}$. The results of these extrapolations are shown in magenta. The blue points are experimental results for $m_{D^0}$, $m_{D^\pm}$ and $m_{D_s}$ \cite{PDG} (note that there are two experimental results in the left plot, because the neutral and the charged $D$ meson differ in mass, $m_{D^\pm} - m_{D^0} = 5 \, \textrm{MeV}$).\footnote{The majority of plots shown in this section, including the two plots in Figure~\ref{fig.A1.0}, follows this style. The right panel of each plot is a zoomed version of the left panel with respect to the vertical axis, but otherwise identical. In the left panels the scale of the vertical axis, which represents the meson mass, always ranges from $0$ to $4.6 \, \textrm{GeV}$. Hence from the left panels one can conveniently read off the relative errors and precision of our results. They also allow to directly compare meson masses from different plots. The right panels are strongly zoomed (individually for each meson) and, hence, show more clearly details regarding the data quality, the absolute size of the errors and the dependence of the meson mass on the $u/d$ quark mass.}

For each ensemble the valence charm quark mass is chosen such that the $(\pm,\mp)$ version of the $D$ meson mass is identical to the experimental result $m_{D^0} = 1865 \, \textrm{GeV}$, as discussed in section~\ref{subsec.massdetermination}. Therefore, these $(\pm,\mp)$ $D$ meson masses should not be considered as predictions and, consequently, a linear extrapolation of these masses (which would trivially be a constant) is neither needed nor shown. The statistical errors of these masses enter an evolved jackknife procedure and are, therefore, not only considered in the errors of the valence quark masses $\mu_{c,\textrm{phys}}$, but also in the errors of all $D$ meson, $D_s$ meson and charmonium masses computed in this work.

The differences between the lattice QCD results obtained with $(\pm,\mp)$ and with $(\pm,\pm)$ twisted mass sign combinations are for both the $D$ and the $D_s$ meson around $50 \, \textrm{MeV}$. These differences, which will vanish in the continuum limit, are a crude estimate of the magnitude of lattice discretization errors associated with our current results obtained at a single lattice spacing, i.e.\ relative errors of around $2.5 \%$. Note that for meson masses obtained with $(\pm,\mp)$ twisted mass sign combinations discretization errors are expected to be significantly smaller \cite{Urbach:2007rt,Frezzotti:2007qv}, i.e.\ the mentioned $50 \, \textrm{MeV}$ are most likely a rather conservative estimate. This is consistent e.g.\ with our $(\pm,\mp)$ lattice result for the $D_s$ meson, which differs by less than $10 \, \textrm{MeV}$ from the experimental result.  

The linear increase of the $D_s$ meson mass for decreasing $u/d$ quark mass is an expected consequence of our procedure for setting the charm valence quark mass $\mu_{c,\textrm{phys}}$. We choose $\mu_{c,\textrm{phys}}$ for each ensemble such that the lattice result for $m_D$ agrees with the experimental result $m_{D^0} = 1865 \, \textrm{MeV}$, i.e.\ independently of the $u/d$ quark mass (cf.\ section~\ref{subsec.massdetermination}). Clearly an increasing $u/d$ quark mass leads to a decreasing $\mu_{c,\textrm{phys}}$ slightly lighter than the physical charm quark mass. This in turn yields the observed $u/d$ quark mass dependence of $m_{D_s}$.

The experimentally observed splitting $m_{D^{\pm}} - m_{D^0} = 5 \, \textrm{MeV}$ indicates the magnitude of electromagnetic and isospin breaking effects. Since the currently estimated discretization errors of $\ltapprox 50 \, \textrm{MeV}$ are much larger, we will at the moment not consider systematic errors due to the neglect of electromagnetism and isospin breaking. We plan to address such errors in a future publication, where we will study the continuum limit.


\subsubsection*{$\mathcal{P} = +$: $D_0^\ast$ and $D_{s0}^\ast$}

\begin{figure}[htb]
\centering
\includegraphics[width=5.55cm,angle=-90]{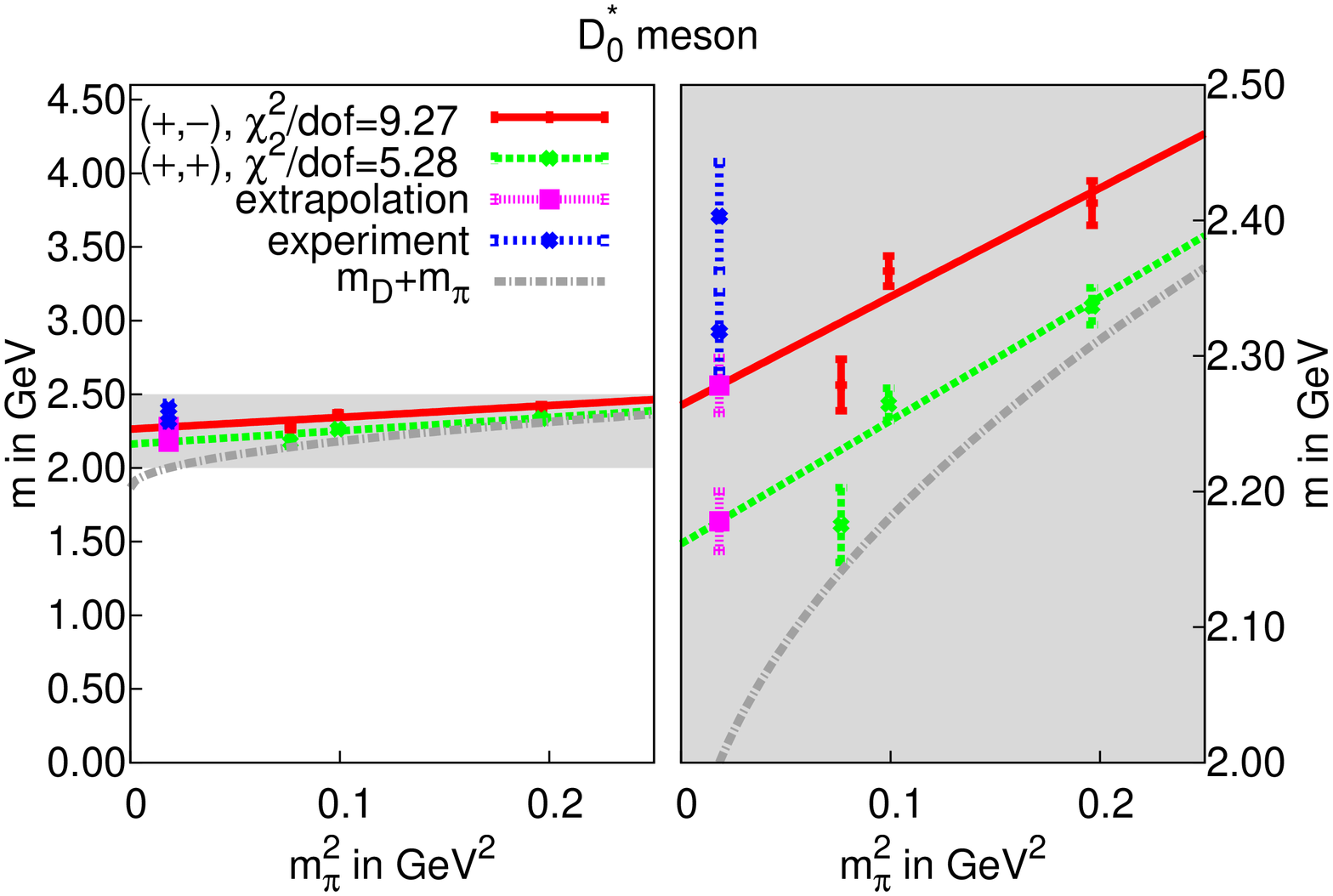}
\includegraphics[width=5.55cm,angle=-90]{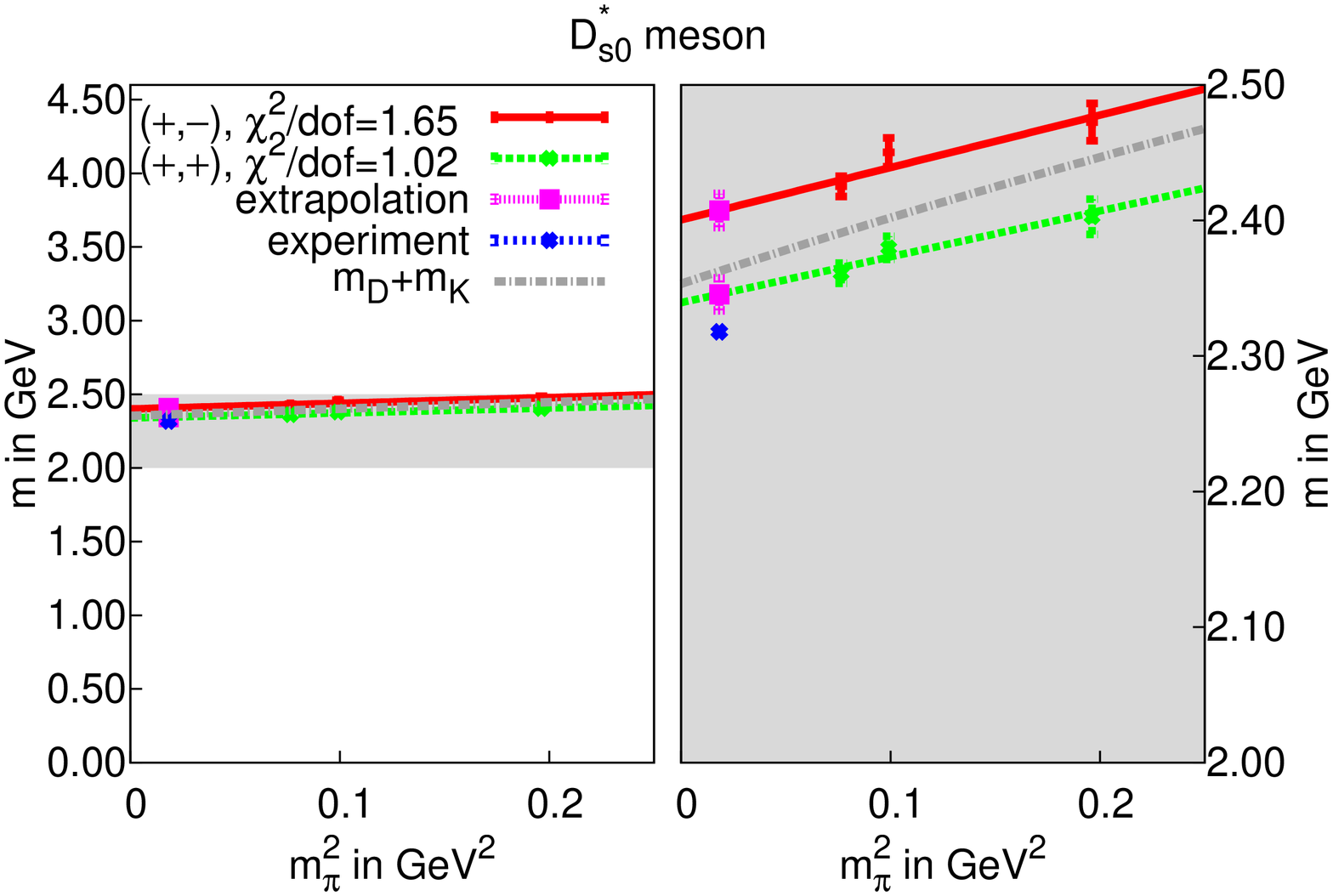}
\caption{\label{fig.A1.1}$A_1$ representation (spin $J=0$), $\mathcal{P} = +$. \textbf{(left):} $D_0^\ast$ meson. \textbf{(right):} $D_{s0}^\ast$ meson.}
\end{figure}

The first excitations in the $A_1$ sectors have positive parity, i.e.\ should correspond to the $D_0^\ast$ meson and the $D_{s0}^\ast$ meson (cf.\ Figure~\ref{fig.A1.1}). While the masses of these states have been extracted from a $2 \times 2$ correlation matrix with creation operators $\Gamma = \gamma_5$ and $\Gamma = \mathds{1}$ (indices $1$ and $3$ in Table~\ref{tab.operators}), we have studied their structure by considering also larger correlation matrices with additional $\mathcal{P} = +$ creation operators (indices $4$, $7$ and $8$ in Table~\ref{tab.operators}). Interestingly, the operator contents of both the $D_0^\ast$ meson and the $D_{s0}^\ast$ meson are mixtures of $\Gamma = \mathds{1},\gamma_0$ and $\Gamma = \gamma_j \mathbf{n}_j,\gamma_0 \gamma_j \mathbf{n}_j$ of roughly the same magnitude. This indicates that the quarks inside these mesons form superpositions of $S$ and $P$ waves and not predominantly $P$ waves as suggested by many quark models.

In both cases we observe decreasing meson masses for decreasing $u/d$ quark mass, a behavior not present for the previously discussed $\mathcal{P} = -$ parity partners. An explanation could be that the extracted $\mathcal{P} = +$ states contain rather light two-meson contributions with the same quantum numbers, $D + \pi$ and $D + K$, respectively. In Figure~\ref{fig.A1.1} this is illustrated by the gray curves, which correspond to the estimated masses of the two-meson states, $m_D + m_\pi$ and $m_D + m_K(m_\pi) \quad$\footnote{Here and in the following we have used $m_K(m_\pi) = \sqrt{(0.477 \, \textrm{GeV}^2 + m_\pi^2) / 2}$, $m_D = 1.865 \, \textrm{GeV}$ and $m_{D^\ast} = 2.007 \, \textrm{GeV}$ for these estimates.}. For the $D_0^\ast$ meson this is also supported by the rather large $\chi^2 / \textrm{dof}$ of the extrapolation to physical $u/d$ quark mass (cf.\ Figure~\ref{fig.A1.1} [left]). Surprisingly such a behavior has not been observed in \cite{Mohler:2011ke}, where a rather similar lattice QCD setup, in particular a similar set of creation operators, has been used.

Since it is unexpectedly light, in the literature the $D_{s0}^\ast$ meson is frequently discussed as a mesonic molecule or tetraquark candidate (cf.\ e.g.\ \cite{Dmitrasinovic:2005gc,Cleven:2014oka}). In principle it is possible to investigate the structure of these states using lattice methods, but this will require the implementation of additional four-quark creation operators of mesonic molecule, of diquark-antidiquark and/or of two-meson type. We are in the process of developing techniques for computing corresponding correlation matrices \cite{Alexandrou:2012rm,Abdel-Rehim:2014zwa,Berlin:2015faa}. Recent lattice papers using four-quark operators and focusing specifically on $D_0^\ast$ and $D_{0s}^\ast$ are \cite{Gong:2011nr,Mohler:2012na,Liu:2012zya,Mohler:2013rwa,Lang:2014yfa}.

This time lattice discretization errors indicated by the differences between results obtained with $(\pm,\mp)$ and $(\pm,\pm)$ twisted mass sign combinations are somewhat larger, of the order of $60 \, \textrm{MeV} \ldots 100 \, \textrm{MeV}$ (i.e.\ relative errors $\approx 3\% \ldots 5\%$). Within this crude conservative estimate there is consistency with experimental results. It is reassuring that the $(\pm,\mp)$ result for $m_{D_0^\ast}$, which is expected to have significantly smaller discretization errors than the $(\pm,\pm)$ result, is within $1 \, \sigma$ of the experimental result for $m_{D_0^\ast(2400)^0}$ without taking discretization errors into account. The $(\pm,\mp)$ $D_{s0}$ result is around $80 \, \textrm{MeV}$ larger than its experimental counterpart. This is similar to what has been found in quark models (cf.\ e.g.\ \cite{Ebert:2009ua}) and other lattice QCD computations using exclusively quark-antiquark meson creation operators (cf.\ e.g.\ \cite{Mohler:2011ke}). It could be an indication that the $D_{s0}^\ast$ meson is not predominantly a quark-antiquark state, but possibly a mesonic $D K$ molecule (e.g.\ supported by \cite{Liu:2012zya}) or a diquark-antidiquark pair.


\subsubsection{\label{SEC429}$T_1$ representation (spin $J=1$)}


\subsubsection*{$\mathcal{P} = -$: $D^\ast$ and $D_s^\ast$}

\begin{figure}[htb]
\centering
\includegraphics[width=5.55cm,angle=-90]{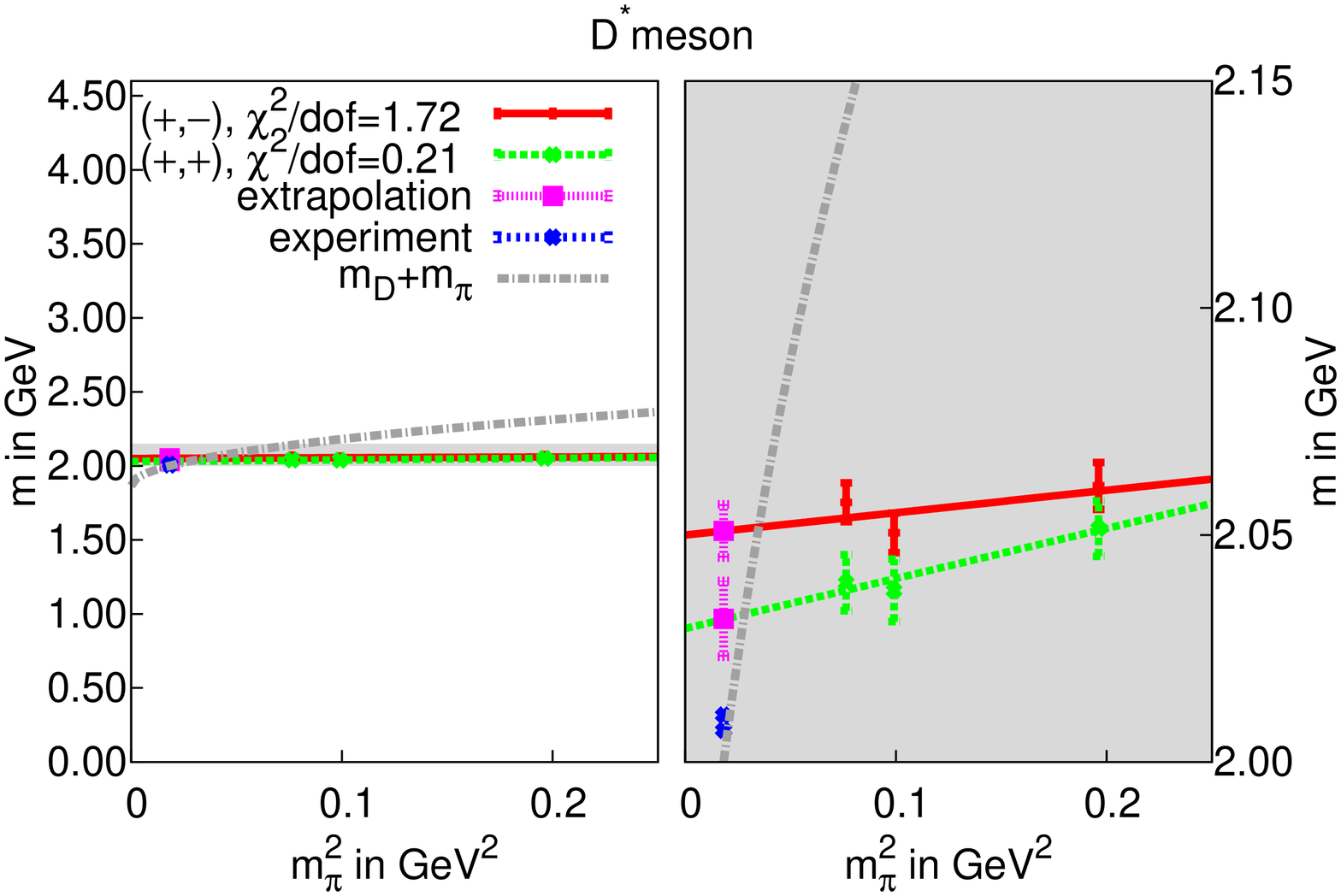}
\includegraphics[width=5.55cm,angle=-90]{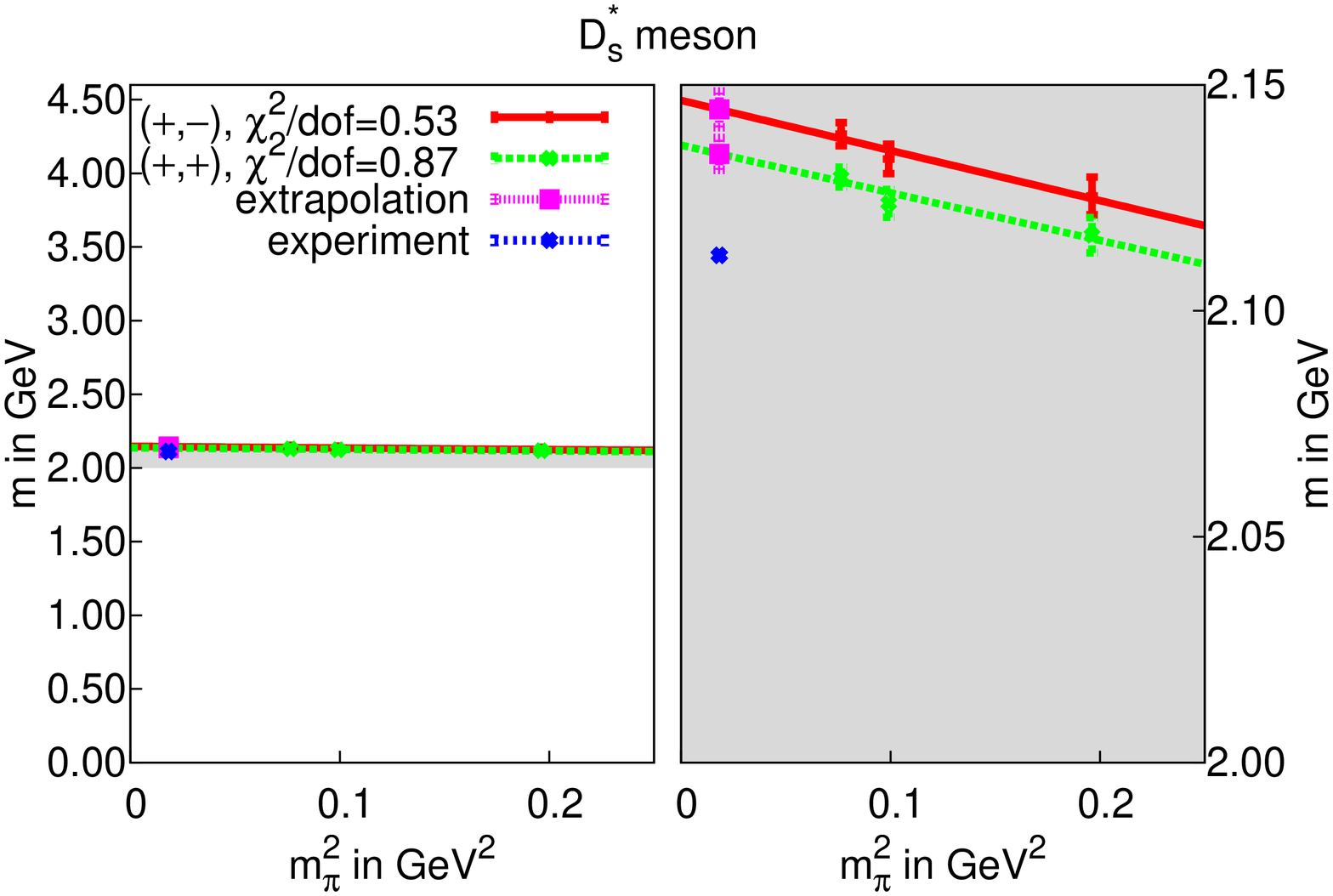}
\caption{\label{fig.T1.0}$T_1$ representation (spin $J=1$), $\mathcal{P} = -$. \textbf{(left):} $D^\ast$ meson. \textbf{(right):} $D_s^\ast$ meson.}
\end{figure}

We proceed as in section~\ref{SEC983}, where we have discussed the $A_1$ representation. This time we have solved generalized eigenvalue problems (\ref{eqn.GEP}) using $4 \times 4$ correlation matrices with creation operator indices $1,2,3,4$ (cf.\ Table~\ref{tab.operators}) including the commonly used $\mathcal{P} = -$ operators $\Gamma \in \{ \gamma_1 , \gamma_0 \gamma_1 \}$. As expected the ground states in the $T_1$ sectors have negative parity, i.e.\ correspond to the $D^\ast$ meson and the $D_s^\ast$ meson (cf.\ Figure~\ref{fig.T1.0}).

While at physically light $u/d$ quark masses the decay of $D^\ast$ to $D + \pi$ is possible, it is excluded in our computations with pion masses $m_\pi \gtapprox 276 \, \textrm{MeV}$. This is consistent with the rather mild $m_\pi^2$ dependence of $m_{D^\ast}$ (cf.\ Figure~\ref{fig.T1.0} [left]).

The differences between the lattice QCD results obtained with $(\pm,\mp)$ and with $(\pm,\pm)$ twisted mass sign combinations are around $10 \, \textrm{MeV} \ldots 20 \, \textrm{MeV}$, i.e.\ lattice discretization errors seem to be smaller than in the $A_1$ representation. There is also a discrepancy of around $20 \, \textrm{MeV} \ldots 40 \, \textrm{MeV}$ to the corresponding experimental results \cite{PDG}. In this context it is interesting to note that the mass differences $m_{D^\ast} - m_D$ and $m_{D_s^\ast} - m_{D_s}$ are rather strongly dependent on scale setting, i.e.\ the lattice spacing in physical units. For example, when using $a = 0.0920(21) \, \textrm{fm}$ (obtained from the nucleon mass \cite{Alexandrou:2013joa}) instead of $a = 0.0885(36) \, \textrm{fm}$ (obtained from the pion decay constant \cite{Carrasco:2014cwa}), the above mentioned discrepancy of $20 \, \textrm{MeV} \ldots 40 \, \textrm{MeV}$ is reduced to $15 \, \textrm{MeV} \ldots 20 \, \textrm{MeV}$. We consider this as an indication that the larger lattice spacing obtained from the nucleon mass might be better suited to determine hadron masses with small discretization errors than the lattice spacing obtained from the pion decay constant (which is more commonly used within the ETM Collaboration). Similar observations have been reported in spectrum computations of $B$ and $B_s$ mesons \cite{Jansen:2008si,Michael:2010aa} and $b$ baryons \cite{Wagner:2011fs}. As mentioned before we plan to investigate such scale setting issues and the continuum limit in detail in a future publication, when we have meson masses available for several values of the lattice spacing.


\subsubsection*{$\mathcal{P} = +$: $D_1(2430)$, $D_1(2420)$ and $D_{s1}(2460)$, $D_{s1}(2536)$}

To determine masses of positive parity states we use larger $8 \times 8$ correlation matrices with creation operator indices $1,2,3,4,7,8,9,10$ (cf.\ Table~\ref{tab.operators}), i.e.\ the previously used $\mathcal{P} = -$ operators $\Gamma \in \{ \gamma_1 , \gamma_0 \gamma_1 \}$ and all available $\mathcal{P} = +$ operators with $L = 0$ and $L = 1$. The first and second excitations in the $T_1$ sectors have positive parity, i.e.\ correspond to $D_1(2430)$, $D_1(2420)$ and $D_{s1}(2460)$, $D_{s1}(2536)$.

The masses of the two states in each sector are rather close. Therefore, it is not obvious, how to correctly assign the obtained lattice QCD results to $D_1(2430)$ and $D_1(2420)$ and to $D_{s1}(2460)$ and $D_{s1}(2536)$, respectively. Such an assignment is not only important for a complete and precise computation of the $D$ and $D_s$ meson spectrum, but also in the context of specific decays, in particular $B^{(\ast)} \rightarrow D^{\ast \ast} + l + \nu$, where $D^{\ast \ast}$ denotes the four positive parity $D$ mesons with $J = 0,1,2$, which include $D_1(2430)$ and $D_1(2420)$. To understand these decays is e.g.\ essential for a precise determination of the CKM matrix element $V_{cb}$. However, there is a long-standing conflict between theory and experiment regarding the corresponding branching ratios \cite{Bigi:2007qp}. While these decays have been studied with lattice QCD in the static limit some time ago \cite{Blossier:2009vy}, recently computations with $b$ and $c$ quarks of finite mass have been started \cite{Atoui:2013sca,Atoui:2013ksa}. The latter computations are, however, restricted to $D^{\ast \ast}$ with $J=0,2$, mainly because the separation of the two $J=1$ states is rather difficult. In the following we demonstrate, how to distinguish those two states and correctly assign their masses to $D_1(2430)$ and $D_1(2420)$ (and similarly to $D_{s1}(2460)$ and $D_{s1}(2536)$ in the $D_s$ sector).

At first it is important to note that even though $D_1(2430)$ and $D_1(2420)$ have similar masses, their structure is quite different. Due to the heavy charm valence quark $D$ mesons are expected to be qualitatively similar to static-light mesons. Since the spin of a static quark is irrelevant, it is appropriate to label static light mesons by the half-integer total angular momentum $j$ of their light degrees of freedom, i.e.\ the light quarks and gluons. One of the two states $D_1(2430)$ and $D_1(2420)$ has $j \approx 1/2$, while the other has $j \approx 3/2$ (this expectation has been confirmed by model calculations, e.g.\ \cite{Ebert:2009ua}). For a detailed discussion cf.\ e.g.\ \cite{Jansen:2008si,Michael:2010aa}.

The experimental results can be classified according to $j \approx 1/2$ and $j \approx 3/2$ as follows:
\begin{itemize}
\item $D$ mesons: \\
Both $D_1(2430)$ and $D_1(2420)$ can decay to $D^\ast + \pi$. $D_1(2430)$ has a rather large width ($\Gamma = 384^{+130}_{-110} \, \textrm{MeV}$), whereas $D_1(2420)$ is comparably stable ($\Gamma = 27.4 \pm 2.5 \, \textrm{MeV}$). This difference in the widths suggests the assignment $j \approx 1/2$ to $D_1(2430)$ and $j \approx 3/2$ to $D_1(2420)$: while $D_1(2430)$ can then readily decay via an $S$ wave, $D_1(2420)$ is protected by angular momentum $j \approx 3/2$, which only allows a less likely $D$ wave decay.

\item $D_s$ mesons: \\
Only $D_{s1}(2536)$ can decay to $D^\ast + K$. $D_{s1}(2460)$ is too light for such a decay, since $m_{D_{s1}(2460)} < m_D^\ast + m_K$. Both states, however, have rather small widths, $\Gamma < 3.5 \, \textrm{MeV}$ and $\Gamma = 0.92 \pm 0.05 \, \textrm{MeV}$, respectively. Consequently, the heavier state $D_{s1}(2536)$ must be protected by angular momentum, i.e.\ have $j \approx 3/2$, while the other state $D_{s1}(2460)$ corresponds to the remaining $j \approx 1/2$.
\end{itemize}

To decide, which of the two lattice QCD results for masses of $J^{\mathcal{P}}=1^+$ $D$ meson states corresponds to $j \approx 1/2$ and which to $j \approx 3/2$, we study the eigenvectors obtained by solving the generalized eigenvalue problem (\ref{eqn.GEP}). We use linear combinations of $T_1$ meson creation operators from Table~\ref{tab.operators}, which excite not only states with definite $J$, but also with definite light total angular momentum $j$. These $16$ linear combinations are collected in Table~\ref{tab.operators2} and sorted into five classes $C_1 , \ldots , C_5$, one with $\mathcal{P} = -$ and four with $\mathcal{P} = +$, where the latter differ in $j = 1/2 , 3/2$ and $L = 0,1,2$. For each extracted mass $n = 0, 1, 2$ ($0$: $\mathcal{P}=-$ ground state; $1,2$: $\mathcal{P}=+$ excitations) and for each class $k = 1,\ldots,5$  we sum over the squared components of the vectors $u_j^{(n)}$ (cf.\ eq.\ (\ref{eqn.Cv})),
\begin{eqnarray}
Z_k^{(n)}(t) \ \ \equiv \ \ \sum_{\{ j \, | \, \hat{O}_j \in C_k \}} \Big|u_j^{(n)}(t)\Big|^2 .
\end{eqnarray}
These quantities $Z_k^{(n)}$ indicate the parity of the state $n$ and in case of $\mathcal{P} = +$ its light total angular momentum $j$ and orbital angular momentum $L$.

\begin{table}[htb]
\centering
\begin{tabular}{|c|c|c|c|c|c|c|}
\hline
\multirow{3}{*}{class} & \multicolumn{4}{c|}{continuum}        &  \multicolumn{2}{c|}{twisted mass lattice QCD} \\
\cline{2-7} 
 & \multirow{2}{*}{$\mathcal{P}$} & \multirow{2}{*}{$j$} & \multirow{2}{*}{$L$} & \multirow{2}{*}{$\Gamma(\mathbf{n})$,pb} & \multirow{2}{*}{tb, $(\pm,\mp)$} & \multirow{2}{*}{tb, $(\pm,\pm)$} \\
 & & & & & & \\
\hline
\hline
\multirow{8}{*}{$C_1$}&\multirow{8}{*}{$-$}&\multirow{4}{*}{$1/2$}&\multirow{2}{*}{$0$}&$\gamma_1$        &$\pm i\gamma_5\times$&pb\\
&&&&$\gamma_0\gamma_1$&pb&$\pm i\gamma_5\times$\\
\cline{4-7} 
&&&\multirow{2}{*}{$1$}&$\gamma_5(\mathbf{n}\times\vec{\gamma})_j+i\gamma_0\mathbf{n}_j$ &pb&$\pm i\gamma_5\times$\\
&&& &$\gamma_0\gamma_5(\mathbf{n}\times\vec{\gamma})_j+ \mathbf{n}_j$ &$\pm i\gamma_5\times$&pb\\
\cline{3-7}
&&\multirow{4}{*}{$3/2$}&\multirow{2}{*}{$1$}&$\gamma_5(\mathbf{n}\times\vec{\gamma})_j-2i\gamma_0\mathbf{n}_j$&$\pm i\gamma_5\times$&pb\\
&&&&$\gamma_0\gamma_5(\mathbf{n}\times\vec{\gamma})_j-2\mathbf{n}_j$&pb&$\pm i\gamma_5\times$\\
\cline{4-7}
&&&\multirow{2}{*}{$2$}&$\gamma_1(2\mathbf{n}^2_1-\mathbf{n}^2_2-\mathbf{n}^2_3)$&$\pm i\gamma_5\times$&pb\\
&&&&$\gamma_0\gamma_1(2\mathbf{n}^2_1-\mathbf{n}^2_2-\mathbf{n}^2_3)$&pb&$\pm i\gamma_5\times$\\ 
\hline\hline
\multirow{2}{*}{$C_2$}&\multirow{2}{*}{$+$}&\multirow{2}{*}{$1/2$}&\multirow{2}{*}{$0$}&$\gamma_5\gamma_1$&$\pm i\gamma_5\times$&pb\\ 
&&&&$\gamma_0\gamma_5\gamma_1$&pb&$\pm i\gamma_5\times$\\
\hline\hline
\multirow{2}{*}{$C_3$}&\multirow{2}{*}{$+$}&\multirow{2}{*}{$1/2$}&\multirow{2}{*}{$1$}&$(\mathbf{n}\times\vec{\gamma})_j-i\gamma_0\gamma_5\mathbf{n}_j$&$\pm i\gamma_5\times$&pb\\ 
&&&&$\gamma_0(\mathbf{n}\times\vec{\gamma})_j-i\gamma_5\mathbf{n}_j$&pb&$\pm i\gamma_5\times$\\ 
\hline\hline
\multirow{2}{*}{$C_4$}&\multirow{2}{*}{$+$}&\multirow{2}{*}{$3/2$}&\multirow{2}{*}{$1$}&$(\mathbf{n}\times\vec{\gamma})_j+2i\gamma_0\gamma_5\mathbf{n}_j$&pb&$\pm i\gamma_5\times$\\ 
&&&&$\gamma_0(\mathbf{n}\times\vec{\gamma})_j+2i\gamma_5\mathbf{n}_j$&$\pm i\gamma_5\times$&pb\\ 
\hline\hline
\multirow{2}{*}{$C_5$}&\multirow{2}{*}{$+$}&\multirow{2}{*}{$3/2$}&\multirow{2}{*}{$2$}&$\gamma_5\gamma_1(2\mathbf{n}^2_1-\mathbf{n}^2_2-\mathbf{n}^2_3)$&$\pm i\gamma_5\times$&pb\\ 
&&&&$\gamma_0\gamma_5\gamma_1(2\mathbf{n}^2_1-\mathbf{n}^2_2-\mathbf{n}^2_3)$&pb&$\pm i\gamma_5\times$\\
\hline
\end{tabular}

\caption{\label{tab.operators2}$J^\mathcal{P} = 1^+$ meson creation operators with definite $j$ and $L$.}
\end{table}

\begin{figure}[htb]
\centering
\includegraphics[width=5.55cm,angle=-90]{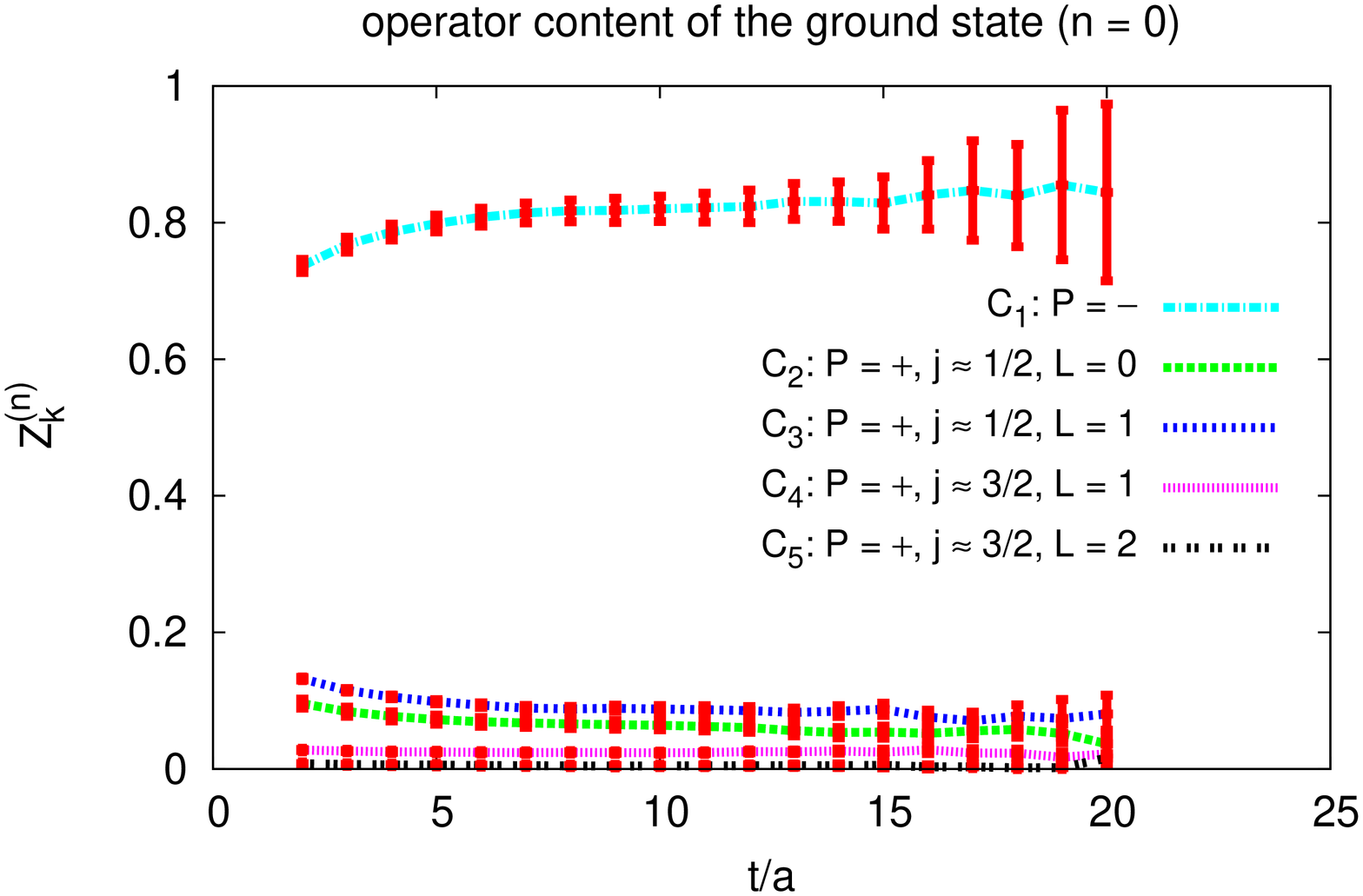}\\
\includegraphics[width=5.55cm,angle=-90]{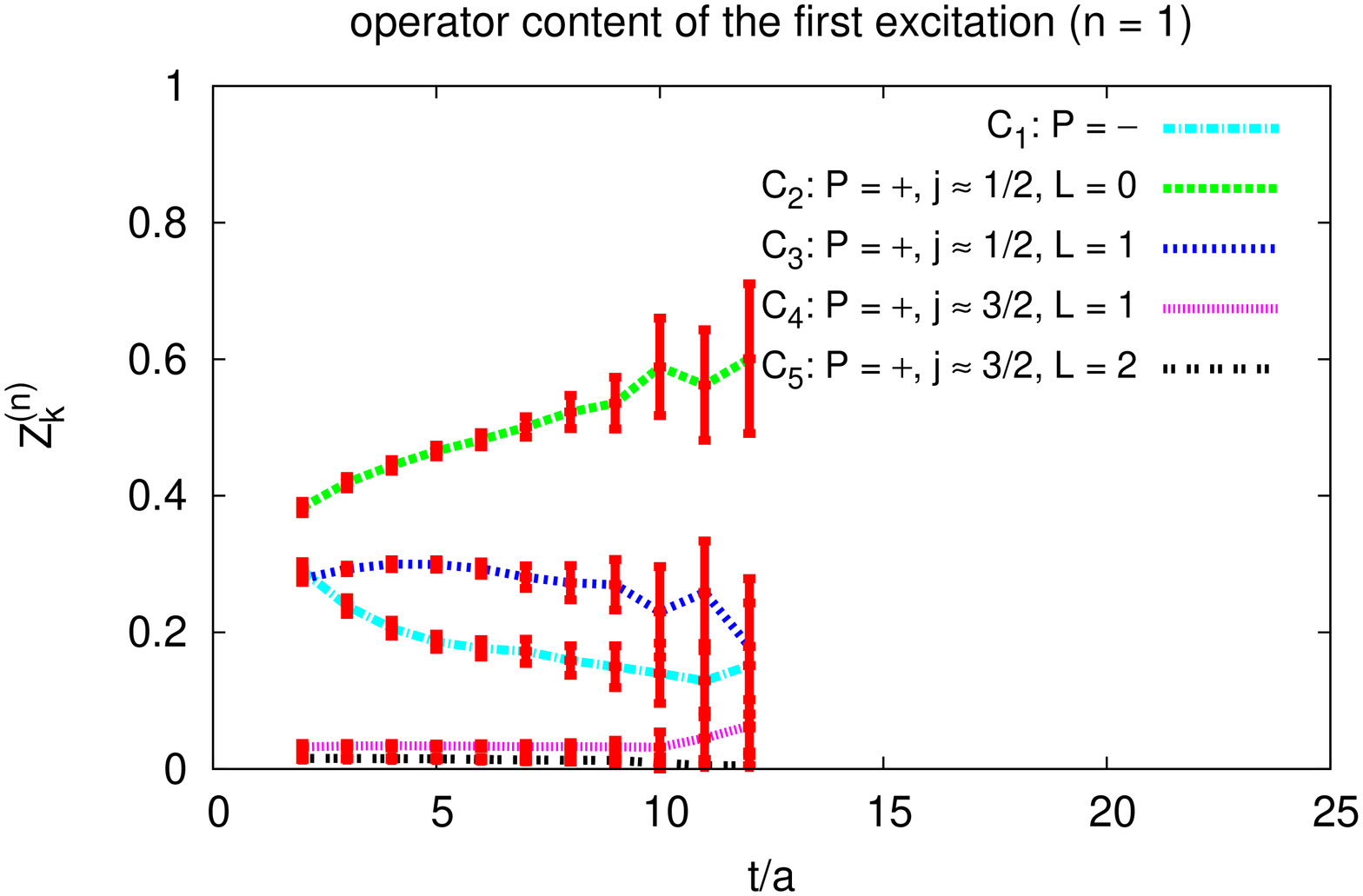}
\includegraphics[width=5.55cm,angle=-90]{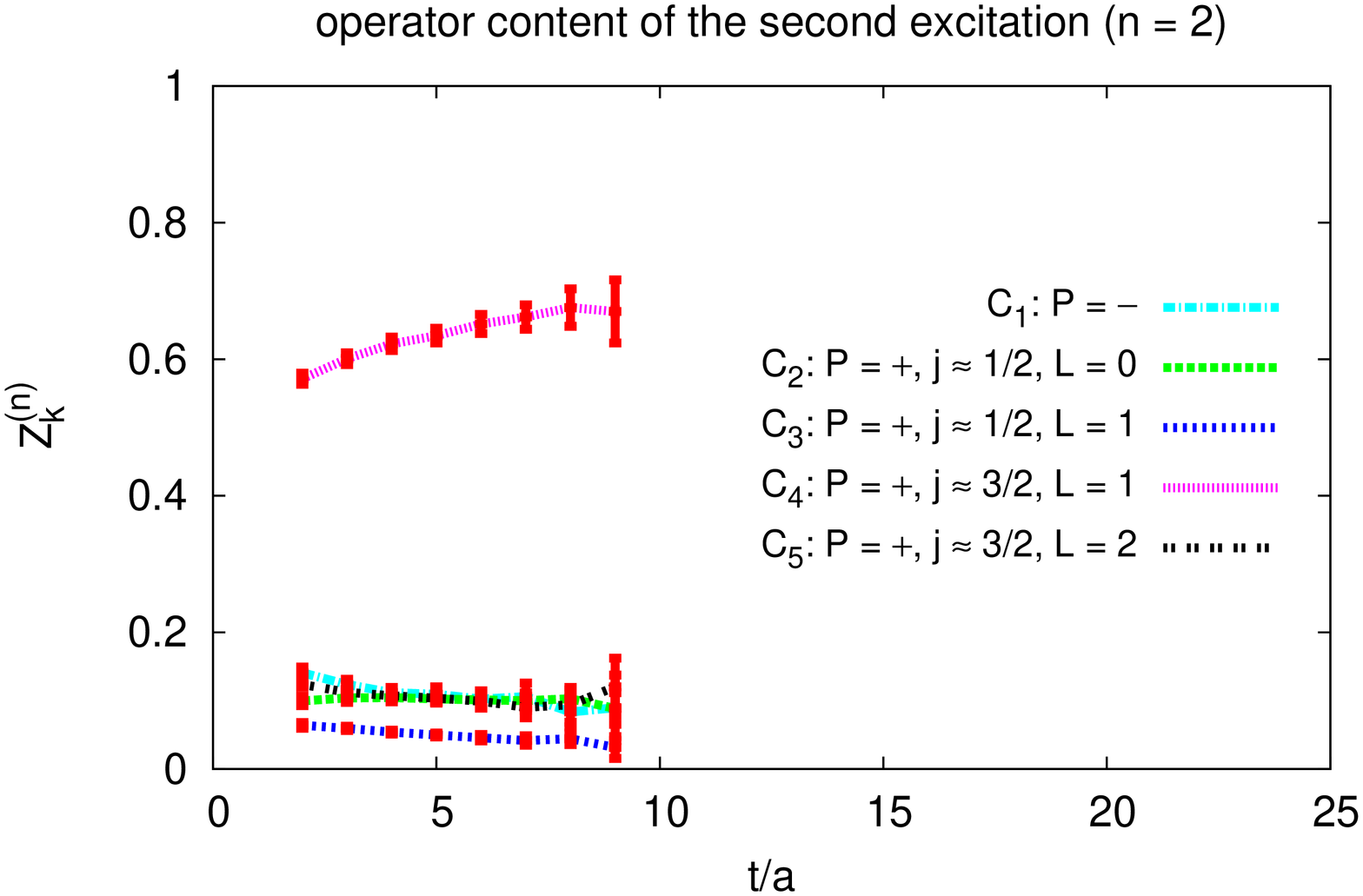}
\caption{\label{fig.T1.1.decomposition_short}Operator content for $J^\mathcal{P} = 1^-$ and $J^\mathcal{P} = 1^+$ $D$ mesons (A30.32 ensemble).}
\end{figure}

$Z_k^{(n)}$ obtained by solving $16 \times 16$ generalized eigenvalue problems are plotted in Figure~\ref{fig.T1.1.decomposition_short} as functions of the temporal separation $t$ for $D$ mesons and the A30.32 ensemble. The upper plot confirms that the ground state has $\mathcal{P} = -$, i.e.\ corresponds to $D^\ast$. The lower left plot corresponding to the first excitation shows a strong dominance of $(\mathcal{P}=+ , j=1/2 , L=0)$ and $(\mathcal{P}=+ , j=1/2 , L=1)$. Clearly, this state should be interpreted as $j \approx 1/2$, i.e.\ as $D_1(2430)$, where $L=0$ has a somewhat larger contribution than $L=1$. The second excitation (lower right plot) is almost exclusively $j \approx 3/2$, i.e.\ corresponds to $D_1(2420)$, with $L=1$ (this is the reason, why we discard $L=2$ meson creation operators in the final determination of $J^\mathcal{P} = 1^+$ $D$ meson masses; cf.\ also the first paragraph of this subsection).

In this context we also refer to the lattice QCD study \cite{Mohler:2012na}, where the identification of $j \approx 1/2$ and $j \approx 3/2$ has been done applying a different strategy, namely including and excluding meson creation operators composed of four quarks. One of the two $J^\mathcal{P} = 1^+$ states is essentially unaffected and, therefore, interpreted as the stable $D_1(2420)$ meson with $j \approx 3/2$, while the other state exhibits a certain sensitivity with respect to the inclusion/exclusion of four quark creation operators and, hence, is interpreted as the less stable $D_1(2430)$ meson with $j \approx 1/2$.

We have carried out a similar analysis for $J^\mathcal{P} = 1^+$ $D_s$ mesons and obtained qualitatively identical results: the lighter of the two extracted states has $j \approx 1/2$ (i.e.\ corresponds to the $D_{s1}(2460)$ meson), while the heavier has $j \approx 3/2$ (i.e.\ corresponds to the $D_{s1}(2536)$ meson). 

\begin{figure}[htb]
\centering
\includegraphics[width=5.55cm,angle=-90]{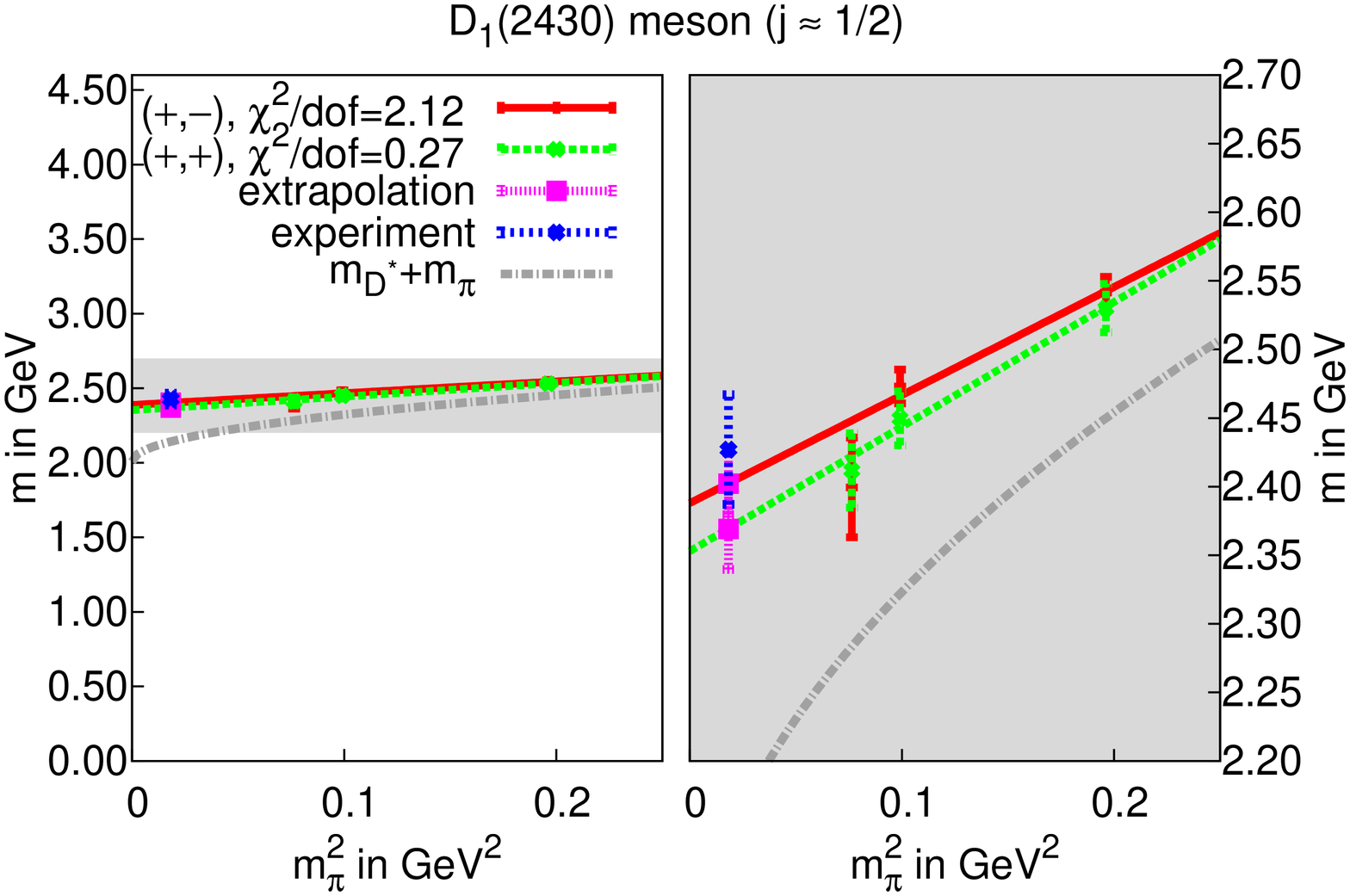}
\includegraphics[width=5.55cm,angle=-90]{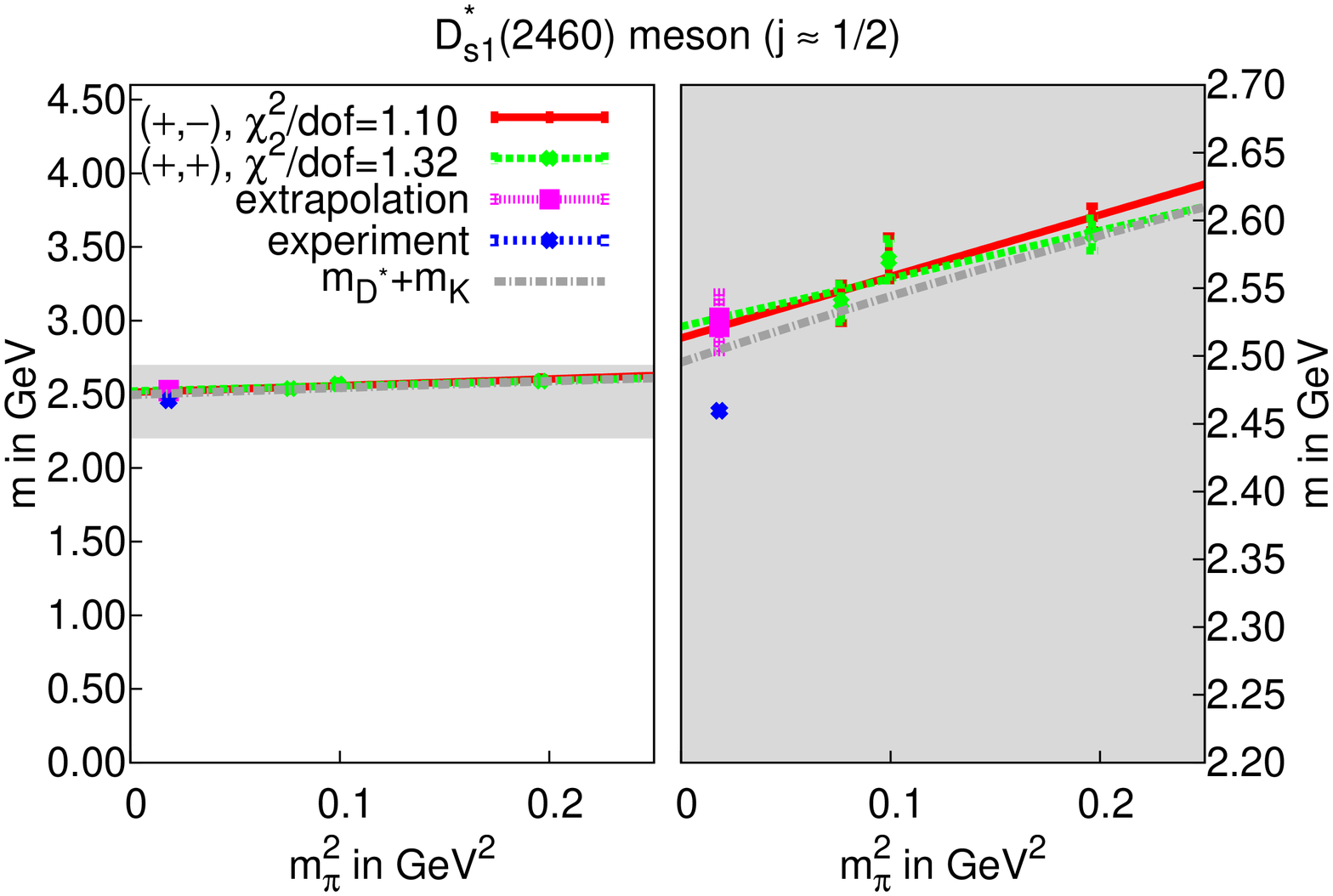} \\
\includegraphics[width=5.55cm,angle=-90]{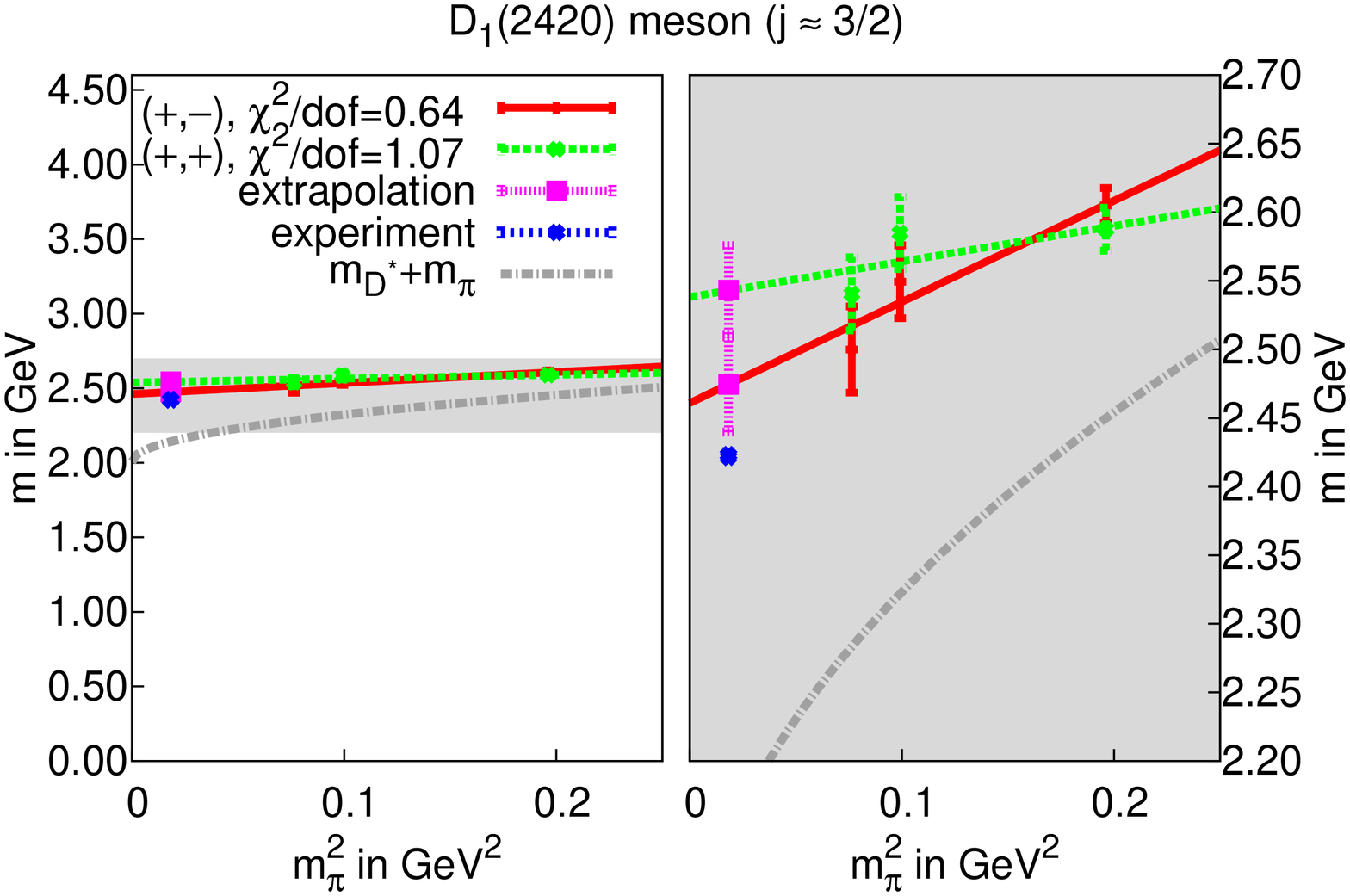}
\includegraphics[width=5.55cm,angle=-90]{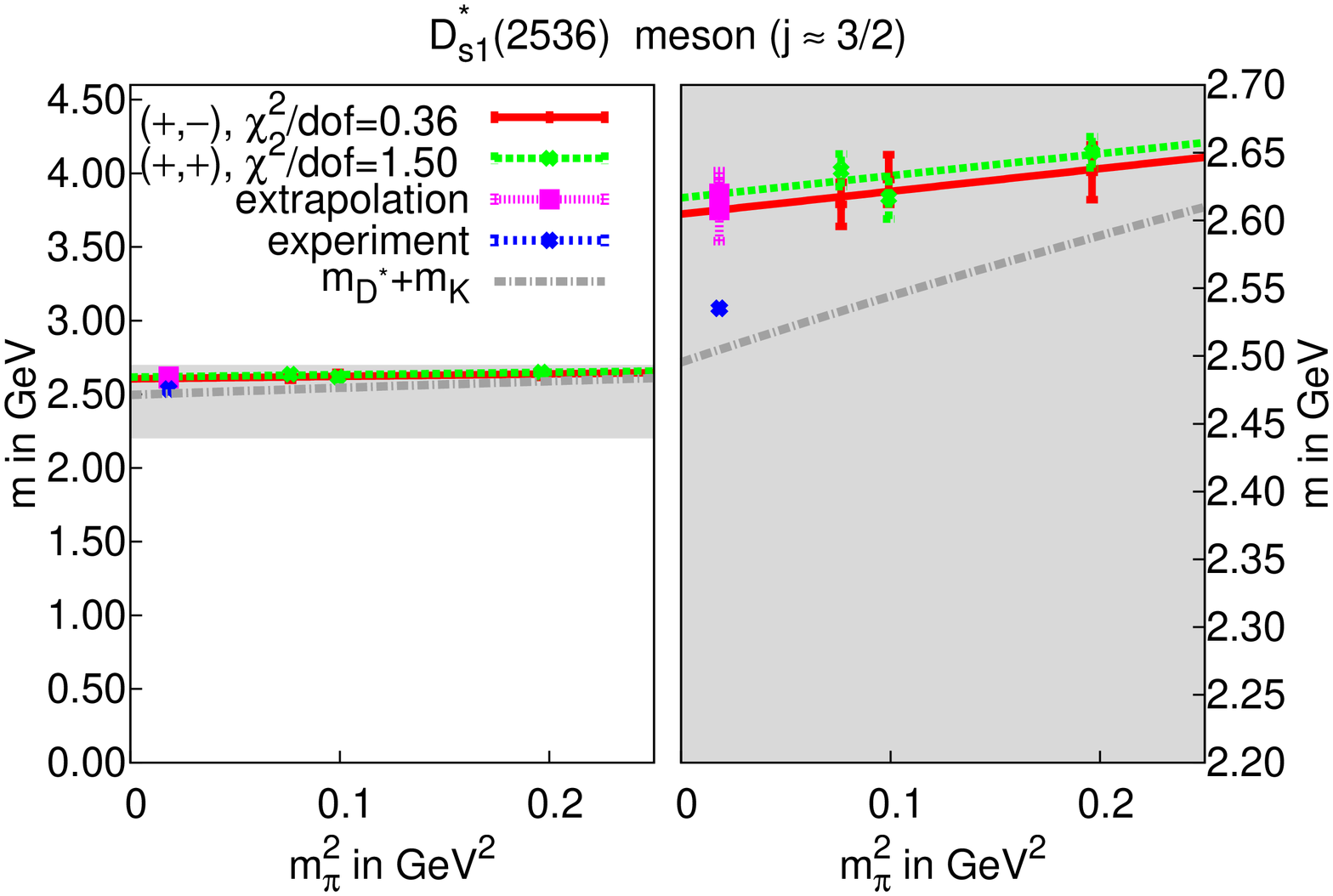}
\caption{\label{fig.T1}$T_1$ representation (spin $J=1$), $\mathcal{P} = +$. \textbf{(top left):} $D_1(2430)$ meson ($j \approx 1/2$). \textbf{(top right):} $D_{s1}(2460)$ meson ($j \approx 1/2$). \textbf{(bottom left):} $D_1(2420)$ meson ($j \approx 3/2$). \textbf{(bottom right):} $D_{s1}(2536)$ meson ($j \approx 3/2$).}
\end{figure}

The $J^\mathcal{P} = 1^+$ $D$ and $D_s$ meson masses are shown in separate plots for $j \approx 1/2$ and $j \approx 3/2$ in Figure~\ref{fig.T1}.

The lattice result for the mass of $D_1(2430)$ ($j \approx 1/2$; top left plot)  is in perfect agreement with the corresponding experimental result. Nevertheless it should be treated with caution. Since the $D_1(2430)$ has a large width and is expected to be rather unstable (it can decay in $D^\ast + \pi$), a solid and more rigorous result will require a proper resonance treatment as e.g.\ pioneered in \cite{Mohler:2012na}.

The corresponding $D_s$ mass with $j \approx 1/2$ ($D_{s1}(2460)$; top right plot) is about $70 \, \textrm{MeV}$ larger than its experimental counterpart. As before for $D_{s0}^\ast$, this is similar to what has been found using quark models, e.g.\ \cite{Ebert:2009ua}. It could be an indication that the $D_{s1}(2460)$ meson is not predominantly a quark-antiquark state, but possibly a mesonic $D^\ast K$ molecule or a diquark-antidiquark pair.

The lattice QCD results for the masses of the $j \approx 3/2$ states, $D_1(2420)$ and $D_{s1}(2536)$ are also around $50 \, \textrm{MeV} \ldots 100 \, \textrm{MeV}$ larger than the corresponding experimental results. This mismatch is somewhat surprising, since both states are rather stable and quark model calculations assuming a straightforward quark antiquark structure are able to reproduce the experimental values quite accurately. We presume that this discrepancy of around $2 \% \ldots 4 \%$ is due to discretization errors. For example when using the lattice spacing $a = 0.0920(21) \, \textrm{fm}$ obtained from the nucleon mass \cite{Alexandrou:2013joa}, the discrepancy is only half as large. It will be interesting to see, whether there will be agreement with experimental results, after performing a continuum extrapolation.


\subsubsection{$E$ and $T_2$ representations (spin $J=2$)}

Effective mass plateaus obtained by solving generalized eigenvalue problems (\ref{eqn.GEP}) are rather short for the $E$ and the $T_2$ representation, i.e.\ the determination of meson masses is more subtle than for the $A_1$ and the $T_1$ representation before.

There seem to be stronger statistical fluctuations for the $E$ representation than for the $T_2$ representation. Therefore, for the $E$ representation we only determine the ground state, which has $\mathcal{P} = +$. We use $4 \times 4$ correlation matrices with creation operator indices $5$ to $8$ (cf.\ Table~\ref{tab.operators}), i.e.\ operators with angular momentum $L = 1$.

For the $T_2$ representation it is possible to extract additionally two $\mathcal{P} = -$ states. The corresponding correlation matrices contain creation operators with indices $1$ to $4$ (operators with $L = 1$; cf.\ Table~\ref{tab.operators}) and for the $D_s$ sector also operators with indices $5$ and $6$ (operators with $L = 2$; cf.\ Table~\ref{tab.operators}).


\subsubsection*{$\mathcal{P} = +$: $D_2^\ast(2460)$ and $D_{s2}^\ast(2573)$}

The ground states in the $E$ and the $T_2$ representations have positive parity and are of similar mass. This strongly indicates that these are $J=2$ states, since total angular momentum $J=2$ is part of both $E$ and $T_2$, in contrast to e.g.\ $J=3$. Consequently, we interpret them as $D_2^\ast(2460)$ and $D_{s2}^\ast(2573)$.

\begin{figure}[htb]
\centering
\includegraphics[width=5.55cm,angle=-90]{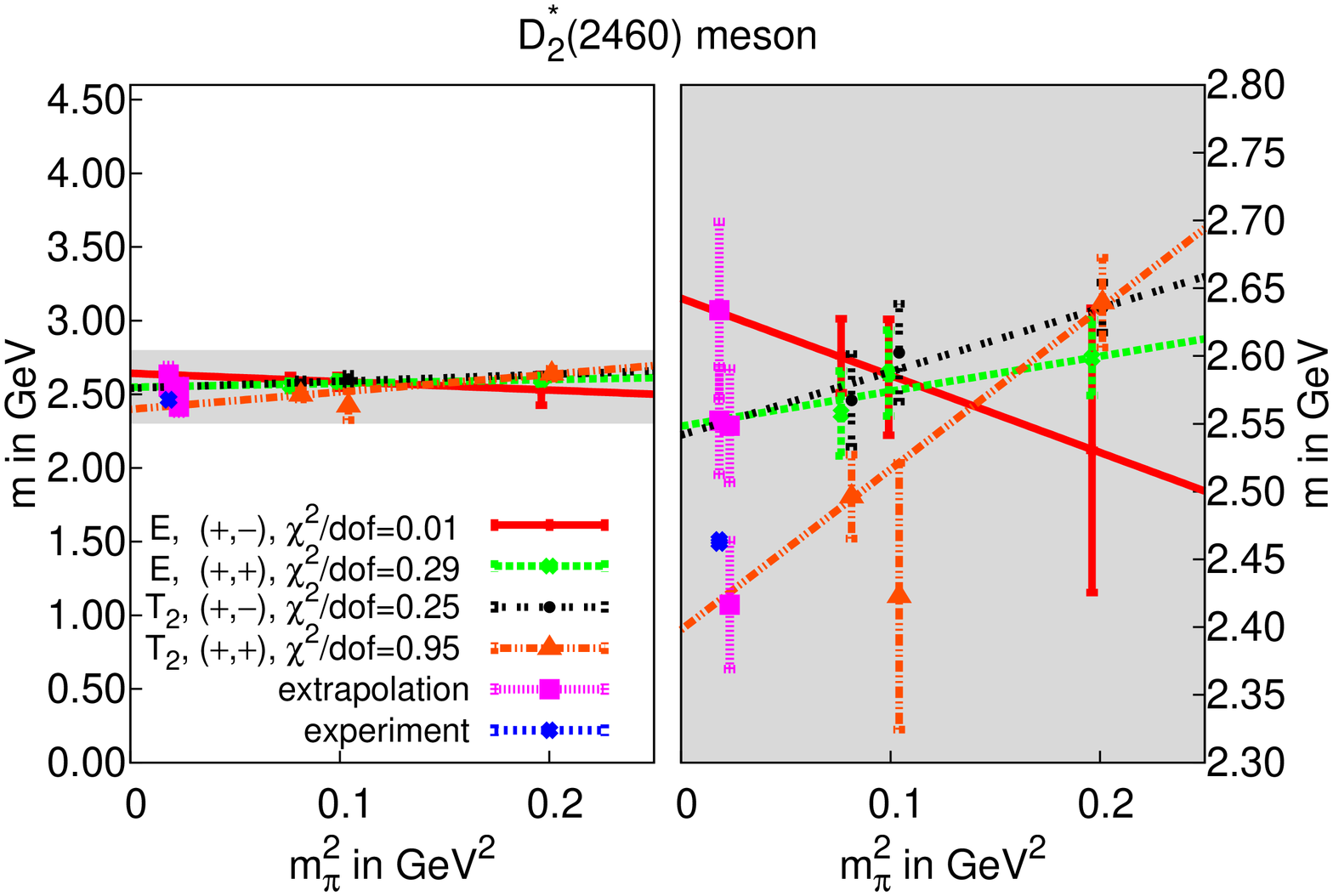}
\includegraphics[width=5.55cm,angle=-90]{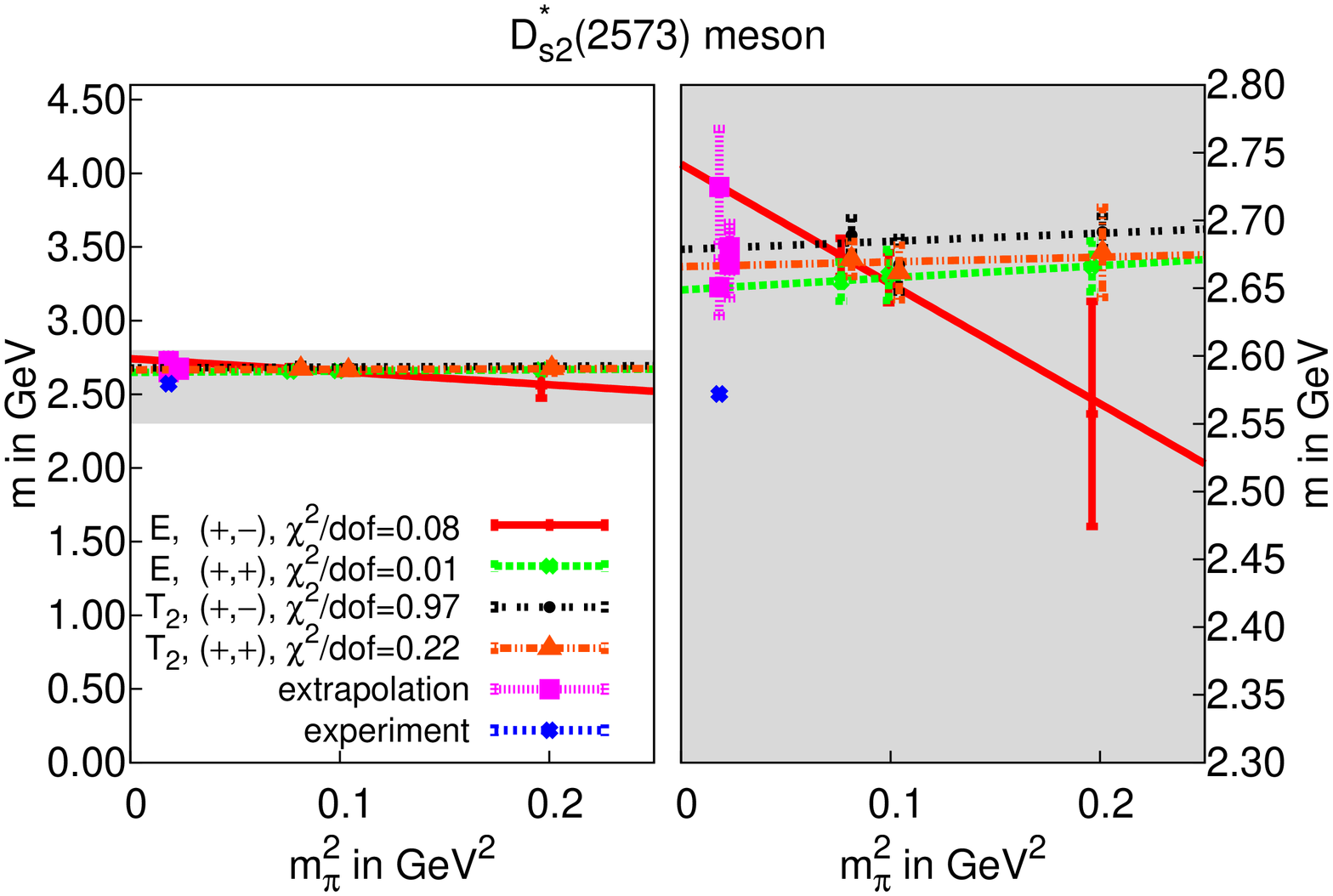}
\caption{\label{fig.E.T2.0}$E$ and $T_2$ representations (spin $J=2$), $\mathcal{P} = +$. \textbf{(left):} $D_2^\ast(2460)$ meson. \textbf{(right):} $D_{s2}^\ast(2573)$ meson.}
\end{figure}

Within statistical errors the meson masses are essentially independent of $m_\pi^2$ (cf.\ Figure~\ref{fig.E.T2.0}). This indicates rather stable states and is expected, because total angular momentum $J=2$ allows decays $D_2^\ast(2460) \rightarrow D + \pi$ and $D_{s2}^\ast(2573) \rightarrow D + K$, respectively, only via $D$ waves. Such decays are strongly suppressed compared to $S$ wave decays, which are possible e.g.\ for corresponding $J^P = 0^+$ and $(J^P = 1^+ , j \approx 1/2)$ states (cf.\ the more detailed discussion in section~\ref{SEC429}).

Discretization errors indicated by the differences between results obtained with $(\pm,\mp)$ and $(\pm,\pm)$ twisted mass sign combinations as well as from the $E$ and the $T_2$ representations are together with statistical errors of the order of $100 \, \textrm{MeV}$ (i.e.\ relative errors $\approx 4\%$). The extrapolation to physically light $u/d$ quark mass yields meson masses, which are also around $100 \, \textrm{MeV}$ larger than the corresponding experimental results. As discussed for the $(J=1,j=3/2)$ states in the previous subsection, discretization errors might be the reason for this discrepancy.


\subsubsection*{$\mathcal{P} = -$: $D(2750)$}

\begin{figure}[htb]
\centering
\includegraphics[width=5.5cm,angle=-90]{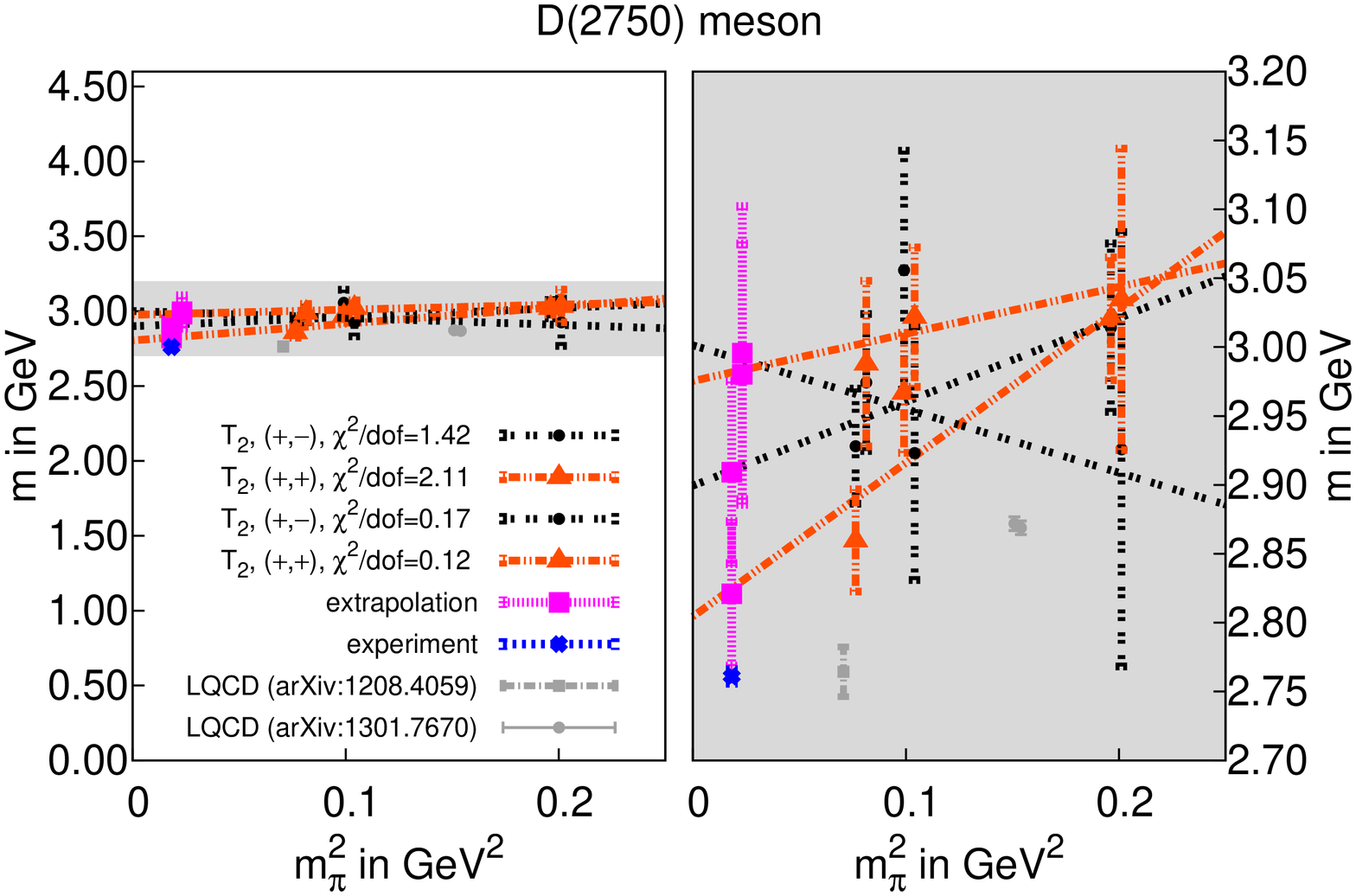}
\includegraphics[width=5.5cm,angle=-90]{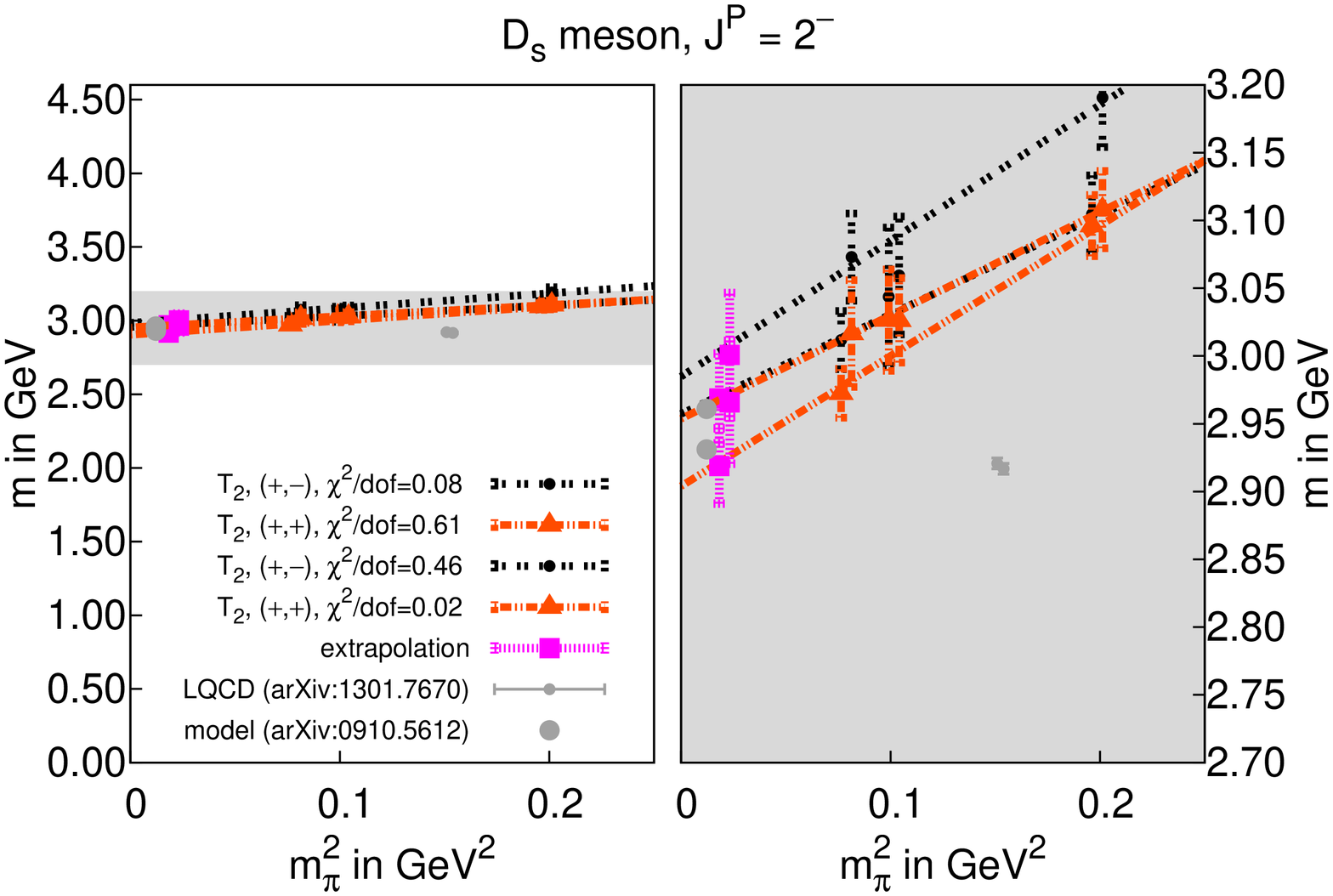}
\caption{\label{fig.E.T2.1}$T_2$ representation (spin $J=2$), $\mathcal{P} = -$. \textbf{(left):} $D(2750)$ meson. \textbf{(right):} $D_s$ sector.}
\end{figure}

Resulting meson masses from the $T_2$ representation are shown in Figure~\ref{fig.E.T2.1}. Crude results with large statistical errors from the $E$ representation (not shown in Figure~\ref{fig.E.T2.1}) are in agreement and, hence, suggest $J = 2$. There seems to be only little dependence on $m_\pi^2$ indicating rather stable states. Note, however, that these negative parity states are less strongly protected by angular momentum than their parity partners $D_2^\ast(2460)$ and $D_{s2}^\ast(2573)$, since $P$ wave decays to $D^\ast + \pi$ and $D^\ast + K$ are possible.

The experimentally observed $D(2750)$ is usually interpreted as a $J^P = 2^-$ state. We find agreement with this experimental result \cite{PDG} within statistical errors as well as with recent lattice QCD computations of the mass of the $D$ meson with $J^\mathcal{P} = 2^-$ (at unphysically heavy $u/d$ quark mass corresponding to $m_\pi \approx 266 \, \textrm{MeV}$ and $m_\pi \approx 391 \, \textrm{MeV}$, respectively) \cite{Mohler:2012na,Moir:2013ub}. The corresponding $J^P = 2^-$ $D_s$ state has experimentally not yet been measured or clearly identified. Therefore, our lattice result can be considered as theoretical prediction. We find meson masses consistent with the lattice QCD computation \cite{Moir:2013ub} and quark model predictions from \cite{Ebert:2009ua}.


\subsection{\label{charmonium}Charmonium}

We compute masses of charmonium states using both twisted mass sign combinations $(\pm,\mp)$ and $(\pm,\pm)$ neglecting disconnected contributions to the correlation matrix elements (\ref{EQN698}). It is expected that this neglect of disconnected diagrams introduces only tiny systematic errors, which are much smaller than our combined statistical uncertainties and lattice discretization errors, because of the OZI suppression of these diagrams. This expectation is quantitatively supported e.g.\ by a quenched lattice QCD computation, where the effect of disconnected diagrams on the charmonium hyperfine splitting is found to be $\approx 1 \ldots 4 \, \textrm{MeV}$ \cite{Levkova:2010ft}. A corresponding consistent perturbative estimate is a shift of charmonium masses by $\approx 2.4 \, \textrm{MeV}$ (cf.\ \cite{Davies:2010ip,Gregory:2010gm,Donald:2012ga} and references therein).

Note that twisted mass quantum numbers are different, when using $(\pm,\mp)$ and $(\pm,\pm)$ twisted mass sign combinations. For $(\pm,\pm)$ charge conjugation $\mathcal{C}$ is a quantum number (but not parity $\mathcal{P}$), while for $(\pm,\mp)$ only the product $\mathcal{P} \circ \mathcal{C}$ is a quantum number (cf.\ also section~\ref{sub.computation}, where we list the sizes of our correlation matrices, or, alternatively, Table~\ref{tab.operators}, columns ``$\mathcal{C}$'' and ``$\mathcal{P C}$'', respectively). All charmonium analyses are based on the maximal sets of available meson creation operators, i.e.\ $2 \times 2$, $4 \times 4$, $6 \times 6$ or $10 \times 10$ correlation matrices as listed in section~\ref{sub.computation}.


\subsubsection{$A_1$ representation (spin $J=0$)}


\subsubsection*{$\mathcal{P} = -$: $\eta_c(1S)$ and $\eta_c(2S)$}

\begin{figure}[htb]
\centering
\includegraphics[width=5.55cm,angle=-90]{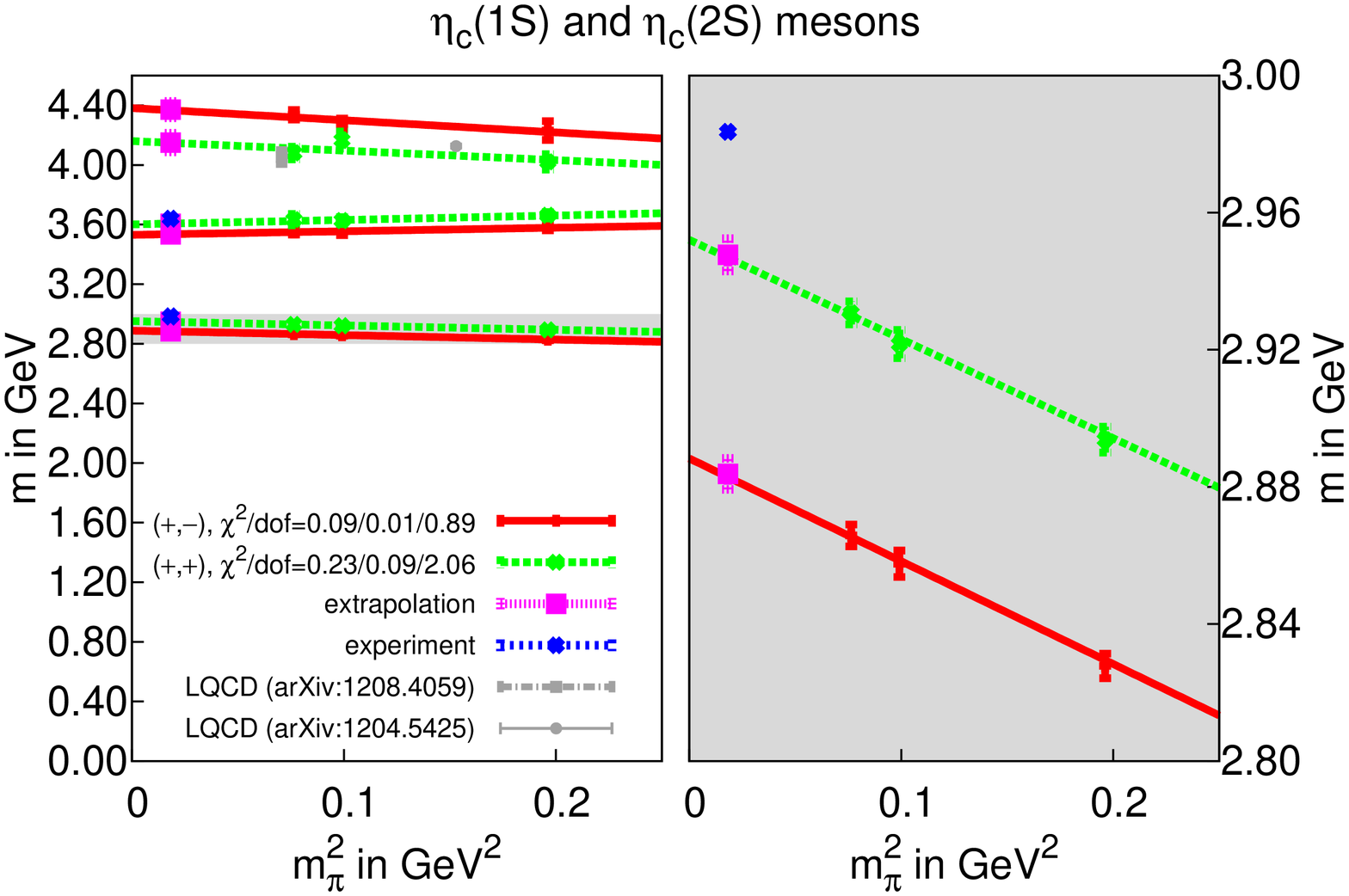}
\includegraphics[width=5.55cm,angle=-90]{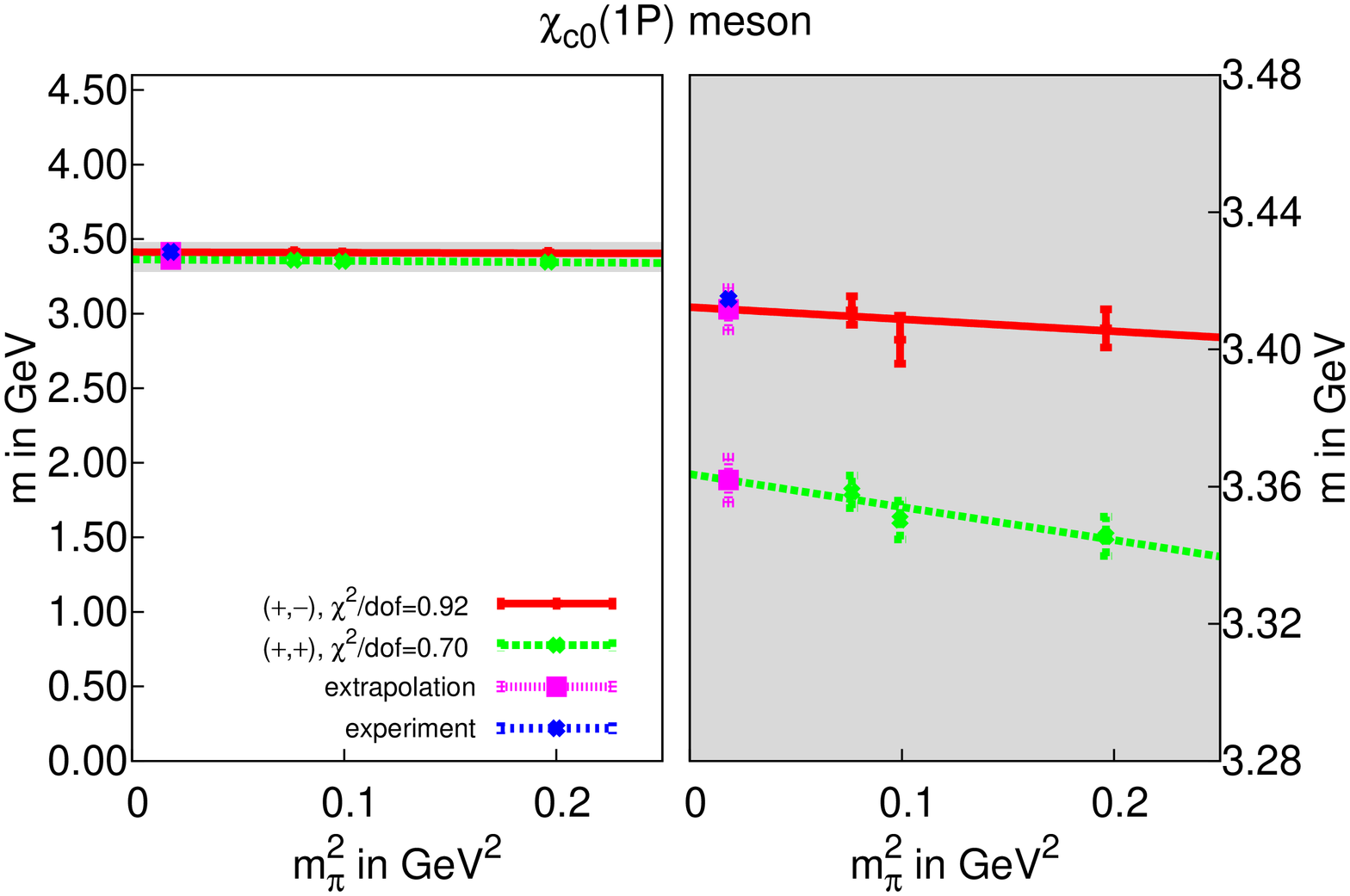}
\caption{\label{fig.cc.A1}$A_1$ representation (spin $J=0$). \textbf{(left):} $\mathcal{P C} = - +$, $\eta_c(1S)$ and $\eta_c(2S)$ mesons. \textbf{(right):} $\mathcal{P C} = + +$, $\chi_{c0}(1P)$ meson.} 
\end{figure}

In Figure~\ref{fig.cc.A1} (left) $J^{\mathcal{P C}} = 0^{- +}$ charmonium masses are shown.

The ground state and the first excitation can clearly be identified with the experimentally known $\eta_c(1S)$ meson and $\eta_c(2S)$ meson. Even though we use a heavy $c$ quark and a heavy $\bar{c}$ antiquark, discretization errors indicated by the differences between results obtained with $(\pm,\mp)$ and $(\pm,\pm)$ twisted mass sign combinations are rather small, around $70 \, \textrm{MeV}$, i.e.\ $\approx 2.5\%$. Moreover, within this uncertainty there is perfect agreement with experimental results \cite{PDG}. The fact that our lattice QCD result for the mass of $\eta_c(1S)$ at finite lattice spacing is slightly below the experimental result is also in agreement with a 2-flavor twisted mass lattice QCD study including the $\eta_c(1S)$ meson \cite{Becirevic:2012dc}.

We are also able to extract a crude signal for a third state, for which there is currently no clearly identified experimental counterpart. Such a state has, however, been observed in previous lattice QCD computations, e.g.\ in \cite{Liu:2012ze,Mohler:2012na} (at unphysically heavy $u/d$ quark mass corresponding to $m_\pi \approx 266 \, \textrm{MeV}$ and $m_\pi \approx 391 \, \textrm{MeV}$, respectively), around $400 \, \textrm{MeV} \ldots 500 \, \textrm{MeV}$ above the mass of the $\eta_c(2S)$ meson. Here we observe an unexpectedly large difference of around $200 \, \textrm{MeV}$ between $(\pm,\mp)$ and $(\pm,\pm)$ results. The reason for this could be the rather bad plateau quality of the corresponding effective masses (short plateaus formed by only three points, before the signal is lost in noise), which might be a sign of contamination by even higher excitations. This is also supported by the fact that our result is around $600 \, \textrm{MeV} \ldots 800 \, \textrm{MeV}$ heavier than the mass of $\eta_c(2S)$, i.e.\ heavier than the above mentioned lattice QCD results from \cite{Liu:2012ze,Mohler:2012na}. Another reason could be mixing with a lighter state of the $\mathcal{P C} = + +$ sector: for the $(\pm,\pm)$ result we observe a rather large contribution of around $30\%$ of the operator $\Gamma(\mathbf{n}) = \gamma_0 \gamma_j \mathbf{n}_j$ corresponding to quantum numbers $\mathcal{P C} = + +$ (index $8$ in Table~\ref{tab.operators}), while for $(\pm,\mp)$ such a mixing is excluded by the symmetry $\mathcal{P} \circ \mathcal{C}$.

Even though we use a meson creation operator with quantum numbers $\mathcal{P C} = - -$ (index $5$ in Table~\ref{tab.operators}), we do not obtain a clear signal, i.e.\ a trustworthy effective mass plateau, to extract a corresponding meson mass.


\subsubsection*{$\mathcal{P} = +$: $\chi_{c0}(1P)$}

In Figure~\ref{fig.cc.A1} (right) the resulting masses for $\chi_{c0}(1P)$ (quantum numbers $J^{\mathcal{P C}} = 0^{+ +}$) are shown. Discretization errors indicated by the differences between results obtained with $(\pm,\mp)$ and $(\pm,\pm)$ twisted mass sign combinations are rather small, around $50 \, \textrm{MeV}$, i.e.\ $\approx 1.5\%$. The $u/d$ quark mass extrapolated $(\pm,\mp)$ result, for which one expects smaller discretization errors, is in perfect agreement with its experimental counterpart \cite{PDG}.

Similarly to the $\mathcal{P C} = - -$ sector we are not able to reliably determine an exotic $\mathcal{P C} = + -$ state.


\subsubsection{\label{SEC_cc_T1}$T_1$ representation (spin $J=1$ and $J=3$)}


\subsubsection*{$\mathcal{P} = -$: $J/\Psi(1S)$, $\Psi(2S)$ and $\Psi(3770)$}

\begin{figure}[htb]
\centering
\includegraphics[width=5.55cm,angle=-90]{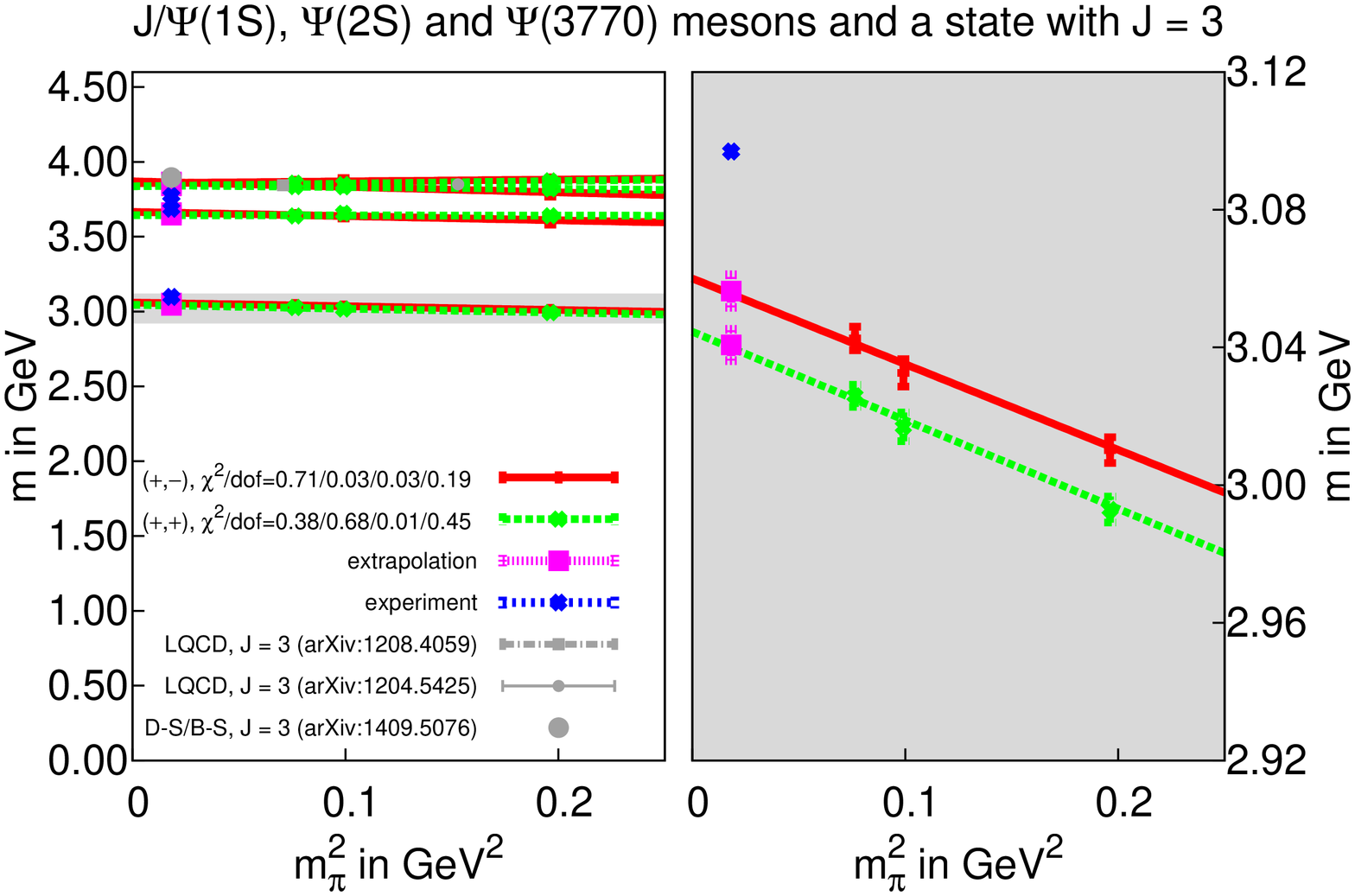} \\
\includegraphics[width=5.55cm,angle=-90]{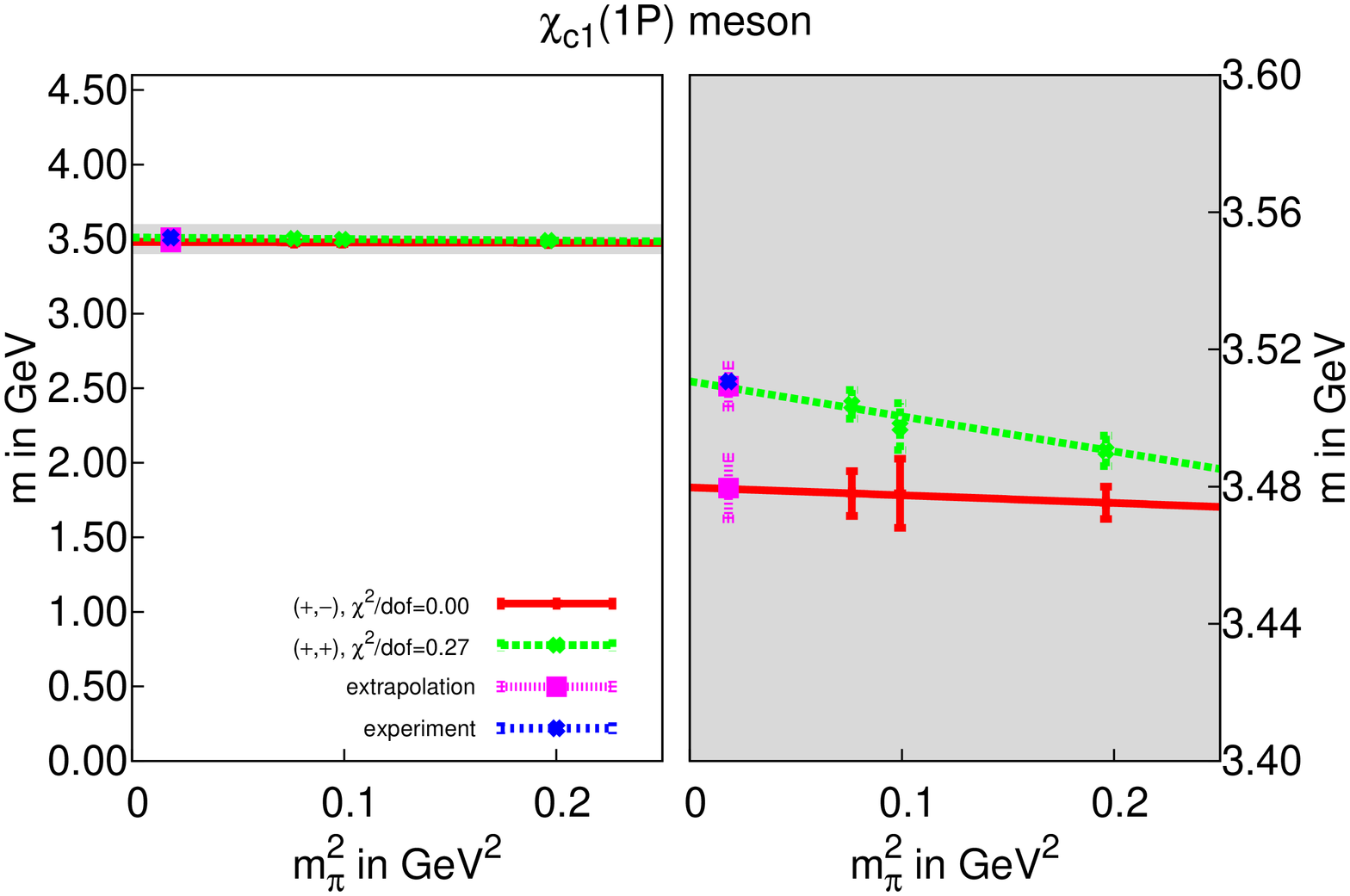}
\includegraphics[width=5.55cm,angle=-90]{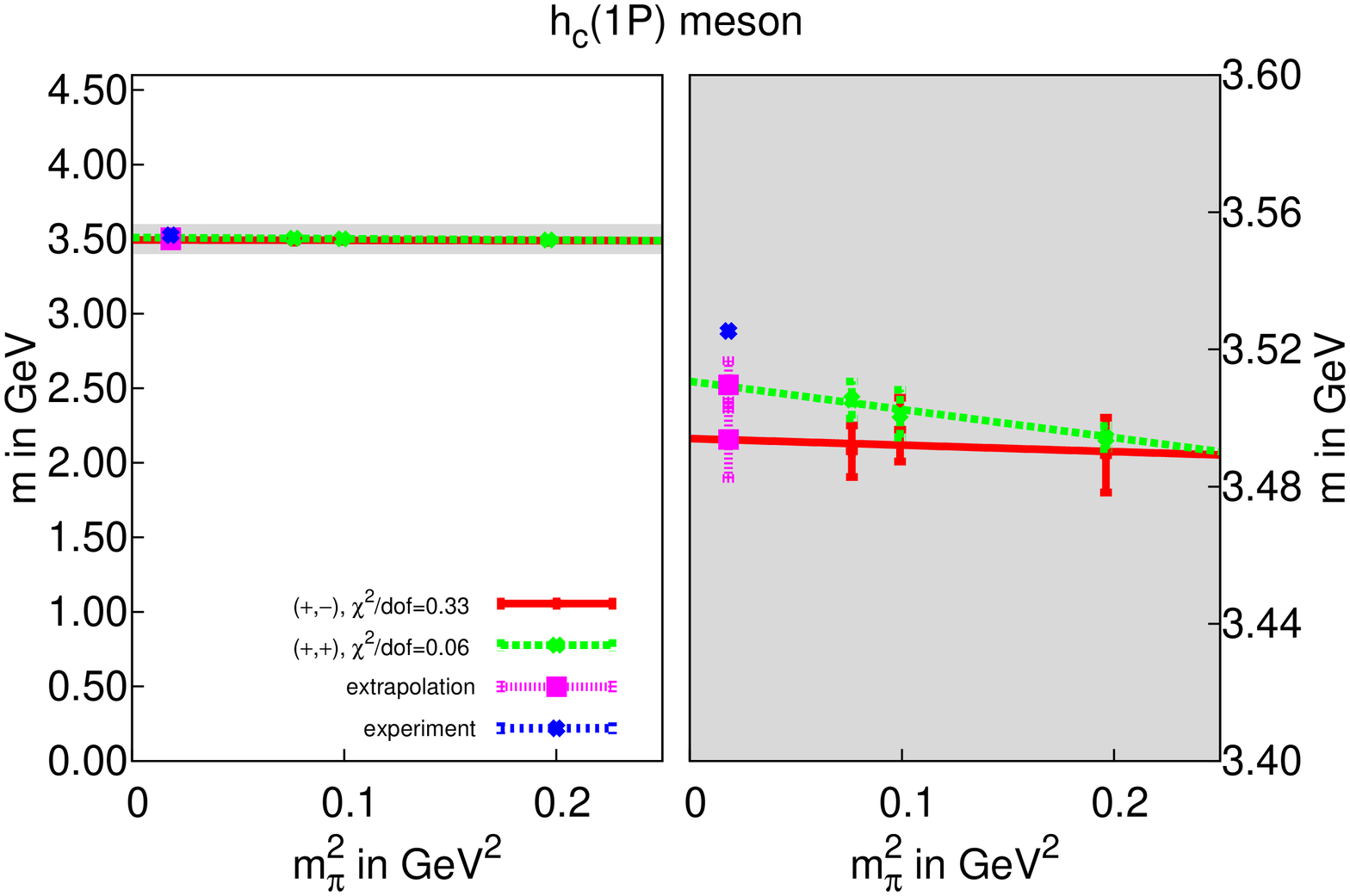}
\caption{\label{fig.cc.T1}$T_1$ representation (spin $J=1$ and $J=3$). \textbf{(top):} $\mathcal{P C} = - -$, $J/\Psi(1S)$, $\Psi(2S)$ and $\Psi(3770)$ mesons and a state with $J = 3$. \textbf{(bottom left):} $\mathcal{P C} = + +$, $\chi_{c1}(1P)$ meson. \textbf{(bottom right):} $\mathcal{P C} = + -$, $h_{c}(1P)$ meson.}
\end{figure}

For $\mathcal{P C} = - -$ we are able to extract the masses of four states, which are shown in in Figure~\ref{fig.cc.T1} (top).

The two lowest states can be identified as the $J/\Psi(1S)$ and the $\Psi(2S)$ meson. The lattice QCD results for both twisted mass discretizations $(\pm,\mp)$ and $(\pm,\pm)$ and the corresponding experimental results \cite{PDG} differ by less than $50 \, \textrm{MeV}$, i.e.\ are in agreement within the expected discretization errors.

The second and third excitation are very close in mass, around $3800 \, \textrm{MeV} \ldots 3900 \, \textrm{MeV}$. One of these two states should correspond to the $\Psi(3770)$ meson. The other state seems to correspond to a $J=3$ state in the continuum (the continuum spins of $T_1$ are $J = 1,3,4,\ldots$), because a state with the same mass is observed in the $T_2$ representation (continuum spins $J = 2,3,4,\ldots$), but not in the $E$ representation (continuum spins $J = 2,4,\ldots$; cf.\ section~\ref{SEC_cc_J2}). A $J^{\mathcal{P C}} = 3^{- -}$ state in this energy region has also been observed in other lattice QCD studies (cf.\ e.g.\ \cite{Liu:2012ze,Mohler:2012na}) and has also been predicted using Dyson-Schwinger and Bethe-Salpeter equations \cite{Fischer:2014cfa}.

Even though we use two creation operators with exotic quantum numbers $\mathcal{P C} = - +$ (indices $6$ and $12$ in Table~\ref{tab.operators}), we do not obtain a clear signal, to reliably extract a corresponding mass.


\subsubsection*{$\mathcal{P} = +$: $\chi_{c1}(1P)$ and $h_{c}(1P)$}

In Figure~\ref{fig.cc.T1} (bottom) the resulting masses for $\chi_{c1}(1P)$ (quantum numbers $J^{\mathcal{P C}} = 1^{+ +}$) and $h_{c}(1P)$ (quantum numbers $J^{\mathcal{P C}} = 1^{+ -}$) are shown. The lattice QCD results for both twisted mass discretizations $(\pm,\mp)$ and $(\pm,\pm)$ and the corresponding experimental results \cite{PDG} differ by less than $40 \, \textrm{MeV}$, i.e.\ are in agreement within the estimated discretization errors.


\subsubsection{\label{SEC_cc_J2}$E$ and $T_2$ representations (spin $J=2$ and $J=3$)}


\subsubsection*{$\mathcal{P} = +$: $\chi_{c2}(1P)$ and $\chi_{c2}(2P)$}

\begin{figure}[htb]
\centering
\includegraphics[width=5.55cm,angle=-90]{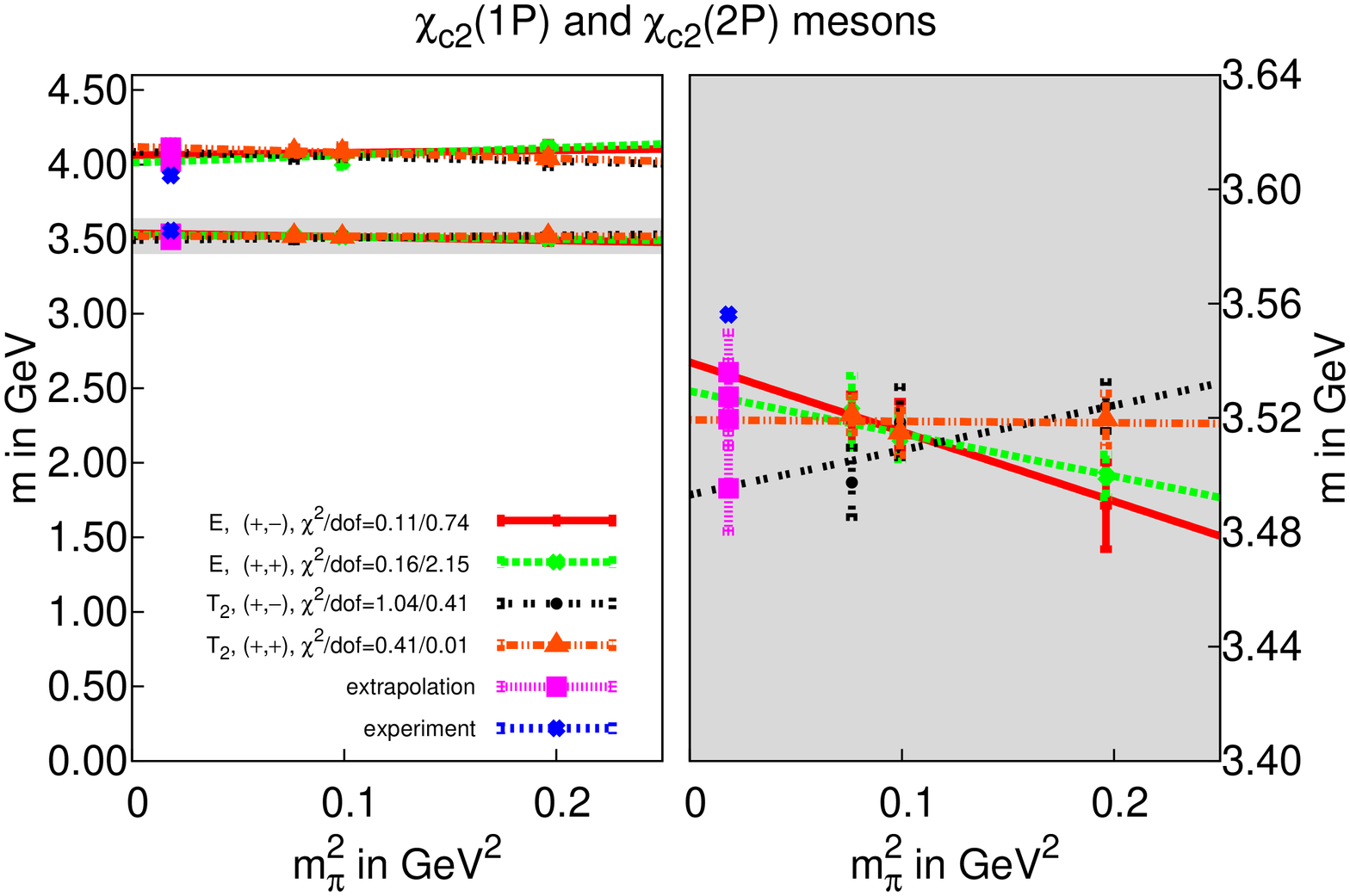} \\
\includegraphics[width=5.55cm,angle=-90]{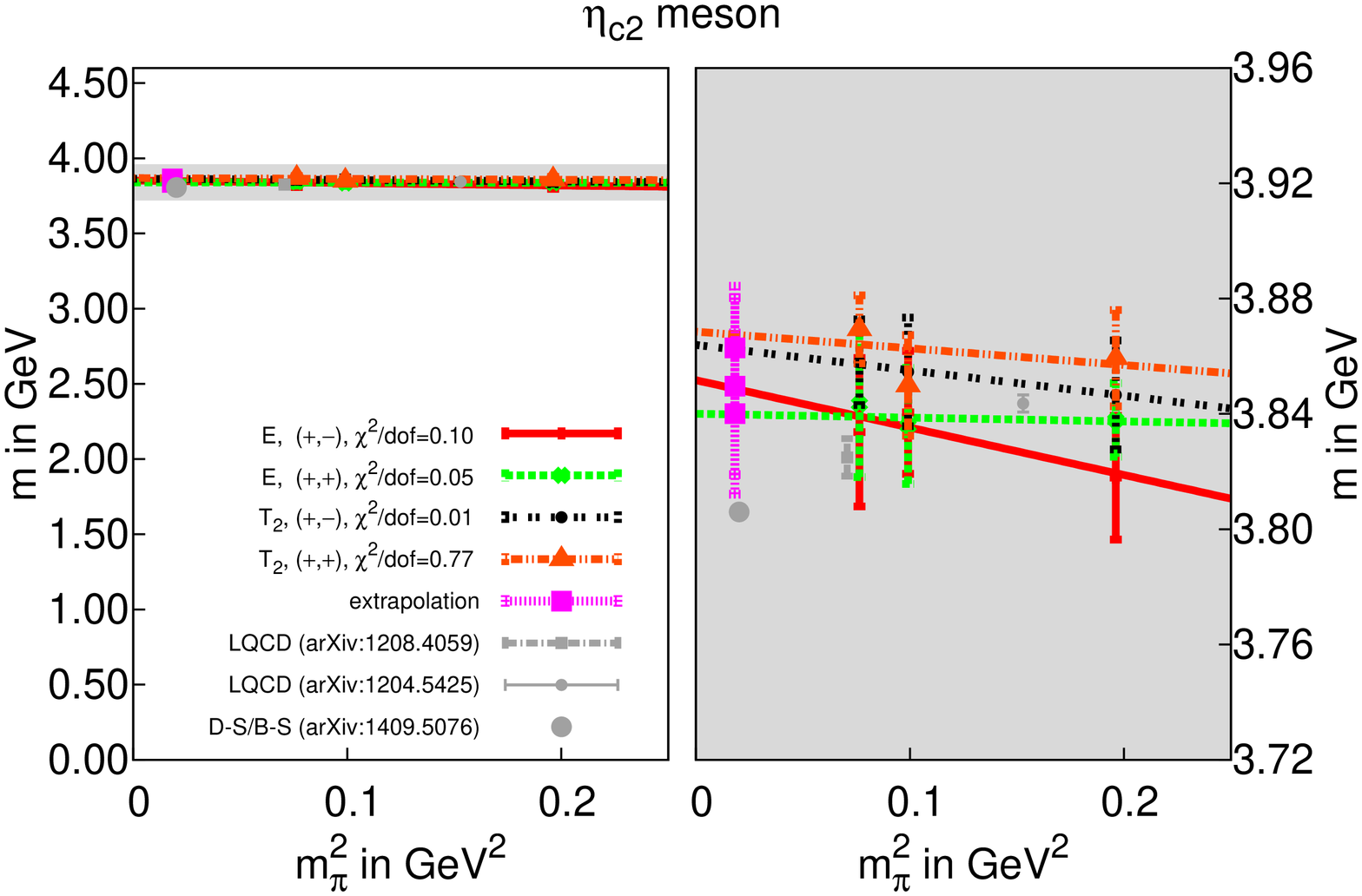}
\includegraphics[width=5.55cm,angle=-90]{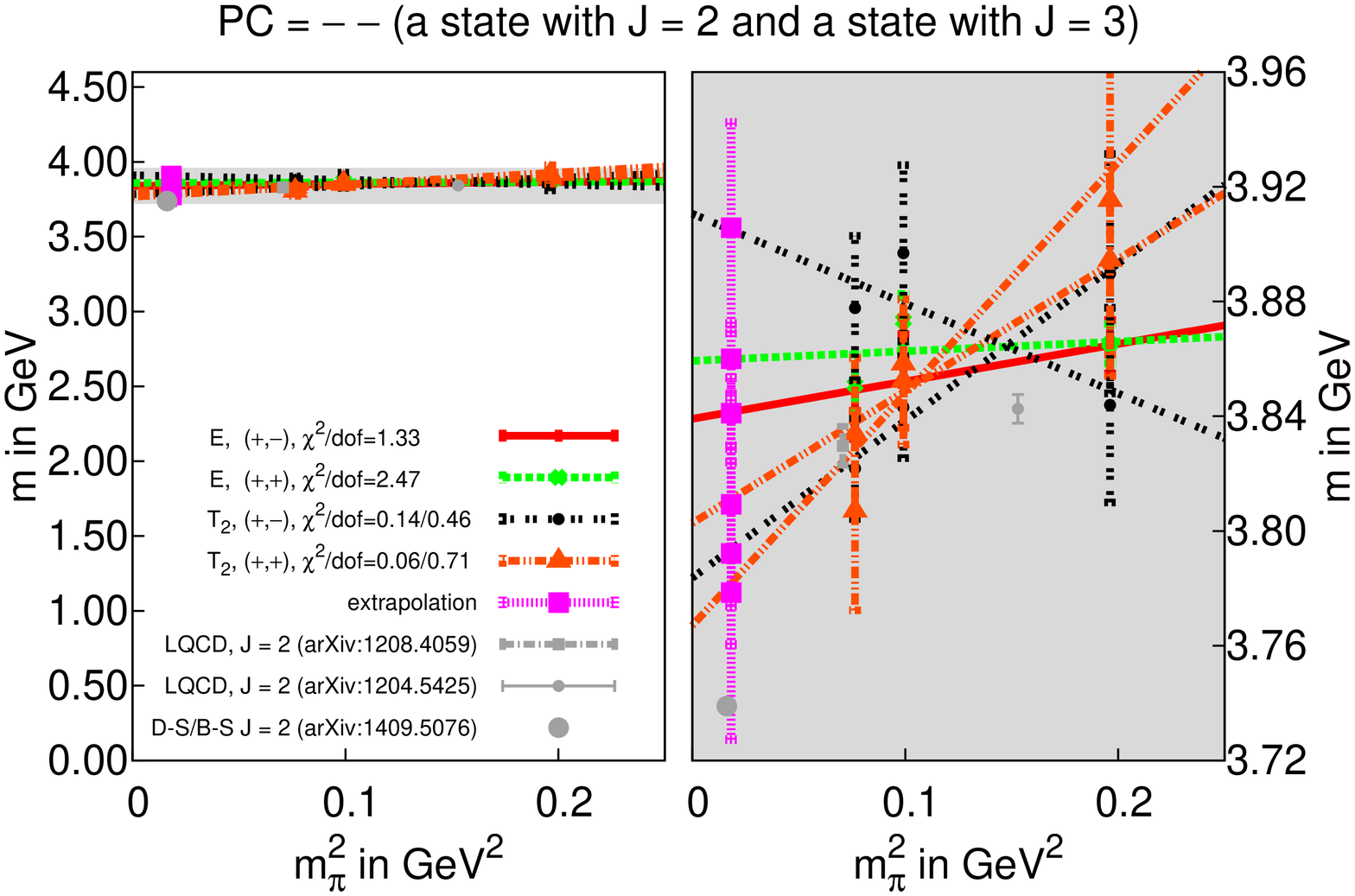}
\caption{\label{fig.cc.ET2}$E$ and $T_2$ representations (spin $J=2$ and $J=3$). \textbf{(top):} $\mathcal{P C} = + +$, $\chi_{c2}(1P)$ and $\chi_{c2}(2P)$ mesons. \textbf{(bottom left):} $\mathcal{P C} = - +$, $\eta_{c2}$ meson. \textbf{(bottom right):} $\mathcal{P C} = - -$, one state with $J = 2$ and (in the $T_2$ representation) another state with $J = 3$.}
\end{figure}

For $\mathcal{P} = +$ we are able to extract two states with $\mathcal{C} = +$ both for the $E$ and the $T_2$ representation (cf.\ Figure~\ref{fig.cc.ET2} [top]). Since the masses of the two ground states as well as the masses of the two excitations agree within errors, we interpret them as $J = 2$ states, i.e.\ as the $\chi_{c2}(1P)$ meson and the $\chi_{c2}(2P)$ meson. Within statistical errors there is essentially no dependence on $m_\pi^2$. Discretization errors indicated by the differences between results obtained with $(\pm,\mp)$ and $(\pm,\pm)$ twisted mass sign combinations as well as from the $E$ and the $T_2$ representations are for the $\chi_{c2}(1P)$ meson together with statistical errors of the order of $50 \, \textrm{MeV}$ (i.e.\ relative errors $\approx 1.5\%$). Within this estimated uncertainty there is excellent agreement with the corresponding experimental result. For the $\chi_{c2}(2P)$ meson the uncertainty is roughly twice as large, i.e.\ around $100 \, \textrm{MeV}$. Our lattice QCD result is also slightly larger (around $150 \, \textrm{MeV}$ i.e.\ $1.5 \sigma$ larger) than the experimentally measured mass, which could be an indication that there is a small contamination of higher excitations (cf.\ also the lattice QCD study \cite{Liu:2012ze}, where a similar trend has been observed).

\subsubsection*{$\mathcal{P} = -$: $\eta_{c2}$}

For $\mathcal{P} = -$ we are able to determine one state with $\mathcal{C} = +$ both for the $E$ and the $T_2$ representation, which should correspond to the $\eta_{c2}$ meson (cf.\ Figure~\ref{fig.cc.ET2} [bottom left]). Again combined discretization and statistical errors are around $50 \, \textrm{MeV}$. Currently there is no established experimental result available, to which we can confront our result. We find, however, agreement with theoretical predictions from other lattice QCD computations \cite{Liu:2012ze,Mohler:2012na} and from a calculation using Dyson-Schwinger and Bethe-Salpeter equations \cite{Fischer:2014cfa}.

For $\mathcal{P} = -$ and $\mathcal{C} = -$ in the mass region $3800 \, \textrm{MeV} \ldots 3900 \, \textrm{MeV}$ we find only a single state for the $E$ representation (continuum spins $J = 2,4,\ldots$), but two states for the $T_2$ representation (continuum spins $J = 2,3,4,\ldots$) (cf.\ Figure~\ref{fig.cc.ET2} [bottom right]). This suggests to interpret the state, which is present both in the $E$ and the $T_2$ representation, as a $J = 2$ state in the continuum and the additional state in the $T_2$ representation as a $J = 3$ state in the continuum. This is supported by our results for the $T_1$ representation (continuum spins $J = 1,3,4,\ldots$), where we have extracted a state in the same mass region (cf.\ section~\ref{SEC_cc_T1}, in particular Figure~\ref{fig.cc.T1} [top]). Again we can only compare to other theoretical predictions, e.g.\ from \cite{Liu:2012ze,Mohler:2012na,Fischer:2014cfa}, which are in agreement with our results.



\section{\label{SEC669}Summary and conclusions}

We have computed masses of low lying $D$ meson, $D_s$ meson and charmonium states with total angular momentum $J = 0,1,2,3$, parity $\mathcal{P} = -,+$ and charge conjugation $\mathcal{C} = -,+$ using Wilson twisted mass lattice QCD. We have used gauge link ensembles generated by the European Twisted Mass Collaboration with three different $u/d$ quark masses corresponding to $m_\pi \in \{ 276 \, \textrm{MeV} \, , \, 315 \, \textrm{MeV} \, , \, 443 \, \textrm{MeV} \}$. After performing computations on these three ensembles, we have extrapolated the resulting meson masses to physically light $u/d$ quark mass.

Our computations are currently limited to a single lattice spacing, $a \approx 0.0885 \, \textrm{fm}$. Therefore, we are not able to study the continuum limit at the moment. In Wilson twisted mass lattice QCD it is, however, possible to compute meson masses using two different discretizations, either $(\pm,\mp)$ or $(\pm,\pm)$ twisted mass sign combinations in the meson creation operators (cf.\ section~\ref{subsec.operators}). The differences between the resulting meson masses can be considered as crude estimates of lattice discretization errors. For the majority of mesons we have found differences of the order of $50 \, \textrm{MeV}$, which we take as an estimate of discretization errors.

We expect these discretization errors to be the currently dominating source of systematic uncertainty. Further sources of error are listed in the following:
\begin{itemize}
\item \textbf{Finite spatial volume:} \\ Since for all three ensembles the spatial volume is rather large, i.e.\ $m_\pi L \gtapprox 4$, we expect that finite volume corrections are negligible compared to the above mentioned discretization errors of $\approx 50 \, \textrm{MeV}$.

\item \textbf{Disconnected diagrams:} \\ When computing correlation matrices for charmonium states, we have omitted disconnected contributions. These contributions are expected to be very small due to OZI suppression. The corresponding systematic errors for charmonium masses have been estimated to be less than $4 \,\textrm{MeV}$ (cf.\ section~\ref{charmonium} and \cite{Levkova:2010ft,Davies:2010ip,Gregory:2010gm,Donald:2012ga}).

\item \textbf{Electromagnetism and isospin breaking:} \\ We estimate the magnitude of electromagnetic corrections and effects due to isospin breaking by comparing experimental results for masses of essentially stable charged and neutral $D$ mesons ($D$, $D^\ast$, $D_1(2420)$, $D_2^\ast$), yielding corresponding systematic errors $\ltapprox 5 \, \textrm{MeV}$. This is consistent with \cite{Davies:2010ip,Gregory:2010gm,Donald:2012ga}, where such effects have been estimated to be $\approx 2.6 \, \textrm{MeV}$ using a potential model.

\item \textbf{Extrapolations in the up/down quark mass:} \\ The extrapolations to physically light $u/d$ quark mass are linear in $m_\pi^2$. To quantify a possibly associated systematic error, one could compare different strategies of extrapolation, e.g.\ also including a quadratic term in $m_\pi^2$ or using parameterizations of meson masses obtained from effective theories. We anticipate the corresponding uncertainty to be smaller than the currently estimated discretization errors of around $50 \, \textrm{MeV}$. We plan to study this issue in more detail in a future publication, where we will address the continuum limit.

\item \textbf{Tuning of strange and charm sea quark mass:} \\ A probably tiny error might arise due to the slight deviation of strange and charm sea quark masses from their physical values (cf.\ section~\ref{subsec.massdetermination}).
\end{itemize}

\begin{figure}[htb]
\centering
\includegraphics[width=5.55cm,angle=-90]{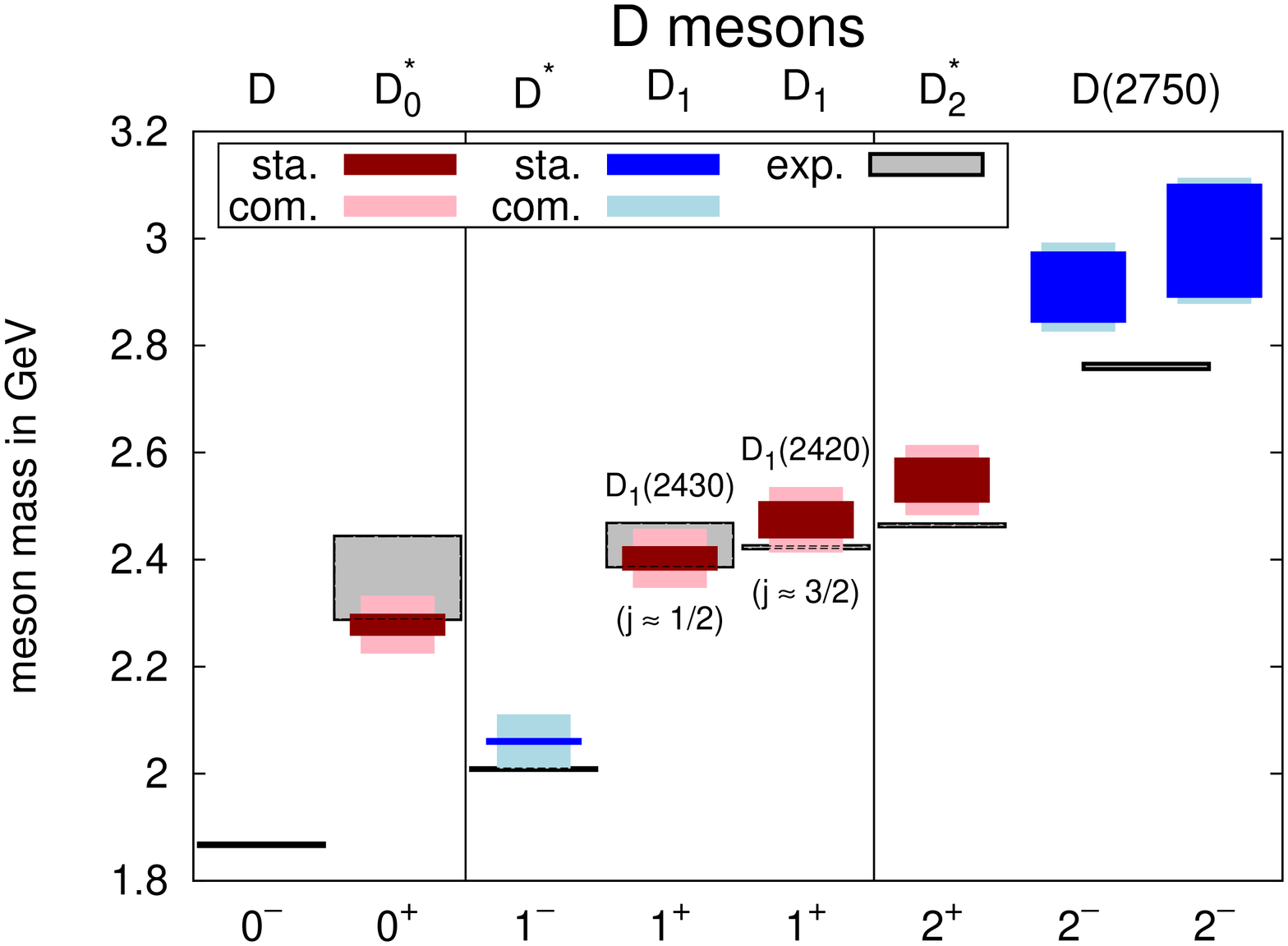}
\includegraphics[width=5.55cm,angle=-90]{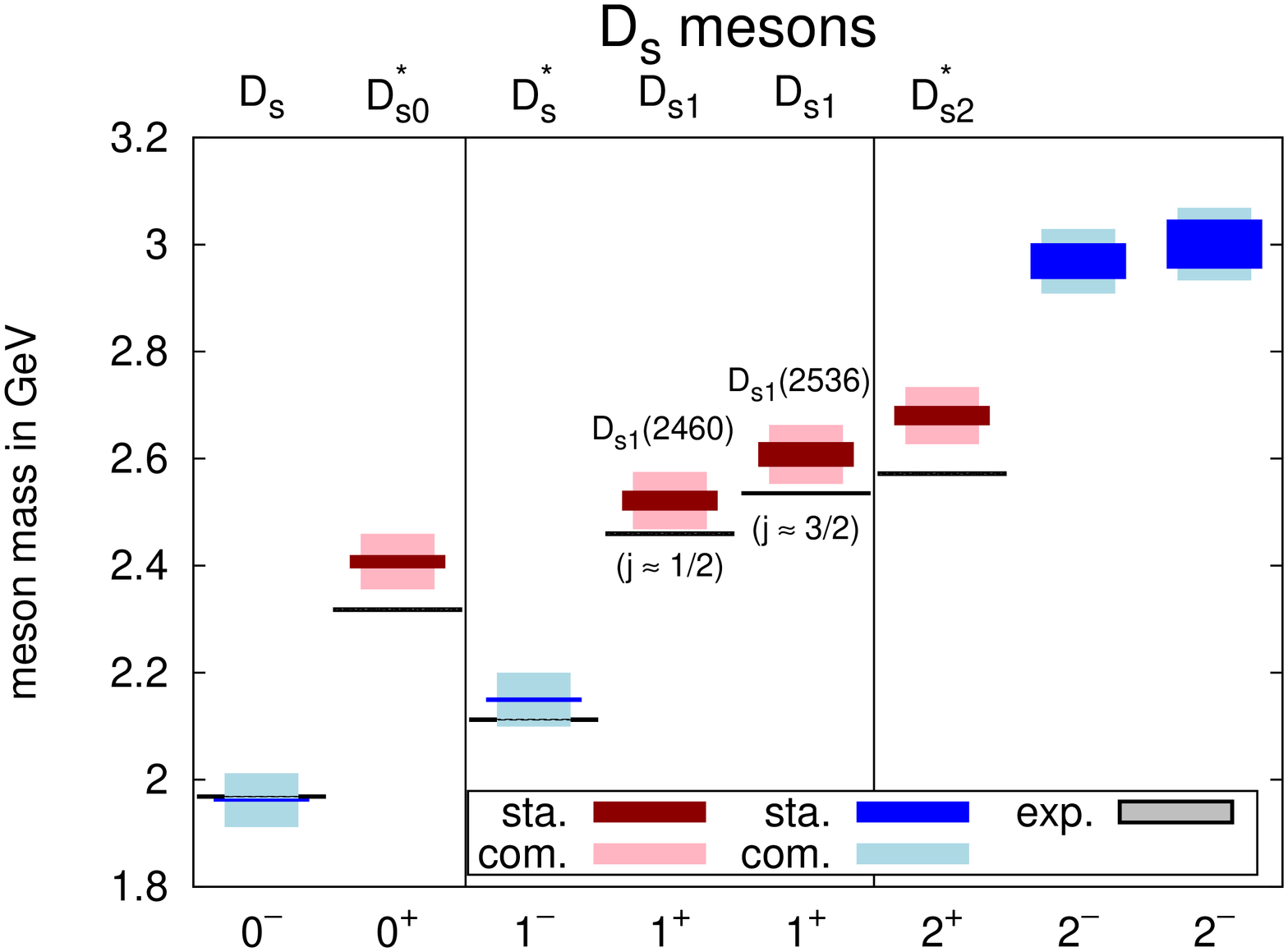}
\includegraphics[width=6.80cm,angle=-90]{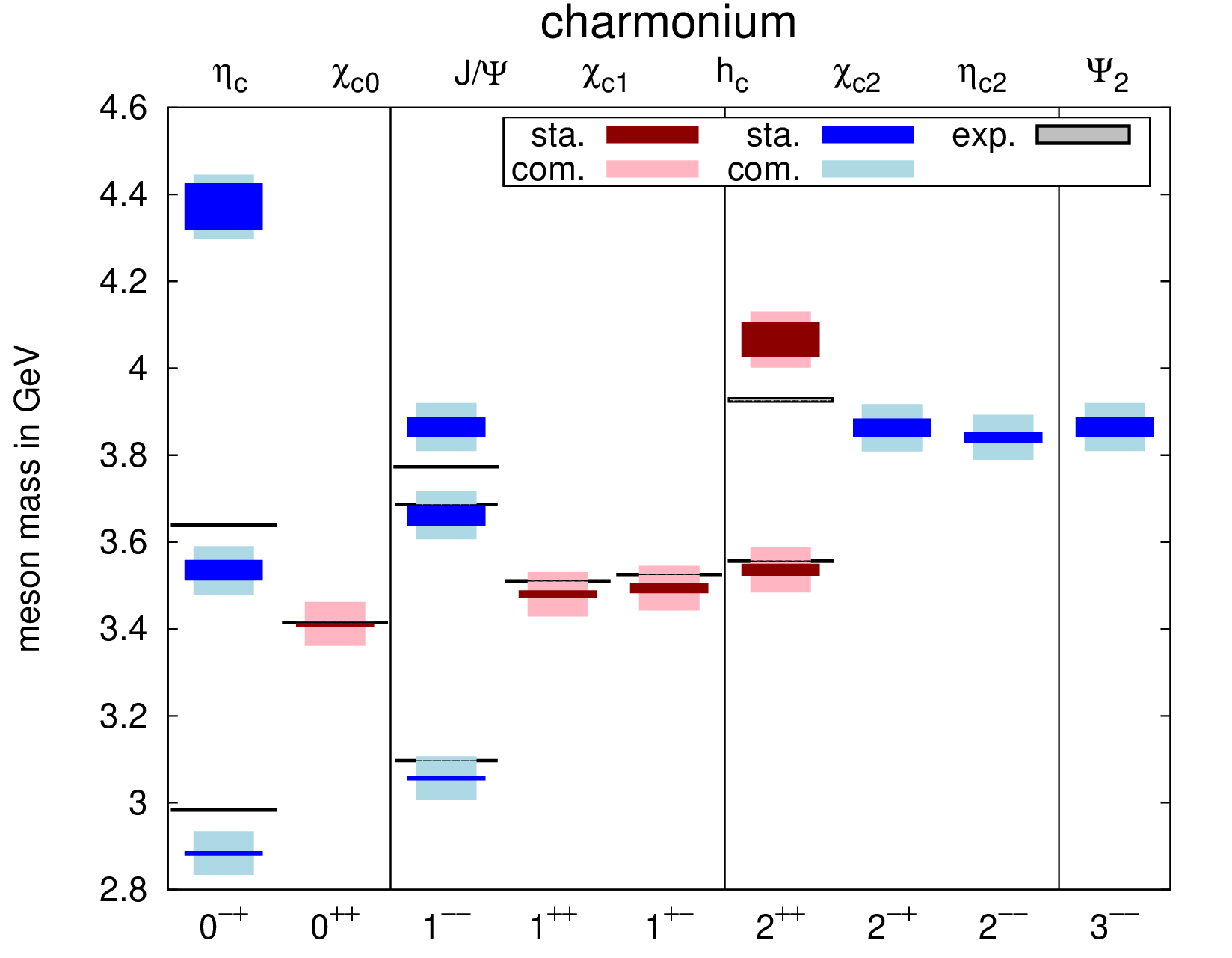}
\caption{\label{FIG140}Summary of lattice QCD results for $D$ meson, $D_s$ meson and charmonium masses ($\mathcal{P} = -$ states are shown in blue, $\mathcal{P} = +$ states in red; statistical errors are represented by dark blue/dark red boxes, combined statistical and systematic errors by light blue/light red boxes). For easy comparison experimental results from \cite{PDG} are shown in black.}
\end{figure}

The final results of this work for $D$ meson, $D_s$ meson and charmonium masses are shown in Figure~\ref{FIG140} and collected in Table~\ref{TAB140}. The numbers we quote have been obtained in the following way:
\begin{itemize}
\item We use the $u/d$ quark mass extrapolated results corresponding to $(\pm,\mp)$ twisted mass sign combinations, which are supposed to exhibit smaller discretization errors than their $(\pm,\pm)$ counterparts \cite{Urbach:2007rt,Frezzotti:2007qv}.

\item At the moment systematic errors are expected to be strongly dominated by discretization errors (cf.\ the detailed discussion above). Therefore, we estimate all systematic errors to be around or less than $50 \, \textrm{MeV}$, the previously mentioned typical difference between meson masses computed with $(\pm,\mp)$ and $(\pm,\pm)$ twisted mass sign combinations. Since we use $(\pm,\mp)$ results, we consider this estimate to be rather conservative.

\item We assume independence of statistical and systematic errors. The total error is, hence, obtained by adding statistical errors and the $50 \, \textrm{MeV}$ representing systematic errors in quadrature.

\item For $J = 2$ charmonium states we take the results from the $E$ representation, where a mixing with and contamination by $J = 3$ states is excluded due to cubic rotational symmetry (in contrast to results from the $T_2$ representation). For $D$ mesons and $D_s$ mesons we take the results from the $T_2$ representation, which exhibit smaller statistical errors than the corresponding results from the $E$ representation.

\item For the $J^{\mathcal{P C}} = 3^{- -}$ charmonium state we take the result from the $T_1$ representation, which has smaller statistical errors than the corresponding result from the $T_2$ representation.
\end{itemize}
In Figure~\ref{FIG140} $\mathcal{P} = -$ states are shown in blue, while $\mathcal{P} = +$ states are shown in red. Statistical errors are represented by dark blue and dark red boxes, while the combined statistical and systematic errors are represented by light blue and light red boxes, respectively. The relative combined statistical and systematic errors of our results are in most cases between $2 \%$ and $3 \%$. Within these errors our lattice QCD results agree with available experimental results \cite{PDG}, which are shown in gray.

\begin{table}[p]
\centering

\begin{tabular}{|c||c|c|c||c|c||c|c|c||c|c|}
\hline
\multirow{3}*{$J^\mathcal{P}$}&\multicolumn{5}{c||}{${{\mathit D}}$ mesons}&\multicolumn{5}{c|}{${{\mathit D}_s}$ mesons}   \\
\cline{2-11}
                              &lat.&\multicolumn{2}{c||}{error}&exp.&PDG&lat.&\multicolumn{2}{c||}{error}&exp.&PDG\\
\cline{3-4}\cline{8-9}
                              &mass&\multicolumn{2}{c||}{sta./com.}&mass&name&mass&\multicolumn{2}{c||}{sta./com.}&mass&name\\
\hline\hline
&&&&&&&&&& \vspace{-0.4cm}\\
\multirow{2}*{$0^-$}&\multirow{2}*{1865}&                  &                   &1870(0)\phantom0&${{\mathit D}^{\pm}}$                                &\multirow{2}*{1962}&\multirow{2}*{ 3}&\multirow{2}*{50}&\multirow{2}*{1968(0)}&\multirow{2}*{${{\mathit D}_{s}}$                     }\\& &&&1865(0)\phantom0&${{\mathit D}^{0}}$                        &&&&&\\&&&&&&&&&& \vspace{-0.3cm}\\
\multirow{2}*{$0^+$}&\multirow{2}*{2278}&\multirow{2}*{\phantom{0}20}          &\multirow{2}*{\phantom{0}54}&2403(40)&${{\mathit D}_{\mathrm 0}^{*}{(2400)}^{\pm}}$        &\multirow{2}*{2407}&\multirow{2}*{12}&\multirow{2}*{51}&\multirow{2}*{2318(1)}&\multirow{2}*{${{\mathit D}_{s\mathrm 0}^{*}{(2317)}}$}\\& &&&2318(29)&${{\mathit D}_{\mathrm 0}^{*}{(2400)}^{0}}$&&&&&\\\hline
&&&&&&&&&& \vspace{-0.4cm}\\
\multirow{2}*{$1^-$}&\multirow{2}*{2051}&\multirow{2}*{\phantom{0}\phantom{0}6}&\multirow{2}*{\phantom{0}50}&2010(0)\phantom0&${{\mathit D}^{*}{(2010)}^{\pm}}$                    &\multirow{2}*{2145}&\multirow{2}*{ 4}&\multirow{2}*{50}&\multirow{2}*{2112(0)}&\multirow{2}*{${{\mathit D}_{s}^{*}}$                    }\\& &&&2007(0)\phantom0&${{\mathit D}^{*}{(2007)}^{0}}$            &&&&&\\&&&&&&&&&& \vspace{-0.4cm}\\
\multirow{2}*{$1^+$}&\multirow{2}*{2402}&\multirow{2}*{\phantom{0}23}          &\multirow{2}*{\phantom{0}55}&2423(2)\phantom{0}&${{\mathit D}_{\mathrm 1}{(2420)}^{\pm}}$            &\multirow{2}*{2521}&\multirow{2}*{18}&\multirow{2}*{53}&\multirow{2}*{2460(1)}&\multirow{2}*{${{\mathit D}_{s\mathrm 1}{(2460)}}$    }\\& &&&2421(1)\phantom0&${{\mathit D}_{\mathrm 1}{(2420)}^{0}}$    &&&&&\\&&&&&&&&&& \vspace{-0.4cm}\\
\multirow{2}*{$1^+$}&\multirow{2}*{2474}&\multirow{2}*{\phantom{0}35}          &\multirow{2}*{\phantom{0}61}&        &                                                     &\multirow{2}*{2608}&\multirow{2}*{23}&\multirow{2}*{55}&\multirow{2}*{2535(0)}&\multirow{2}*{${{\mathit D}_{s\mathrm 1}{(2536)}}$    }\\& &&&2427(40)&${{\mathit D}_{\mathrm 1}{(2430)}^{0}}$    &&&&&\\\hline
&&&&&&&&&& \vspace{-0.4cm}\\
\multirow{2}*{$2^+$}&\multirow{2}*{2549}&\multirow{2}*{\phantom{0}42}          &\multirow{2}*{\phantom{0}65}&2464(2)\phantom{0}&${{\mathit D}_{\mathrm 2}^{*}{(2460)}^{\pm}}$        &\multirow{2}*{2680}&\multirow{2}*{18}&\multirow{2}*{53}&\multirow{2}*{2572(1)}&\multirow{2}*{${{\mathit D}_{s\mathrm 2}^{*}{(2573)}}$      }\\& &&&2463(1)\phantom0&${{\mathit D}_{\mathrm 2}^{*}{(2460)}^{0}}$&&&&&\\&&&&&&&&&& \vspace{-0.4cm}\\
\multirow{2}*{$2^-$}&\multirow{2}*{2909}&\multirow{2}*{\phantom{0}66}          &\multirow{2}*{\phantom{0}83}&\multirow{4}*{2761(5)\phantom{0}}&\multirow{4}*{${{\mathit D}{(2750)}}$}&\multirow{2}*{2969}&\multirow{2}*{33}&\multirow{2}*{60}&\multirow{2}*{        }&\multirow{2}*{                                              }\\& &&&&                                                   &&&&&\\&&&&&&&&&& \vspace{-0.4cm}\\
\multirow{2}*{$2^-$}&\multirow{2}*{2996}&\multirow{2}*{106}                    &\multirow{2}*{118}          &        &                                                     &\multirow{2}*{3001}&\multirow{2}*{46}&\multirow{2}*{68}&\multirow{2}*{        }&\multirow{2}*{                                              }\\& &&&&                                                   &&&&&\\\hline
\end{tabular}

\vspace{0.3cm}
\begin{tabular}{|c||c|c|c||c|c|}
\hline
\multirow{3}*{$J^\mathcal{P}$}&\multicolumn{5}{c|}{charmonium}\\
\cline{2-6}
                              &lattice&\multicolumn{2}{c||}{error}& exp. &PDG\\
\cline{3-4}
                              &mass&\multicolumn{2}{c||}{sta./com.}                & mass &name\\
\hline\hline
&&&&&\vspace{-0.4cm}\\
$0^{-+}$&2884&\phantom{0}4&50&2984(1)&${{\mathit \eta}_{c}{(1S)}}$           \\
$0^{-+}$&3535&23&55&3639(1)&${{\mathit \eta}_{c}{(2S)}}$                     \\
$0^{-+}$&4372&54&74&&                                                        \\&&&&&\vspace{-0.3cm}\\
$0^{++}$&3412&\phantom{0}6&50&3415(0)&${{\mathit \chi}_{c\mathrm 0}{(1P)}}$  \\\hline
&&&&&\vspace{-0.4cm}\\
$1^{--}$&3056&\phantom{0}5&50&3097(0)&${{\mathit J / \psi}{(1S)}}$           \\
$1^{--}$&3662&24&56&3686(0)&${{\mathit \psi}{(2S)}}$                         \\
$1^{--}$&3865&23&55&3773(0)&${{\mathit \psi}{(3770)}}$                       \\&&&&&\vspace{-0.3cm}\\
$1^{++}$&3480&\phantom{0}9&51&3511(0)&${{\mathit \chi}_{c\mathrm 1}{(1P)}}$            \\&&&&&\vspace{-0.2cm}\\
$1^{+-}$&3494&11&51&3525(0)&${{\mathit h}_{c}{(1P)}}$              \\\hline
&&&&&\vspace{-0.4cm}\\
$2^{++}$&3536&14&52&3556(0)&${{\mathit \chi}_{c\mathrm 2}{(1P)}}$            \\
$2^{++}$&4066&41&64&3927(3)&${{\mathit \chi}_{c\mathrm 2}{(2P)}}$            \\&&&&&\vspace{-0.3cm}\\
$2^{-+}$&3863&21&54&&                                                        \\&&&&&\vspace{-0.2cm}\\
$2^{--}$&3841&12&52&&                                                        \\\hline
$3^{--}$&3865&23&55&&                                                        \\
\hline
\end{tabular}

\caption{\label{TAB140}Summary of lattice QCD results for $D$ meson, $D_s$ meson and charmonium masses (error sta.: statistical error; error com.: combined statistical and systematic error) and comparison to experimental results from \cite{PDG}.}
\end{table}

One of the next steps will be a computation of the same meson masses on several gauge link ensembles at finer lattice spacings. This will enable us to perform a continuum extrapolation. Since the combined statistical and systematic errors are currently dominated by lattice discretization errors (crudely estimated by $50 \, \textrm{MeV}$), we expect that a continuum extrapolation will lead to significantly more precise results.

Our strategy of computing meson masses using quark-antiquark creation operators allows to obtain solid and accurate results for states, which are mainly composed of a quark and an antiquark and which are quite stable. States, which might not fulfill these requirements, e.g.\ the rather unstable $D_0^\ast$, which readily decays into $D + \pi$, or $D_{s0}^\ast$, which is frequently discussed as a tetraquark candidate, should finally be treated with more advanced lattice techniques. Further creation operators composed of four quarks (e.g.\ of mesonic molecule type, of diquark-antidiquark type and of two-meson type) have to be included in the correlation matrices. In case of unstable mesons, corresponding resonance parameters (mass, width) can then be extracted from the volume dependence of the spectrum of scattering states. We are currently in the process of developing and implementing such methods using a similar lattice setup \cite{Alexandrou:2012rm,Abdel-Rehim:2014zwa,Berlin:2015faa}.

Another important aspect of this work is the separation and classification of the two $J^\mathcal{P} = 1^+$ $D$ meson states. Even though they have identical quantum numbers, their structure is quite different: one of them, $D_1(2430)$, has $j \approx 1/2$, while the other, $D_1(2420)$, has $j \approx 3/2$, where $j$ denotes the spin and angular momentum of the light quark and gluons. Extracting those two states unambiguously from a single $J^\mathcal{P} = 1^+$ correlation matrix (both masses and eigenvector components, where the latter provide suitable linear combinations of $D_1(2430)$ and $D_1(2420)$ creation operators) constitutes an important first step to study decays $B^{(*)} \rightarrow D_1 + l + \nu$ using lattice QCD. Such a study is of particular interest, because there is a long standing conflict between theory and experiment regarding the corresponding branching ratios (``$1/2$ versus $3/2$ puzzle''). A solid understanding of these decays is in turn necessary for a precise determination of the standard model parameter $V_{c b}$ (cf.\ \cite{Bigi:2007qp} for a detailed discussion). The results and corresponding techniques discussed in section~\ref{SEC429} can be used to extend existing lattice computations of decays $B^{(*)} \rightarrow D^{\ast \ast} + l + \nu$, where $D^{\ast \ast}$ is currently limited to $J^\mathcal{P} = 0^+$ and $J^\mathcal{P} = 2^+$, but does not include the two $D_1$ states \cite{Atoui:2013sca,Atoui:2013ksa}.


\appendix


\section{\label{sec.computation}Computation of correlation matrices using the one-end trick}

To compute the elements of the correlation matrices defined in (\ref{EQN698}) we first insert the definition of the meson creation operators (\ref{EQN507}),
\begin{eqnarray}
\nonumber & & \hspace{-0.7cm} C_{\Gamma_j;\Gamma_k;\bar{\chi}^{(1)} \chi^{(2)}}(t) \ \ = \\
\nonumber & & = \ \ \frac{1}{V/a^3} \langle \Omega | \bigg(\sum_\mathbf{r} (\bar{\chi}^{(1)} S)(\mathbf{r},t) \sum_{\Delta \mathbf{r} = \pm \mathbf{e}_x , \pm \mathbf{e}_y , \pm \mathbf{e}_z} U(\mathbf{r};\mathbf{r}+\Delta \mathbf{r};t) \Gamma_j(\Delta \mathbf{r}) (S \chi^{(2)})(\mathbf{r}+\Delta \mathbf{r},t)\bigg)^\dagger \\
\nonumber & & \hspace{0.675cm} \sum_\mathbf{s} (\bar{\chi}^{(1)} S)(\mathbf{s},0) \sum_{\Delta \mathbf{s} = \pm \mathbf{e}_x , \pm \mathbf{e}_y , \pm \mathbf{e}_z} U(\mathbf{s};\mathbf{s}+\Delta \mathbf{s};0) \Gamma_k(\Delta \mathbf{s}) (S \chi^{(2)})(\mathbf{s}+\Delta \mathbf{s},0) | \Omega \rangle ,
\end{eqnarray}
where $U$ denote APE smeared gauge links. Gaussian smearing is a linear operation on the quark fields and can, therefore, be written in terms of a matrix $S$ in color and position space, where $S = S^\dagger$. After writing the vacuum expectation value $\langle \Omega | \ldots | \Omega \rangle$ as a path integral and integrating over the quark fields, one obtains an average over gauge link configurations (denoted by $\langle \ldots \rangle$), which includes quark propagators $(D^{(f)})^{-1}$ ($f \in \{ u,d,s^+,s^-,c^+,c^- \}$ is the quark flavor),
\begin{eqnarray}
\nonumber & & \hspace{-0.7cm} C_{\Gamma_j;\Gamma_k;\bar{\chi}^{(1)} \chi^{(2)}}(t) \ \ = \\
\nonumber & & = \ \ -\frac{1}{V/a^3} \sum_\mathbf{r} \sum_{\Delta \mathbf{r} = \pm \mathbf{e}_x , \pm \mathbf{e}_y , \pm \mathbf{e}_z} \sum_\mathbf{s} \sum_{\Delta \mathbf{s} = \pm \mathbf{e}_x , \pm \mathbf{e}_y , \pm \mathbf{e}_z} \\
\nonumber & & \hspace{0.675cm} \bigg\langle \textrm{Tr}_\textrm{spin,color}\bigg(\Gamma_k(\Delta \mathbf{s}) \gamma_5 U(\mathbf{s};\mathbf{s}+\Delta \mathbf{s};0) (S (D^{(\bar{2})})^{-1,\dagger} S)(\mathbf{s}+\Delta \mathbf{s},0;\mathbf{r}+\Delta \mathbf{r},t) U(\mathbf{r}+\Delta \mathbf{r};\mathbf{r};t) \\
 & & \hspace{0.675cm} \gamma_5 \gamma_0 \Gamma_j^\dagger(\Delta \mathbf{r}) \gamma_0 (S (D^{(1)})^{-1} S)(\mathbf{r},t;\mathbf{s},0)\bigg)\bigg\rangle .
\end{eqnarray}
We have also used twisted mass $\gamma_5$ hermiticity, $(D^{(f)})^{-1} = \gamma_5 (D^{(\bar{f})})^{-1,\dagger} \gamma_5$, where the bar on top of the flavor index indicates a flip of the sign in front of the twisted mass term (e.g.\ if $2$ denotes $u$, $\bar{2}$ denotes $d$, if $2$ denotes $s^+$, $\bar{2}$ denotes $s^-$, etc.). Finally, it is convenient to rearrange the expression and to write spin indices $A,B,C,D$ explicitly,
\begin{eqnarray}
\nonumber & & \hspace{-0.7cm} C_{\Gamma_j;\Gamma_k;\bar{\chi}^{(1)} \chi^{(2)}}(t) \ \ = \\
\nonumber & & = \ \ -\frac{1}{V/a^3} \sum_{\Delta \mathbf{r} = \pm \mathbf{e}_x , \pm \mathbf{e}_y , \pm \mathbf{e}_z} (\gamma_5 \gamma_0 \Gamma_j^\dagger(\Delta \mathbf{r}) \gamma_0)_{AB} \sum_{\Delta \mathbf{s} = \pm \mathbf{e}_x , \pm \mathbf{e}_y , \pm \mathbf{e}_z} (\Gamma_k(\Delta \mathbf{s}) \gamma_5)_{CD} \\
\nonumber & & \hspace{0.675cm} \sum_\mathbf{r} \sum_\mathbf{s} \bigg\langle \textrm{Tr}_\textrm{color}\bigg(U(\mathbf{s};\mathbf{s}+\Delta \mathbf{s};0) (S (D^{(\bar{2})})^{-1,\dagger} S)_{DA}(\mathbf{s}+\Delta \mathbf{s},0;\mathbf{r}+\Delta \mathbf{r},t) U(\mathbf{r}+\Delta \mathbf{r};\mathbf{r};t) \\
\label{EQN586} & & \hspace{0.675cm} (S (D^{(1)})^{-1} S)_{BC}(\mathbf{r},t;\mathbf{s},0)\bigg)\bigg\rangle .
\end{eqnarray}

To estimate (\ref{EQN586}) stochastically, we generate for each gauge link configuration $4$ spin diluted stochastic timeslice sources (index $B = 1,2,3,4$) on a randomly chosen timeslice (here w.l.o.g.\ at time $t=0$),
\begin{eqnarray}
\label{EQN677} \xi_A^{a,(B)}(\mathbf{r},t) \ \ = \ \ \delta_{t,0} \delta_{A B} N^a(\mathbf{r})
\end{eqnarray}
($a$ is a color index, $A$ is a spin index), where the entries of the noise vector $N^a(\mathbf{r})$ are randomly and uniformly chosen numbers $\pm 1/\sqrt{2} \pm i / \sqrt{2}$. Then we solve the $4$ linear systems (index $C = 1,2,3,4$)
\begin{eqnarray}
\nonumber & & \hspace{-0.7cm} \sum_y D^{(1)}(x;y) \phi^{(1,C)}(x) \ \ = \ \ (S \xi^{(C)})(y) \\
 & & \hspace{-0.7cm} \rightarrow \quad \phi^{(1,C)}(\mathbf{r},t) \ \ = \ \ \sum_x ((D^{(1)})^{-1} S)(\mathbf{r},t;x) \xi^{(C)}(x)
\end{eqnarray}
and the $4 \times 6$ linear systems (indices $D = 1,2,3,4$ and $\Delta \mathbf{s} = \pm \mathbf{e}_x , \pm \mathbf{e}_y , \pm \mathbf{e}_z$)
\begin{eqnarray}
\nonumber & & \hspace{-0.7cm} \sum_y D^{(\bar{2})}(x;y) \phi^{(\bar{2},D)}(y) \ \ = \ \ \sum_y S(x;y+(0,\Delta \mathbf{s})) U(y+(0,\Delta \mathbf{s});y) \xi^{(D)}(y) \\
\nonumber & & \hspace{-0.7cm} \rightarrow \quad \phi^{(\bar{2},D,\Delta \mathbf{s})}(\mathbf{r},t) \ \ = \ \ \sum_y ((D^{(\bar{2})})^{-1} S)(\mathbf{r},t;y+(0,\Delta \mathbf{s})) U(y+(0,\Delta \mathbf{s});y) \xi^{(D)}(y) \\
 & & \hspace{-0.7cm} \rightarrow \quad (\phi^{(\bar{2},D,\Delta \mathbf{s})})^\dagger(\mathbf{r},t) \ \ = \ \ \sum_y (\xi^{(D)})^\dagger(y) U(y;y+(0,\Delta \mathbf{s})) (S (D^{(\bar{2})})^{-1,\dagger})(y+(0,\Delta \mathbf{s});\mathbf{r},t)
\end{eqnarray}
with respect to $\phi$ using \cite{Jansen:2009xp} (to minimize the required computation time, we always choose the quark flavors such, that the lighter quark corresponds to flavor $1$ and the heavier to quark to flavor $\bar{2}$). Using these results we compute the quantities
\begin{eqnarray}
X_{\bar{\chi}^{(1)} \chi^{(2)}}^{A B C D;\Delta \mathbf{r} \Delta \mathbf{s}}(t) \ \ \equiv \ \ \sum_\mathbf{r} \bigg\langle ((\phi^{(\bar{2},D,\Delta \mathbf{s})})^\dagger S)_A(\mathbf{r}+\Delta \mathbf{r},t) U(\mathbf{r}+\Delta \mathbf{r};\mathbf{r};t) (S \phi^{(1,C)})_B(\mathbf{r},t) \bigg\rangle_\textrm{MC}
\end{eqnarray}
($\langle \ldots \rangle_\textrm{MC}$ denotes the average over the finite number of Monte Carlo generated gauge link configurations [cf.\ Table~\ref{tab.ensembles}]), for which one can show
\begin{eqnarray}
\nonumber & & \hspace{-0.7cm} X_{\bar{\chi}^{(1)} \chi^{(2)}}^{A B C D;\Delta \mathbf{r} \Delta \mathbf{s}}(t) \ \ = \\
\nonumber & & = \ \ \sum_\mathbf{r} \sum_\mathbf{s} \bigg\langle \textrm{Tr}_\textrm{color}\bigg(U(\mathbf{s};\mathbf{s}+\Delta \mathbf{s};0) (S (D^{(\bar{2})})^{-1,\dagger} S)_{DA}(\mathbf{s}+\Delta \mathbf{s},0;\mathbf{r}+\Delta \mathbf{r},t) \\
\label{EQN733} & & \hspace{0.675cm} U(\mathbf{r}+\Delta \mathbf{r};\mathbf{r};t) (S (D^{(1)})^{-1} S)_{BC}(\mathbf{r},t;\mathbf{s},0)\bigg) \bigg\rangle + \textrm{noise} ,
\end{eqnarray}
where
\begin{eqnarray}
\Big\langle N^a(\mathbf{r}) (N^b)^\ast(\mathbf{s}) \Big\rangle_\textrm{MC} \ \ = \ \ \delta^{a b} \delta_{\mathbf{r},\mathbf{s}} + \textrm{noise}
\end{eqnarray}
has been used and ``noise'' denotes unbiased stochastic noise, which decreases proportional to $1/\sqrt{N}$, where $N$ is the number of gauge link configurations used to compute $\langle \ldots \rangle_\textrm{MC}$. This technique of stochastic estimation is referred to as one-end trick (cf.\ e.g.\ \cite{Foster:1998vw,McNeile:2006bz}). It is most efficient for large spatial volumes and light quark masses. For the computations of $D$ mesons, $D_s$ mesons and charmonium states done in this work it has been found to be superior compared to the traditional technique of using point sources and point-to-all propagators \cite{Abdel-Rehim:2014zwa}.

To obtain arbitrary elements of the correlation matrices (\ref{EQN698}), one simply has to combine (\ref{EQN586}) and (\ref{EQN733}),
\begin{eqnarray}
\nonumber & & \hspace{-0.7cm} C_{\Gamma_j;\Gamma_k;\bar{\chi}^{(1)} \chi^{(2)}}(t) \ \ = \\
\nonumber & & = \ \ -\frac{1}{V/a^3} \sum_{\Delta \mathbf{r} = \pm \mathbf{e}_x , \pm \mathbf{e}_y , \pm \mathbf{e}_z} (\gamma_5 \gamma_0 \Gamma_j^\dagger(\Delta \mathbf{r}) \gamma_0)_{AB} \sum_{\Delta \mathbf{s} = \pm \mathbf{e}_x , \pm \mathbf{e}_y , \pm \mathbf{e}_z} (\Gamma_k(\Delta \mathbf{s}) \gamma_5)_{CD} X_{\bar{\chi}^{(1)} \chi^{(2)}}^{A B C D;\Delta \mathbf{r} \Delta \mathbf{s}}(t) . \\
 & &
\end{eqnarray}



\section*{Acknowledgments}

It is a pleasure to thank V.~O.~Galkin for many useful discussions. Moreover, we acknowledge useful conversations with J.~Berlin, B.~Blossier, K.~Cichy, F.~Giacosa, M.~F.~M.~Lutz, O.~P\`ene and D.~H.~Rischke.

We acknowledge support by the Emmy Noether Programme of the DFG (German Research Foundation), grant WA 3000/1-1.

This work was supported in part by the Helmholtz International Center for FAIR within the framework of the LOEWE program launched by the State of Hesse.

Calculations on the LOEWE-CSC and on the on the FUCHS-CSC high-performance computer of the Frankfurt University were conducted for this research. We would like to thank HPC-Hessen, funded by the State Ministry of Higher Education, Research and the Arts, for programming advice.




\end{document}